\documentclass[pra,10pt,twocolumn,superscriptaddress,showpacs]{revtex4-2}

\usepackage{epsfig}
\usepackage{amsmath}
\usepackage{graphicx}
\usepackage{graphics}
\usepackage{float}
\usepackage{amssymb}
\usepackage{natbib}
\usepackage{epstopdf}
\usepackage{wrapfig}
\usepackage{mathtools}
\usepackage{xcolor}
\RequirePackage{tcolorbox}
\RequirePackage[outline]{contour}
\usepackage{amsfonts}
\usepackage{dsfont}
\usepackage{bm}
\usepackage{physics}
\usepackage{verbatim}
\usepackage{microtype}
\renewcommand{\bra}[1]{\left\langle #1\right|}
\renewcommand{\ket}[1]{\left| #1\right\rangle}

\usepackage[english]{babel}

\usepackage{hyperref}
 \hypersetup{
     colorlinks=true,
     linkcolor=blue,
     filecolor=magenta,      
     urlcolor=blue,
     citecolor=blue}

\usepackage{subfigure}
\usepackage{empheq}

\definecolor{fore}{RGB}{249,242,215}
\definecolor{myblue}{rgb}{.8, .8, 1}
\definecolor{forshading}{RGB}{185,217,255}
\newcommand*{\boxedcolor}{blue}
\renewcommand{\boxed}[1]{\textcolor{\boxedcolor}{\fbox{\normalcolor\m@th$\displaystyle#1$}}}
\makeatother

\begin{document}

\title{Enhancement of non-Markovianity due to environment-induced indirect interaction}

\author{Asif Zaman}
\affiliation{Department of Physics, Lahore University of Management Sciences, Lahore 54792, Pakistan}

\author{Muhammad Faryad}
\affiliation{Department of Physics, Lahore University of Management Sciences, Lahore 54792, Pakistan}

\author{Adam Zaman Chaudhry}
\email{adam.zaman@lums.edu.pk}
\affiliation{Department of Physics, Lahore University of Management Sciences, Lahore 54792, Pakistan}

\begin{abstract}
Non-Markovian effects are often significant when the system-environment coupling is not weak. Indeed, we find that the non-Markovianity is negligible for a single two-level system undergoing pure dephasing via a weak interaction with a harmonic-oscillator environment. In this paper, we show that, within the framework of pure dephasing, the non-Markovianity displayed by a two-level system can, in fact, be far more pronounced. To demonstrate that this is indeed the case, we consider a pure dephasing model where a collection of two-level systems interacts with a common environment. We obtain analytically the dynamics of the collection of the two-level systems, and then take a partial trace over all the two-level systems except one. This remaining single two-level system exhibits markedly non-Markovian dynamics, even when the system-environment coupling is weak. This is due to the indirect interaction between the two-level systems, induced by their interaction with the common environment. In fact, this indirect interaction can not only increase the non-Markovianity by orders of magnitude, but also qualitatively change the characteristics of the non-Markovian behavior. For instance, for a single two-level system undergoing pure dephasing, the dynamics are Markovian for both Ohmic and sub-Ohmic environments. This is markedly not the case when we consider multiple two-level systems. These findings provide insights into controlling decoherence in multi-qubit quantum systems and have implications for quantum technologies where non-Markovianity can be a resource rather than a limitation.

\end{abstract}

\pacs{03.65.Yz,03.65.Ta,03.65.-w}

\maketitle

\section{Introduction}
The study of open quantum systems is crucial to the advancement of modern quantum technologies \cite{nielsen2010quantum,haroche2014exploring}. Any realistic quantum system unavoidably interacts with its environment, leading to decoherence as it builds correlations, such as entanglement, with its surroundings \cite{schlosshauer2007decoherence,breuer2002theory}. Understanding and quantifying memory effects in such systems is essential to mitigate decoherence in quantum devices \cite{RevModPhys.89.015001,RevModPhys.86.1203,bylicka2014non,PhysRevLett.109.170402,PhysRevA.104.012213,10.3389/frqst.2023.1134583,10.3389/frqst.2023.1207552,Mouloudakis_2023,doi:10.1021/acsomega.3c09720,CANGEMI20241,Link_2023}. For weak system-environment coupling, the environment has negligible memory, and information generally flows irreversibly from the system to the environment, leading to Markovian dynamics. In contrast, non-Markovian dynamics arise when information flows back from the environment to the system for some duration, causing memory effects that make future system states dependent on past interactions \cite{PhysRevLett.109.170402,PRXQuantum.6.020316}. Non-Markovianity plays a crucial role in various physical phenomena, including quantum biology \cite{Cao:2020pup}, quantum metrology \cite{PhysRevLett.127.060501}, quantum computing \cite{PhysRevResearch.6.033215}, and quantum communication \cite{white2020demonstration,caruso2014quantum,PhysRevResearch.6.043127,Kamin_2020,e26090742,Roy_2024}. Specifically, in the context of quantum computing, errors such as leakage from the computational subspace, fluctuations in control fields and qubit properties, as well as imperfections in the physical sample, lead to quantum dynamics that break the assumptions of the Markovian approximation \cite{PhysRevB.109.014315}. This type of behavior, referred to as non-Markovian, necessitates dedicated techniques for its correction, modeling, and analysis \cite{PRXQuantum.2.040351,PhysRevLett.132.200401}.
As such, several measures have been proposed to quantify non-Markovianity, based on, for example, the trace distance \cite{PhysRevLett.103.210401,PhysRevA.106.042212,PhysRevA.98.012120}, relative entropy changes \cite{PhysRevA.81.062115,Rivas_2014,PhysRevA.106.042212,PhysRevLett.127.030401}, moments of the Choi state as well as uncertainty relations \cite{PhysRevA.109.022247,Maity_2020}, and the quantum Fisher information (QFI) flow \cite{PhysRevA.82.042103,Rashid_2024,abiuso2023characterizing,el2020different}. Many studies have explored non-Markovianity for one as well as two two-level systems \cite{Rashid_2024,PhysRevA.90.052103,addis2013two}, investigating how it depends on system-environment parameters. A particularly interesting finding is that coupling two two-level systems to a common environment can significantly enhance QFI if one qubit is traced out \cite{mirza2024improving}. This enhancement arises from an indirect qubit-qubit interaction mediated by the shared environment. Other studies have highlighted the role played by structured environments in enhancing non-Markovianity \cite{SerraPRA2020, AllatiPRE2024, AllatiPRA2025}. Such structured environments are often modelled as a small quantum system interacting with a large environment; the system of interest interacts only with the small quantum system. Yet another interesting work has highlighted that a Markovian noise term in the Hamiltonian can in fact lead to non-Markovian dynamics \cite{KurtEntropy2023}.

Motivated by these works, our objective in this work is simple. We expect that for a single two-level system (TLS) interacting weakly with its environment, non-Markovian effects will be weak. If we instead consider a collection of two-level systems interacting with a common environment, is the non-Markovianity enhanced? To answer this question, we consider $N$ TLSs undergoing pure dephasing due to their coupling with a common environment. We analytically compute the density matrix for $N$ TLSs, then trace out $(N-1)$ TLSs, leaving a single TLS state for analysis. Due to the interaction of the TLSs with the common environment, an indirect interaction between the TLSs is induced. Our primary goal is to systematically investigate how two different measures of non-Markovianity - based on the trace distance and the relative entropy - are affected by this indirect interaction. We find that the presence of multiple TLSs drastically changes the non-Markovianity. To highlight a few significant differences, we note that for a single two-level system undergoing pure dephasing, the non-Markovianity is zero for sub-Ohmic and Ohmic environments; it is small in the weak system-environment coupling regime; and it saturates as time increases. The non-Markovianity when we have multiple TLSs is markedly different: it is decidedly non-zero for Ohmic and sub-Ohmic environments, it can be non-negligible even in the weak system-environment coupling regime, and it can keep on increasing with time. In fact, the non-Markovianity generally increases by orders of magnitude due to the indirect interaction. These effects can be further enhanced by tuning the environment parameters, and may also be increased by increasing the number of TLSs. We also note that our work differs from related work done using structured reservoirs \cite{SerraPRA2020, AllatiPRE2024, AllatiPRA2025} - in our study, the central TLS does not interact directly with any other TLS but instead interacts only with the environment of harmonic oscillators.

This paper is organized as follows. In Sec.~II, we present a simple yet general formalism to calculate two different non-Markovianity measures for a TLS undergoing pure dephasing, taking into account the possible effect of an indirect interaction due to coupling with a common environment. In Sec.~III, we apply this formalism to a model of $N$ TLSs interacting with a common environment of harmonic oscillators and conclusively show the drastic influence of the indirect interaction via two non-Markovianity measures. We then conclude in Sec.~IV. Some technical derivations and details, as well as results obtained using a third measure of non-Markovianity, are given in the appendixes.

\section{Non-Markovianity in pure dephasing models}

We start by considering an arbitrary initial state of a two-level system (TLS) undergoing pure dephasing due to its interaction with an environment. This means that the TLS Hamiltonian commutes with the system-environment Hamiltonian. Then, in the eigenbasis of the Hamiltonian of the TLS, only the off-diagonal elements of the density matrix of the TLS evolve. Given this constraint on the time evolution, we discuss in the main text the evaluation of the BLP (Breuer, Laine, Piilo) measure \cite{PhysRevLett.103.210401,PhysRevA.106.042212,PhysRevA.98.012120}, and the non-Markovianity found via the relative entropy \cite{PhysRevA.81.062115,Rivas_2014,PhysRevA.106.042212,PhysRevLett.127.030401} since these two measures will turn out to show similar behavior. In Appendix \ref{RHPappendix}, we also discuss the RHP (Rivas, Huelga, Plenio) measure \cite{RivasPRL2010,Rivas_2014}.

\subsection{BLP Measure}
The BLP measure of non-Markovianity quantifies the evolution of the maximum possible trace distance between the density matrices corresponding to two different initial states. An increasing trace distance is a signature of non-Markovian evolution. For a TLS undergoing pure dephasing, the maximum trace distance is obtained if, in the Bloch sphere picture, the two initial states for this TLS lie in the equatorial plane and have opposite Bloch vectors. Without loss of generality, we can choose the initial states to have Bloch vectors $(1,0,0)$ and $(-1,0,0)$. The time-evolved density matrices must then be of the form 
\[
\rho_1(t) = \frac{1}{2}\begin{pmatrix} 
1 & e^{-i\omega_0 t} f(t) \\
e^{i\omega_0 t} f(t) & 1
\end{pmatrix},
\]
\[
\rho_2(t) = \frac{1}{2}\begin{pmatrix} 
1 & -e^{-i\omega_0 t} f(t)  \\
-e^{i\omega_0 t} f(t)  & 1
\end{pmatrix},
\]
where $f(t)$ is typically an exponentially decaying factor that describes the dephasing (without loss of generality, we are assuming $f(t)$ to be real and positive), and $\omega_0$ is the natural frequency of the TLS.

Our key insight is that if we have two TLS, interacting with a common environment, then the evolution of one of these TLS is affected by the indirect interaction between them that arises due to the fact that they are interacting with a common environment. We still assume pure dephasing. The effect of the indirect interaction is to then modify the time evolution of the off-diagonal elements. In particular, we now write the time-evolved quantum states as 
\[
\rho_1(t) = \frac{1}{2}\begin{pmatrix} 
1 & e^{-i\omega_0 t} f(t) g(t) e^{-i\chi(t)} \\
e^{i\omega_0 t} f(t) g(t) e^{i\chi(t)} & 1
\end{pmatrix},
\]
\[
\rho_2(t) = \frac{1}{2}\begin{pmatrix} 
1 & -e^{-i\omega_0 t} f(t) g(t) e^{-i\chi(t)} \\
- e^{i\omega_0 t} f(t) g(t) e^{i\chi(t)} & 1
\end{pmatrix}.
\]
The off-diagonals now contain an additional factor $g(t)$ (which we take to be real and positive) as well as a possible time-dependent phase factor. These encapsulate the possible effect of the indirect interaction. Our idea, then, is that this indirect interaction may lead to much stronger non-Markovian effects than in the case where it is absent. To investigate this, let us calculate the BLP measure, accounting for the indirect interaction. Defining 
$$\alpha(t) = f(t) g(t) e^{-i[\omega_0 t + \chi(t)]}, $$ 
we can write
\[
\rho_1 - \rho_2 = \begin{pmatrix}
0 & \alpha(t) \\
\alpha^*(t) & 0
\end{pmatrix}.
\]
The eigenvalues of this Hermitian matrix are \( \pm |\alpha(t)| \). Therefore, the trace distance becomes
\[
D(\rho_1, \rho_2) = |\alpha(t)| = f(t) g(t).
\]
Notice that this result is independent of the phase $\chi(t)$. To proceed, we now set, without loss of generality, $f(t) = e^{-\Gamma(t)}$.
Then the time derivative of the trace distance is
\begin{align}
\sigma (t)  = e^{-\Gamma(t)} \left[ \dot{g}(t) - \dot{\Gamma}(t)\, g(t) \right]
\label{trace}
\end{align} 
Since the BLP measure is
\begin{align*}
\mathcal{N}_{\text{BLP}}= \underset{\rho_{1,2}(0)}{\text{max}}\int_{\sigma (t) > 0} dt \, \sigma(t,\rho_{1,2}(0)), \label{BLP}
\end{align*}
we see from Eq.~\eqref{trace} that if
\begin{equation}
\dv{g}{t} > \dv{\Gamma}{t}\, g(t),
\label{nonmarkovianitycondition}
\end{equation}
then non-Markovian effects arise. For the scenario without any indirect interaction, $g(t) = 1$, leading to the simpler condition $\frac{d\Gamma}{dt} < 0$. The presence of a non-trivial $g(t)$ means that the evolution can possibly be non-Markovian for a longer duration. To compute the BLP measure, we find the time intervals for which the condition in Eq.~\eqref{nonmarkovianitycondition} holds and calculate the change in the trace distance over these intervals.

\subsection{Relative Entropy}
We now look at a different measure of non-Markovianity, namely one that utilizes the relative entropy \cite{PhysRevA.81.062115,Rivas_2014,PhysRevA.106.042212,PhysRevLett.127.030401}. An increasing relative entropy between the states $\rho_1(t)$ and $\rho_2(t)$ is a signature of non-Markovianity. By using a unitary transformation, we compute the relative entropy between these states as (taking into account the effect of the indirect interaction) 
\[
S[\rho_1(t)\|\rho_2(t)] = D(t) \log \left( \frac{1 + D(t)}{1 - D(t)} \right),
\]
where $D(t) = f(t)g(t)$ is the previously computed trace distance. 
As before, taking $f(t) = e^{-\Gamma(t)}$, we find that 
\begin{align}
\dv{S}{t} &= e^{-\Gamma(t)} \left[ \frac{dg}{dt} - \frac{d\Gamma}{dt}\, g(t) \right] \notag  \\
&\left[ \log \left( \frac{1 + D}{1 - D} \right) + \frac{2 D}{1 - D^2} \right]. \label{entropy}    
\end{align}
For a non-Markovian process, 
\begin{align} 
    \mathcal{N}_{\text{Entropy}}= \int_{\frac{dS}{dt} > 0}\frac{dS}{dt} dt. 
\end{align}
Then from Eq.~\eqref{entropy} we see that if
\[
\dv{g}{t} > \dv{\Gamma}{t}\, g(t),
\]
then non-Markovian effects arise. This is exactly the same condition as the one that we obtained via the trace distance. Note, however, that the two measures of non-Markovianity will generally not be numerically equal to each other.

\section{Enhanced non-Markovianity in a system of multiple two-level systems}
\label{model}

To apply our simple yet general formalism to see whether or not the indirect interaction can boost non-Markovianity, we consider $N$ TLSs interacting with a common harmonic oscillator environment. We refer to the $N$ TLSs as our `system', find the dynamics of this system, and then take the partial trace over $(N-1)$ TLSs to find the dynamics of a single TLS. From these dynamics, we can read off $f(t)$ and $g(t)$ and thereby compute the non-Markovianity measures. These can then be contrasted with the scenario in which a single TLS interacts with the harmonic oscillator environment (in which case $g(t) = 1$). 

To work out the dynamics, we write the total Hamiltonian as $H  = H_S + H_B + H_{SB}$, where (we take $\hbar = 1$ throughout)
$$
\begin{aligned}
H_{S} &= \frac{\omega_0}{2} \sum_{i=1}^{N} \sigma_z^{(i)}, \\
H_{B} &=\sum_{k} \omega_{k} b_{k}^{\dagger} b_{k}, \\
H_{SB} &= \sum_{i=1}^{N} \sigma_z^{(i)} \sum_{k} \left( g_{k}^* b_{k} + g_{k}b_{k}^{\dagger} \right).
\end{aligned}
$$
Here $H_S$ corresponds to the Hamiltonian for the TLSs alone, and we have assumed that each TLS has the same energy spacing $\omega_0$. They are labeled via the superscript $i$, and $\sigma_z$ is the standard Pauli operator. $H_B$ is the harmonic oscillator environment Hamiltonian with the usual creation and annihilation operators (we have dropped the zero-point energy for convenience). Finally, $H_{SB}$ corresponds to the system-environment interaction. Note that this is a pure dephasing model, which is applicable when, as is usually the case, the decoherence timescales are much shorter than the relaxation timescales \cite{breuer2002theory,ChaudhryPRA2013,Rashid_2024}. To proceed, it is convenient to first transform to the interaction picture via the unitary operator $U_0(t)=e^{-i \left(H_S+H_B\right) t}$. We then obtain
\[
H_{SB}(t) = \sum_{i=1}^{N} \sigma_z^{(i)} \sum_{k} \left( g_{k}^{*} b_{k} e^{-i \omega_k t} + g_{k} b_{k}^{\dagger} e^{i \omega_k t} \right).
\]
Using the Magnus expansion, this leads to the total exact unitary time-evolution operator in the original frame (see Appendix \ref{derivation} for a detailed derivation)
\begin{widetext}
\begin{equation}
    U(t) =  \exp[-i (H_S + H_B)t]  \exp \left[\frac{1}
    {2}\sum_{i=1}^{N}\sigma_z^{(i)}\sum_k ( \alpha_k b_k^\dagger - \alpha_k^* b_k) - \frac{i}{2} \Delta(t) \sum_{\substack{i,j \\ (i < j)}}^{N} \sigma_z^{(i)} \sigma_z^{(j)}  \right],
\end{equation} 
\end{widetext}
with 
\begin{equation}
\alpha_k\left( t \right)
    =\frac{2g_k\left(1-e^{i\omega_k t} \right)}{\omega_k},
    \end{equation} 
    and  
    \begin{equation}
    \displaystyle{\Delta(t) 
    =\sum_k   \frac{4\abs{g_k}^2}{\omega_k^2}\left[\sin(\omega_k t) - \omega_k t\right]}.\label{delta}
\end{equation} 
We will see that $\alpha_k(t)$ will lead to decoherence, while the term $\Delta(t)$ describes an indirect interaction between the TLSs induced by the common environment.   

We now obtain the reduced density operator of the system of $N$ TLSs via $\rho_S \left(t\right) 
        =\text{Tr}_B \left[ U(t) \rho\left(0\right)U^\dagger(t) \right]$, where $\text{Tr}_B$ denotes a partial trace over the environment. To perform this partial trace, it is convenient to consider the joint eigenbasis of the $\sigma_z^{(i)}$ operators. We first define the states $\ket{n_i}$ as $\sigma_z^{(i)}\ket{n_i} = n_i \ket{n_i}$, with $n_i = \pm 1$. It then follows that 
\begin{equation}
\sum_{i=1}^{N} \sigma_z^{(i)} \ket{n}  = \left[ \sum_{i = 1}^N n_i \right] \ket{n}, 
\label{basisstates}
\end{equation}
where
\[
\ket{n} = \ket{n_1 n_2 \ldots n_N}.
\]
We also define $s_n = \sum_{i = 1}^N n_i$. Given a particular joint eigenbasis $\ket{n}$, we can find the corresponding number $s_n$. To find the matrix elements of the reduced density matrix in the joint eigenbasis, we now introduce the operator $P_{nn'} = \ket{n}\bra{n'}$. We then find that (see Appendix \ref{derivation} for the detailed derivation) 
\begin{align}
\left[\rho_S(t)\right]_{n'n} &= e^{-i\omega_0 t (s_{n'} - s_n)/2} e^{-\frac{i}{4} \Delta(t) (s_{n'}^2 - s_n^2)} \notag \\
&\times \text{Tr}\left(\rho(0) e^{-R_{n'n}} P_{nn'} \right)
\label{Gen.state} 
\end{align}
where
\[
R_{n'n} = \frac{1}{2}(s_n - s_{n'})\sum_{k} \left(\alpha_k(t)  b_k^\dagger - \alpha_k^*(t) b_k\right).
\]
To proceed further, we need to specify the total initial state. For simplicity, we assume that the total initial state is an uncorrelated product state. In other words, denoting the total initial state as $\rho(0)$, we have
\begin{align}
    \rho(0) 
    = \rho_S(0) \otimes \rho_B,
\end{align}
with the environment in a thermal equilibrium state given by $\rho_B 
= \frac{e^{-\beta H_B}}{Z_B} \  \text{and}  \  Z_B = \text{Tr}_B \left[ e^{-\beta H_B} \right].$ From Eq.~\eqref{Gen.state}, we then have 
\begin{align}
\left[\rho_S(t)\right]_{n'n} &= \left[\rho_S(0)\right]_{n'n} 
e^{-i\omega_0 t (s_{n'} - s_n)/2} \nonumber e^{-\frac{i}{4} \Delta(t) \left(s_{n'}^2 - s_n^2\right)} \nonumber \\
&\times e^{-\frac{1}{4} (s_{n'} - s_n)^2 \Gamma(t)}
\label{ch5-4}
\end{align}
where
\begin{equation}
\label{gammafactor}
\Gamma(t)= \sum_{k} \frac{4|g_k|^2}{\omega_k^2} [1 - \cos (\omega_k t)] \coth \left( \frac{\beta \omega_k}{2} \right),
\end{equation}
and $\Delta(t)$ has been defined before in Eq.~\eqref{delta}. Note that $\Gamma\left(t\right)$ describes decoherence, while $\Delta(t)$ describes the indirect interaction between the TLSs. 

With the density matrix for the $N$ TLSs available, we now turn our attention to the density matrix of only one of these TLSs. Since we are looking at the non-Markovianity, the most obvious choice of the initial state of the TLSs is to choose $\rho_S\left(0\right)=\ket{+++\dots +}\bra{+\dots +++},$ where $\ket{+} = \frac{1}{\sqrt{2}}(\ket{1} + \ket{-1})$ (remember that $\sigma_z \ket{1} = \ket{1}$ and $\sigma_z\ket{-1} = -\ket{-1}$). Now taking a partial trace over $(N-1)$ TLSs, the state of the remaining TLS is (see Appendix \ref{derivation})
\begin{align}
\rho_1(t)
= \frac{1}{2}
\begin{pmatrix}
    1 
    & \eta \\
    \eta^*
    & 1
\end{pmatrix},\label{evoldstat}
\end{align}
where $\eta = e^{-i\omega_0 t - \Gamma\left(t\right)}\left(\cos\left[\Delta\left(t\right)\right]\right)^{N-1}$.
On the other hand, if every TLS starts from the initial state $\ket{-} = \frac{1}{\sqrt{2}}(\ket{1} - \ket{-1})$, the state of the single TLS (after taking the partial trace over $(N - 1)$ TLSs) is 
\begin{align}
\label{evoldstat2}
    \rho_2(t)     
    = \frac{1}{2}
\begin{pmatrix}
    1 
    & - \eta \\
   - \eta^*     & 1
\end{pmatrix}.
\end{align}
\begin{figure}[t!]
\centering
\subfigure[]
{\includegraphics[width=0.48\textwidth]{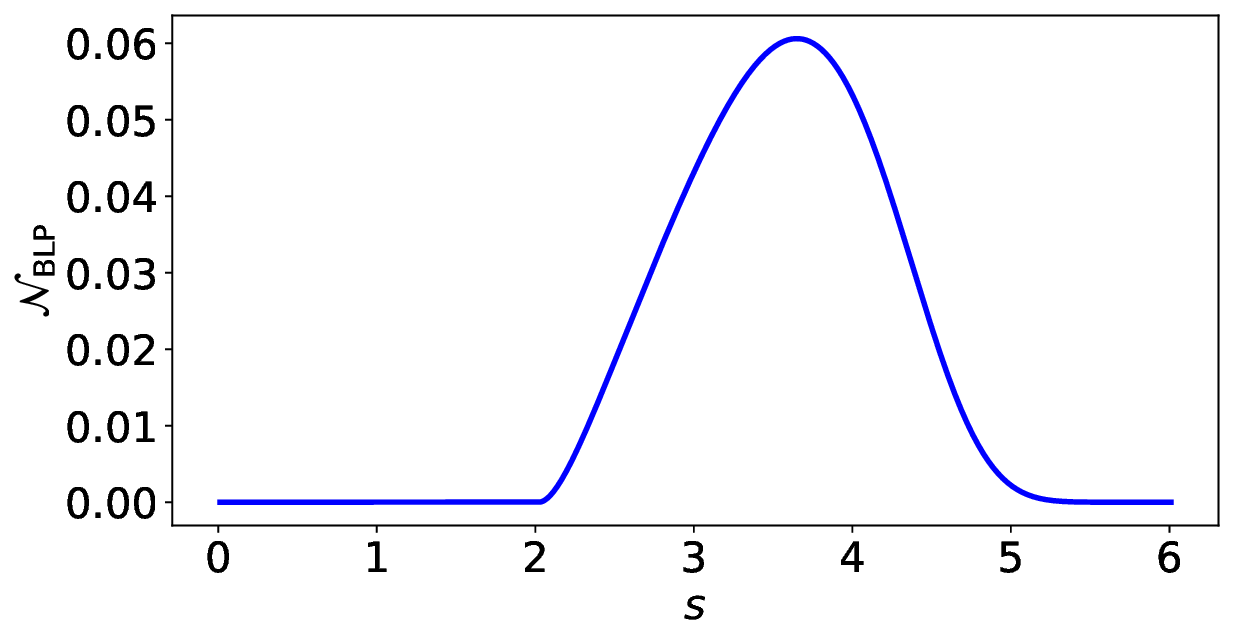}}
\subfigure[]{\includegraphics[width=0.48\textwidth]{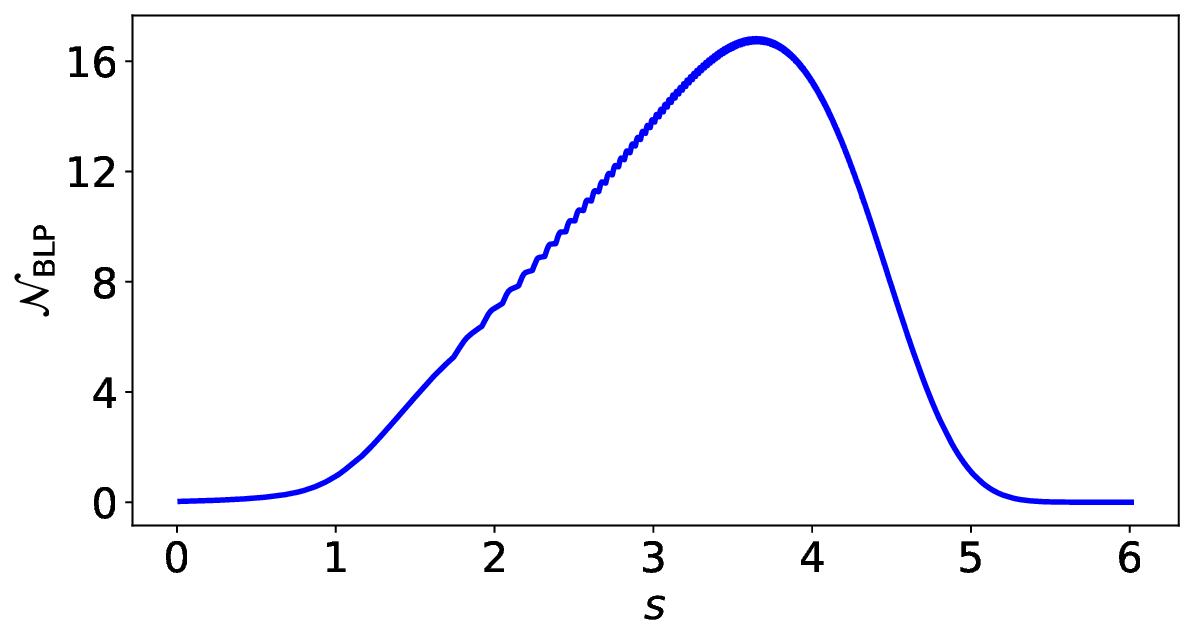}}
\caption{(a) Plot of the BLP measure of non-Markovianity, $\mathcal{N}_{\text{BLP}}$, as a function of the Ohmicity parameter $s$ for a single TLS (that is, $N = 1$). We are working in dimensionless units with $\hbar = 1$ throughout, and here we have set $G = 1$ and $\omega_c = 3$. Throughout this paper, unless stated otherwise, we are in the zero temperature regime. To calculate the non-Markovianity measure $\mathcal{N}_{\text{BLP}}$, we need to integrate over time; here we have set the upper limit of this integral to be $T = 20$. (b) Same as (a), but using two TLSs and then tracing out one of these. The dynamics are now influenced by an indirect interaction.}
\label{BLP vs s}
\end{figure}
Note that the same $\rho_2(t)$ is obtained if the initial state of the TLSs is $\ket{-++\cdots +}$ - see Appendix \ref{derivation}. It is then clear that, for this model, $f(t) = e^{-\Gamma(t)}$ and $g(t) = \left|\left(\cos\left[\Delta(t)\right]\right)^{N-1}\right| = \sqrt{\left[\frac{1}{2}\left(1 + \cos[2\Delta(t)]\right)\right]^{N-1}}$. Consequently, to compute the non-Markovianity measures, we need to compute $\Gamma(t)$ and $\Delta(t)$. This can be done by introducing the spectral density $J(\omega)$, which encapsulates the effect of the environment. This function effectively converts the sum over the environment modes to an integral via $\displaystyle{\sum_k 4 \abs{g_k}^2 (\hdots) \rightarrow \int_0^\infty d\omega \, J\left(\omega\right) (\hdots)}$. The spectral density is often assumed to be of the form 
$J\left(\omega\right)
= G\omega^s \omega_c^{1-s}e^{-\omega/\omega_c}$ \cite{schlosshauer2007decoherence}. Here $G$ is the coupling strength, $\omega_c$ is the cutoff frequency, and $s$ is the Ohmicity parameter with $s<1$, $s=1$, and $s>1$ representing sub-Ohmic, Ohmic, and super-Ohmic spectral densities, respectively. In the zero temperature limit that we will be focusing on, 
\begin{align*}
    \Gamma(t) &= \frac{G}{2}\ln\left(1+ \omega^2_c t^2\right)
\end{align*}
if $s = 1$, and 
\begin{align*}
&\Gamma(t) = G \, \times \, \text{Gamma($s-1$)} \, \times \, \\
&\left[1 - \left(1 + \omega_c^2 t^2 \right)^{\frac{1-s}{2}} \cos \left\{ (s-1) \tan^{-1} (\omega_c t) \right\} \right]   
\end{align*}
if $s \neq 1$. Also, 
\begin{align*}
\Delta(t) = G \left[\tan^{-1}(\omega_c t) - \omega_c t \right] 
\end{align*}
for $s = 1$, and 
\begin{align*}
&\Delta(t)= G\,\times \text{Gamma}(s-1) \, \times \\
&\Bigg[(1+\omega_c^{2} t^{2})^{-s/2}
\sin\!\left[s\arctan(\omega_c t)\right] - 
(1+\omega_c^{2} t^{2})^{-s/2} \times \\
&\omega_c t \cos\!\left[s\arctan(\omega_c t)\right]
-
(s-1)\omega_c t
\Bigg]
\end{align*}
for $s \neq 1$. Here $\text{Gamma}(z)$ is the usual gamma function defined as Gamma(z)$=\int_{0}^{\infty} t^{z-1} e^{-t} dt$. Before moving on, let us also note that if we have a single TLS interacting with the harmonic oscillator environment, then $g(t) = 1$ and $f(t) = e^{-\Gamma(t)}$.

\begin{figure}[t!]
\centering
    \subfigure[]
{\includegraphics[width=0.48\textwidth]{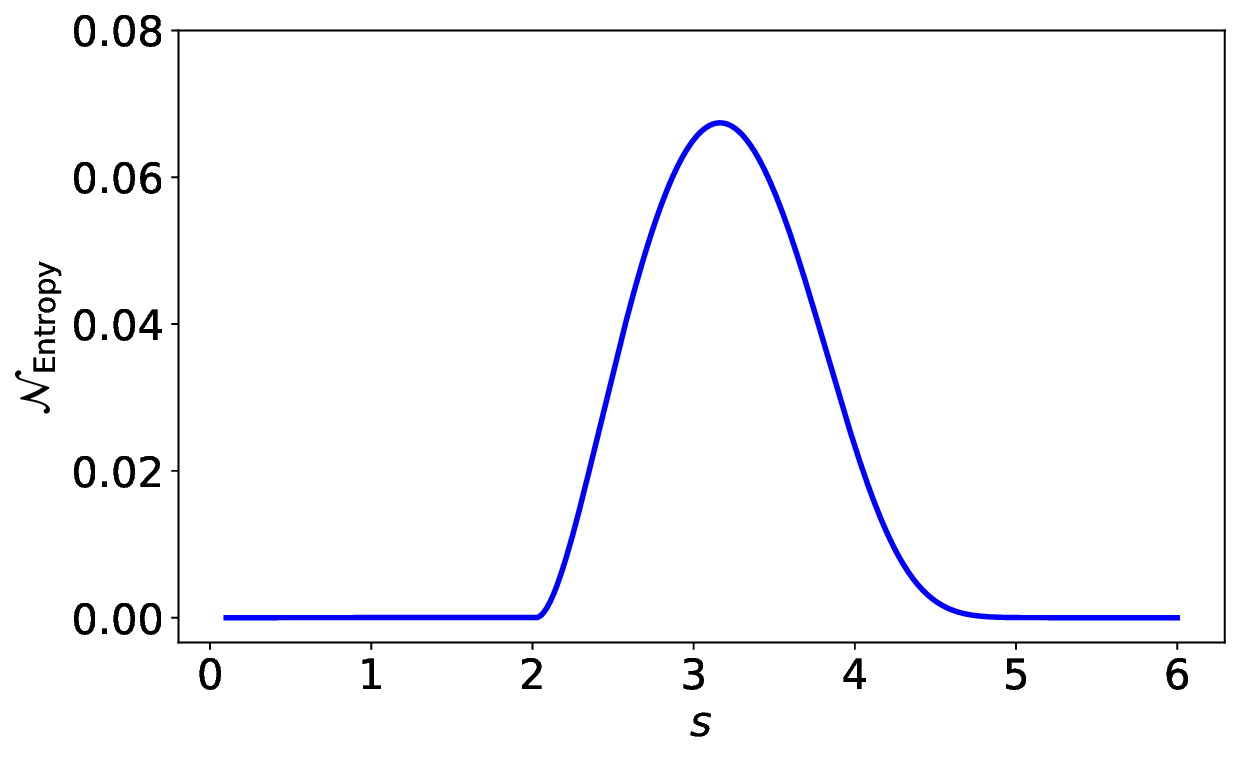}}
\subfigure[]{\includegraphics[width=0.48\textwidth]{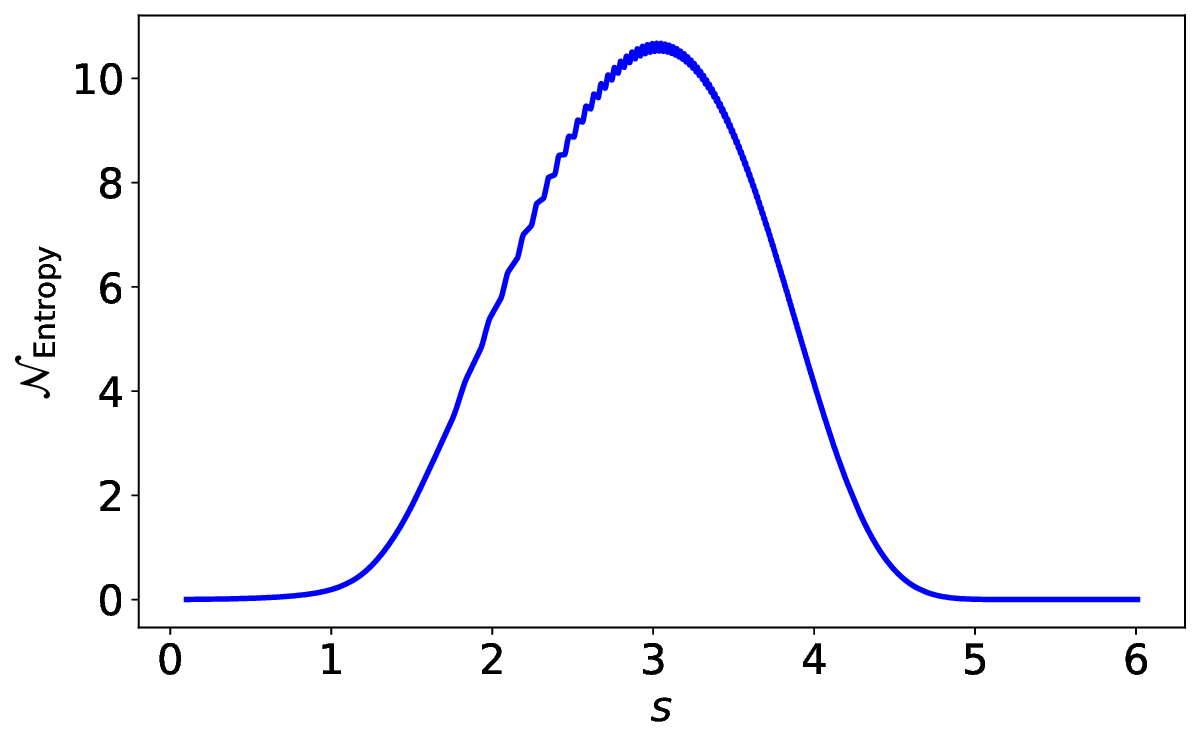}}
\caption{Same as Fig.~\ref{BLP vs s}, except that we are now looking at the non-Markovianity measure $\mathcal{N}_{\text{Entropy}}$. Once again, (a) shows the non-Markovianity measure for a single TLS, while (b) includes the effect of the indirect interaction.}
\label{entropy vs s}
\end{figure}

We now have everything we need to compute the non-Markovianity measures. We first investigate the behavior of $\mathcal{N}_{\text{BLP}}$ as a function of the Ohmicity parameter $s$. Typical results are illustrated in Fig.~\ref{BLP vs s}. In Fig.~\ref{BLP vs s}(a), we show $\mathcal{N}_{\text{BLP}}$ for a single TLS interacting with the environment, while Fig.~\ref{BLP vs s}(b) considers two TLSs interacting with a common environment (we trace out one TLS). We can immediately see the extremely large effect of the indirect interaction - the non-Markovianity measure has increased by orders of magnitude. This is not a peculiarity of the non-Markovianity measure $\mathcal{N}_{\text{BLP}}$; qualitatively similar behavior is obtained using the non-Markovianity measure $\mathcal{N}_{\text{Entropy}}$ (see Fig.~\ref{entropy vs s}). Moreover, for both measures, the values of the Ohmicity parameter $s$ for which the non-Markovianity is non-zero change due to the presence of the indirect interaction. For example, for a single TLS, the non-Markovianity for an Ohmic environment is zero, but this is not the case when the indirect interaction is present. We also note that prior work, such as Ref.~\cite{addis2013two}, does not focus on a single TLS within multiple TLSs interacting with a common environment. By doing so, we are able to find much more pronounced non-Markovian behavior.

\begin{figure}[t]
\centering
\includegraphics[width=0.48\textwidth]{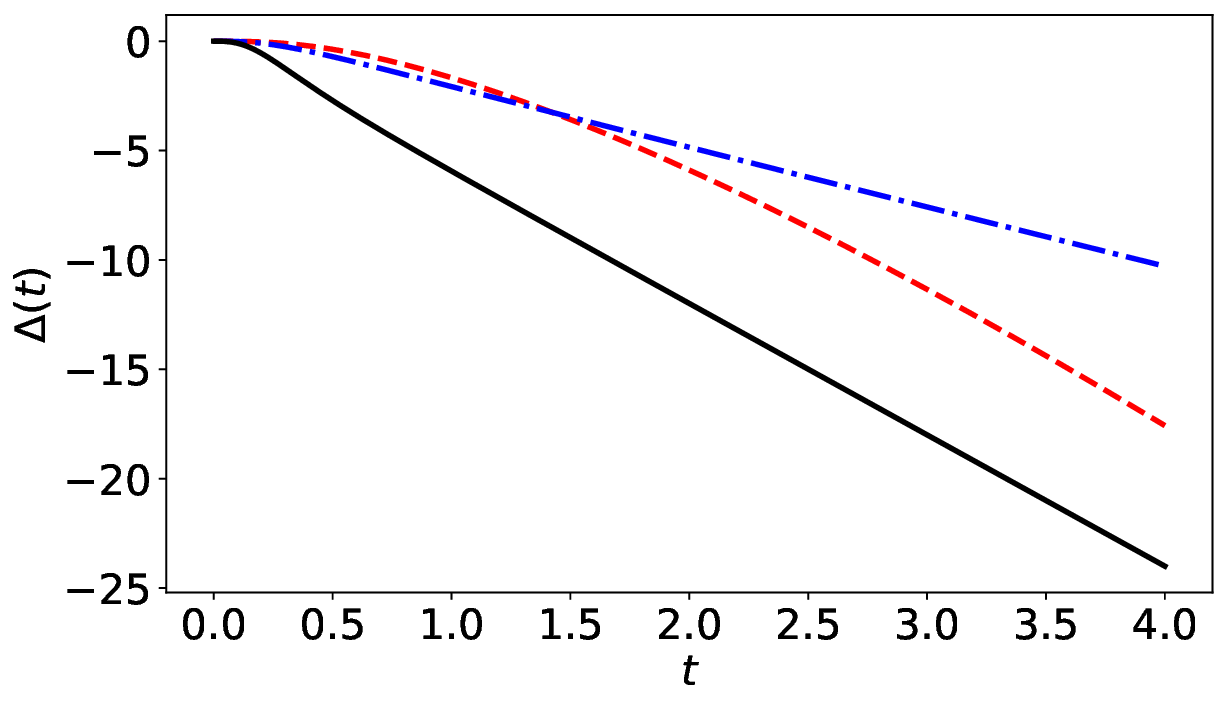}
\caption{Time evolution of the indirect-interaction induced factor $\Delta(t)$ for $s=0.1$ (red dashed), $s=1.5$ (blue dashed-dotted), and $s=3$ (black solid), with $G=1.0$ and $\omega_c=3$.
For $s = 0.1$, $\Delta(t)$ does not become a linear function at long times, but it does become linear for $s = 1.5$ and $s = 3$.}
\label{Delta-vs-t}
\end{figure}

\begin{figure}[t]
\centering
\includegraphics[width=0.48\textwidth]{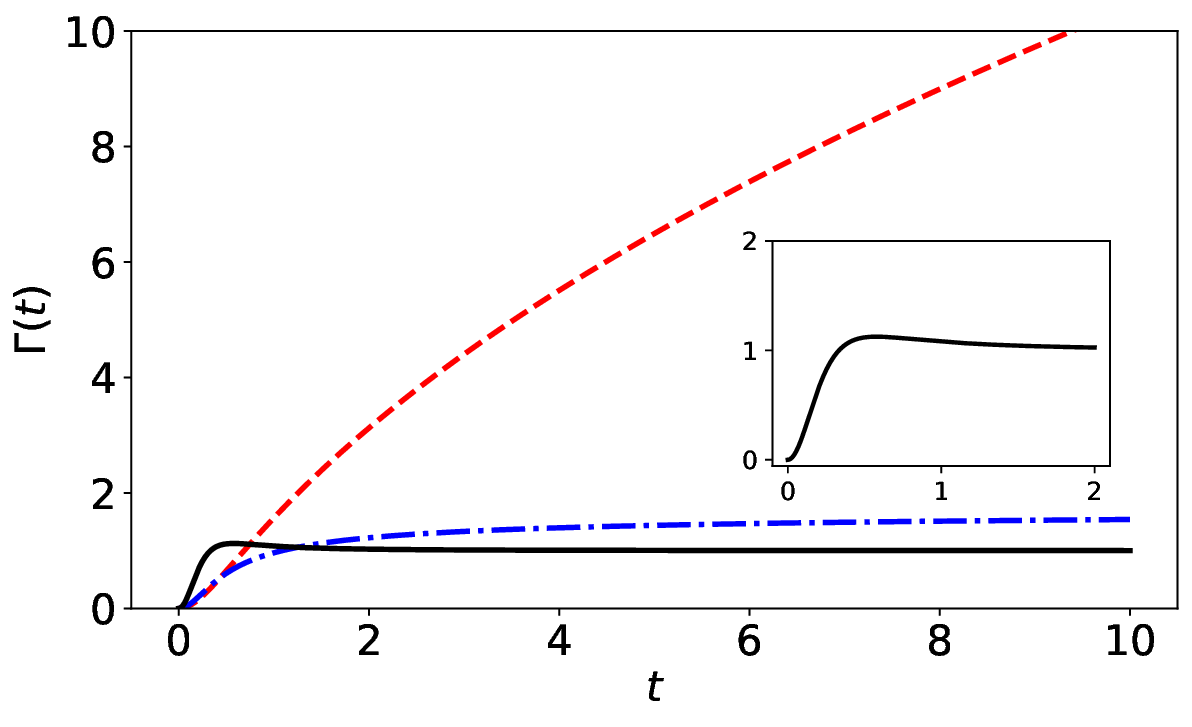}
\caption{Time evolution of the decoherence function $\Gamma(t)$ with $s=0.1$ (red dashed), $s=1.5$ (blue dashed-dotted), and $s=3$ (solid black), for coupling strength $G=1$ and cutoff frequency $\omega_c=3$. For $s = 0.1$, $\Gamma(t)$ increases without bound, while $\Gamma(t)$ approaches a constant value for $s = 1.5$ and $s = 3$ at longer times. The inset shows the behavior of $\Gamma(t)$ for $s = 3$ at smaller times to emphasize that $\Gamma(t)$ can decrease over some time interval for $s = 3$. It is precisely this decrease that leads to non-Markovianity for a single TLS without any indirect interaction. As $s$ increases further, the behavior of $\Gamma(t)$ remains qualitatively the same as for $s = 3$.}
\label{Gamma-vs-t}
\end{figure}

\begin{figure}[t!]
    \centering
    \subfigure[]{\includegraphics[width=0.48\textwidth]{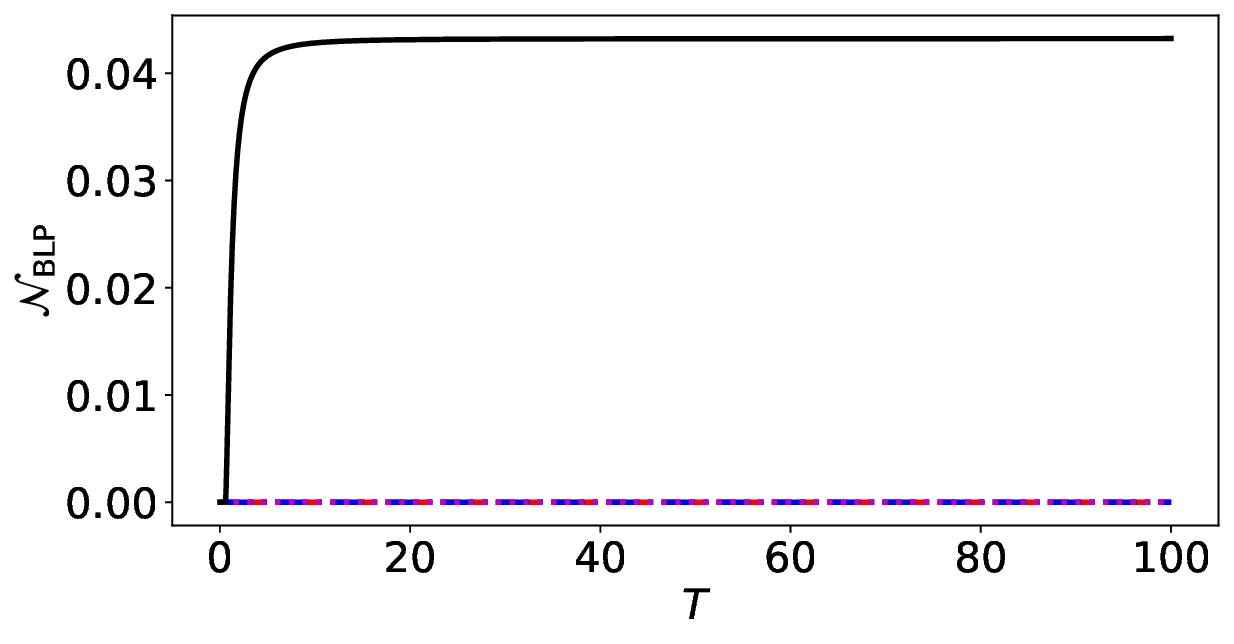}}
    \subfigure[]
    {\includegraphics[width=0.48\textwidth]{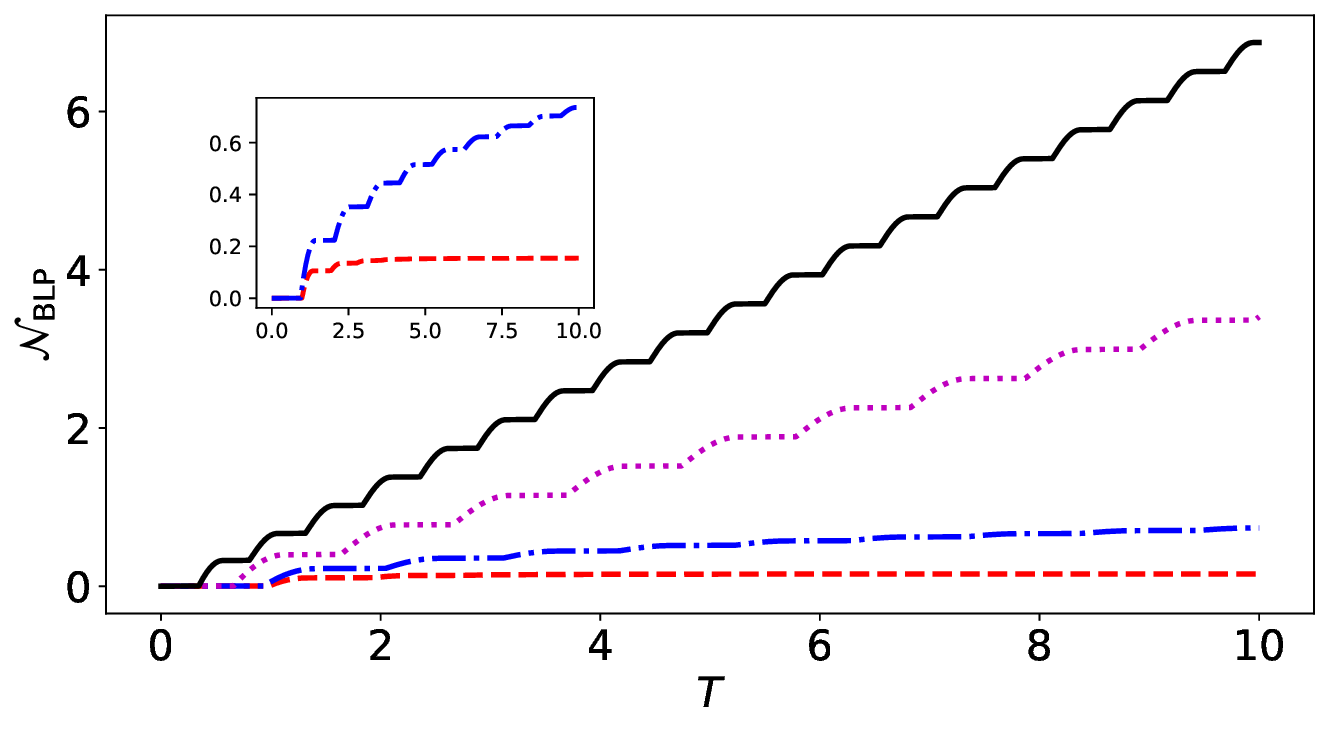}}    
    \caption{Non-Markovianity measure $\mathcal{N}_{\text{BLP}}$ as a function of the total evolution time $T$. In (a), we have only a single TLS, so there is no indirect interaction. In (b), we have taken $N = 2$, so indirect interaction is present.  As before, we are working in dimensionless units with $\hbar = 1$. We have $G=1$ and $\omega_c=3$. The solid black curve is for $s=3$, the dotted-magenta curve is for $s=2$, the dashed-dotted blue curve is for $s=1$, and the red-dashed curve is for $s=0.5$. In the inset, the same dashed-dotted blue curve is for $s=1$, and the red-dashed curve is for $s=0.5$. The non-Markovianity for the Ohmic environment saturates very slowly since the decoherence factor $\Gamma(t)$ grows relatively slowly; this saturation is clearer if we use a larger value of $G$.}
    \label{fig:NMBLP_vs_T}
\end{figure}

\begin{figure}[t!]
    \centering
    \subfigure[]{\includegraphics[width=0.48\textwidth]{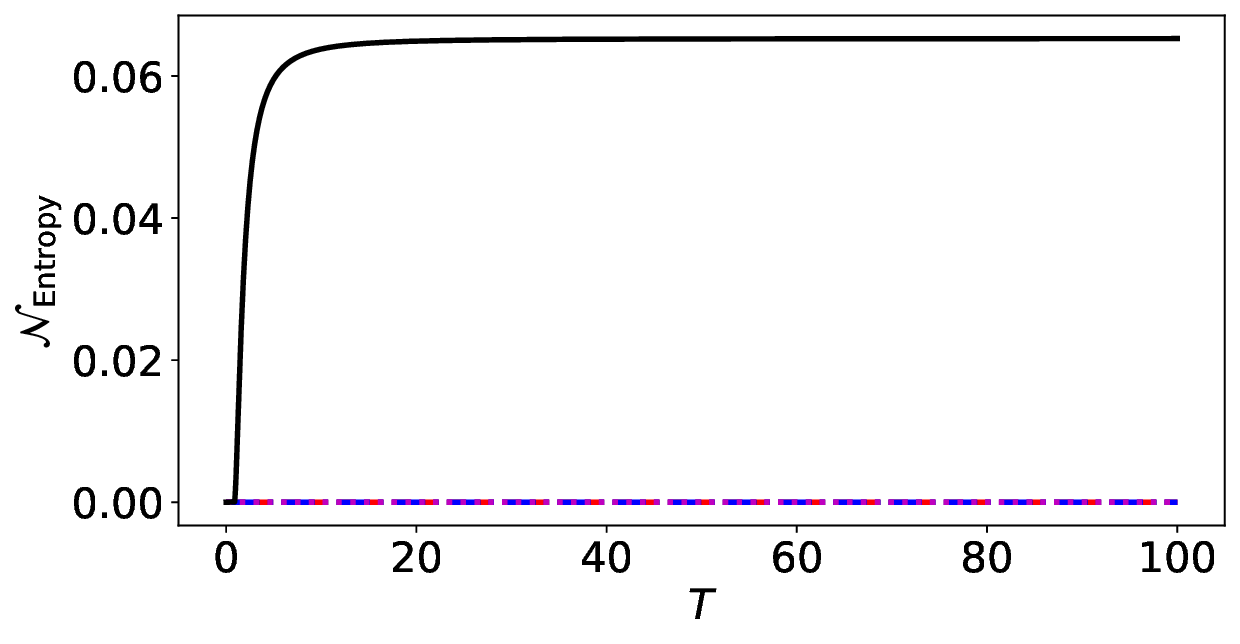}}
    \subfigure[]
    {\includegraphics[width=0.48\textwidth]{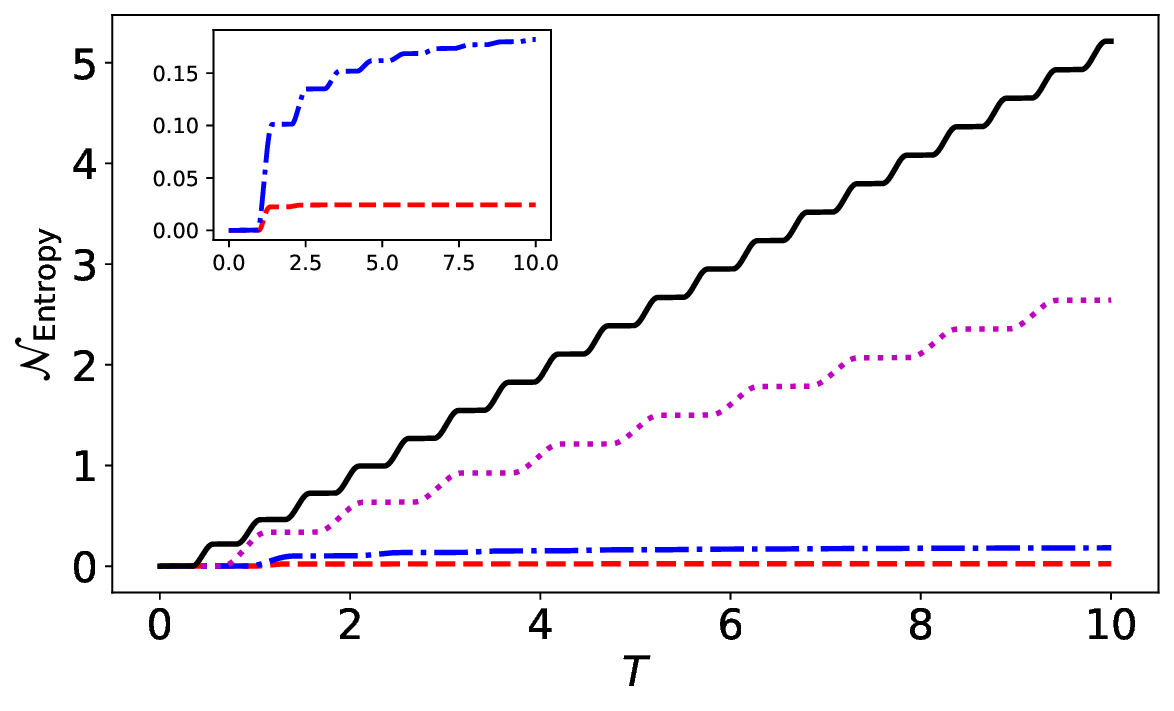}}
    \caption{Same as Fig.~\ref{fig:NMBLP_vs_T}, except that we are now looking at the non-Markovianity measure $\mathcal{N}_{\text{Entropy}}$.}
    \label{fig:NMentropy_vs_T}
\end{figure}

To understand these results, it is useful to look at what happens to the decoherence factor and the indirect interaction at long times. Essentially, at zero temperature, we are in the regime of incomplete decoherence for $s > 1$ - the off-diagonals do not decay completely, while the indirect interaction repeatedly leads to intervals where there is a significant backflow of information from the environment to the system. Let us analyze this in more detail. To derive the long-time limits of $\Gamma(t)$ and $\Delta(t)$, we use 
$$ \lim_{t \to \infty} f(t) = \lim_{\varepsilon \to +0} \varepsilon \int_0^\infty dt \, e^{-\varepsilon t} f(t). $$
We then find
\begin{align*}
\lim_{t \to \infty} \Gamma(t) &= \int_0^\infty d\omega \frac{J(\omega)}{\omega^2}, \\
\lim_{t \to \infty} \Delta(t) &= \frac{\pi}{2} \lim_{\omega \to 0} \frac{J(\omega)}{\omega} - t \int_0^\infty d\omega \, \frac{J(\omega)}{\omega}.
\end{align*}
Given these, it is clear that at long times, for $s \geq 1$, $\Delta(t)$ approaches a linear function of time. On the other hand, for $s > 1$, $\Gamma(t)$ approaches a constant (this constant depends, of course, on the parameters in the spectral density). These behaviors are illustrated in Figs.~\ref{Delta-vs-t} and \ref{Gamma-vs-t}. Consequently, for $s > 1$, the off-diagonal elements of the reduced density matrix [see Eq.~\eqref{evoldstat}] do not decay with time and contain a simple oscillating factor due to the indirect interaction. As such, there is a repeated backflow of information, and we expect that the non-Markovianity measures will increase without bound when $s > 1$. This is exactly what we observe. Remember that, in the definition of the non-Markovianity measures, we are integrating over time. If we denote the upper limit of this integration by $T$, we expect that, for $s > 1$, the indirect interaction causes the non-Markovianity measure to increase as we increase $ T$. We illustrate such behavior in Figs.~\ref{fig:NMBLP_vs_T} and \ref{fig:NMentropy_vs_T}. Let us slowly unpack what is being presented in these figures. We first focus on Fig.~\ref{fig:NMBLP_vs_T}(a). Notice that, in the absence of any indirect interaction, the non-Markovianity is zero for any $T$ if we have $s \leq 2$ - this is consistent with the results presented in, for example, Ref.~\cite{PhysRevA.90.052103}. Moreover, the non-Markovianity is significantly non-zero only for a single time-interval near the initial time. This is consistent with the behavior of $\Gamma(t)$ as illustrated in Fig.~\ref{Gamma-vs-t}. As such, the non-Markovianity measure $\mathcal{N}_{\text{BLP}}$ quickly reaches a constant value as $T$ increases. On the other hand, the behavior is extremely different and far richer when we do have the indirect interaction. The time evolution is now repeatedly non-Markovian due to the oscillatory nature of $g(t)$; physically, the indirect interaction causes a repeated backflow of information. This is most evident in the `staircase-like' structure shown in the solid, black and dotted, magenta curves in Fig.~\ref{fig:NMBLP_vs_T}(b) - the non-Markovianity measure does not increase whenever the non-Markovianity condition is not met, while it increases continuously with $T$ when the condition is being met. Very interestingly, the non-Markovianity condition is repeatedly met even for an Ohmic environment [see in particular the inset of Fig.~\ref{fig:NMBLP_vs_T}(b)]. However, the contribution to the non-Markovianity measure during each time-interval when we have non-Markovianity is much smaller. In fact, even a sub-Ohmic environment can show non-Markovianity. In this case, while the non-Markovianity condition is repeatedly met, the contribution to the non-Markovianity measure is too small, and so the staircase-like structure is not very clear. Simply put, $\Gamma(t)$ quickly becomes too dominant. Finally, as can be seen in Figs.~\ref{fig:NMentropy_vs_T}(a) and (b), the behavior of the non-Markovianity measure $\mathcal{N}_{\text{Entropy}}$ is very similar. 

\begin{figure}[h!]
\centering
\subfigure[]{\includegraphics[width=0.48\textwidth]{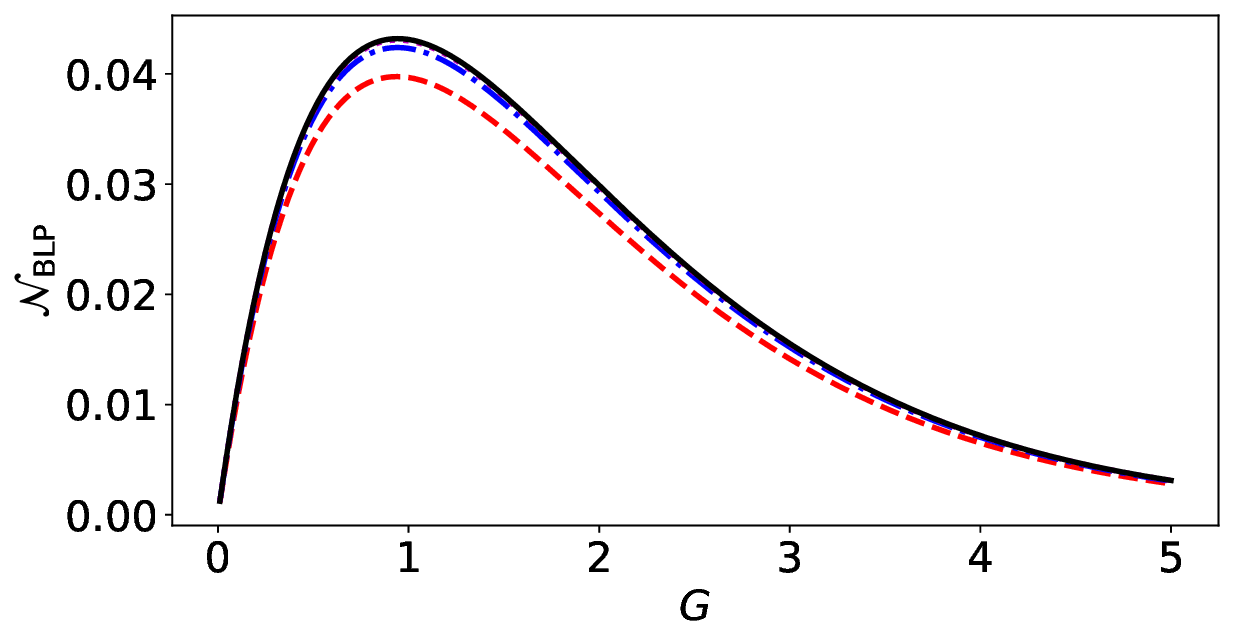}}
\subfigure[]{\includegraphics[width=0.48\textwidth]{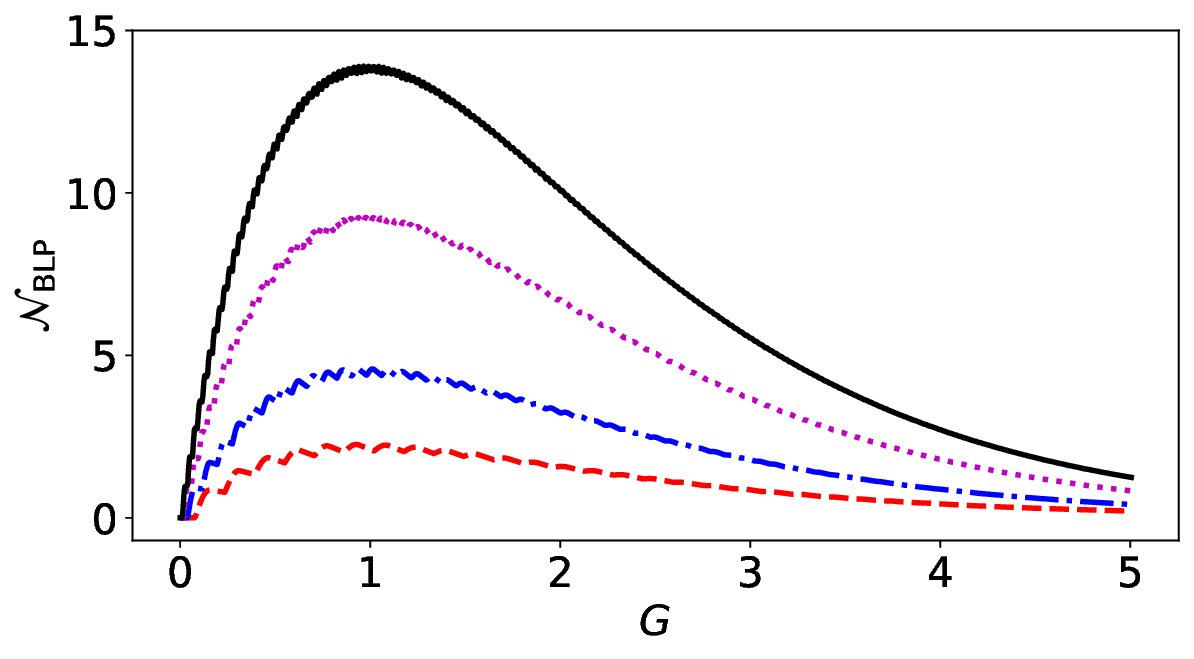}}
\caption{(a) BLP measure of non-Markovianity $\mathcal{N}_{\text{BLP}}$ for a single two-level system plotted as a function of the coupling strength $G$ for four different values of the cutoff frequency $\omega_c$. The red-dashed, blue dashed-dotted, magenta-dotted, and black-solid curves correspond to $\omega_c = 0.5$, $1$, $2$, and $3$, respectively, while $s=3$ for all.  (b) Same as (a), but with the indirect interaction included.}
\label{BLPvsG}
\end{figure}

\begin{figure}[h!]
\centering
\subfigure[]{\includegraphics[width=0.48\textwidth]{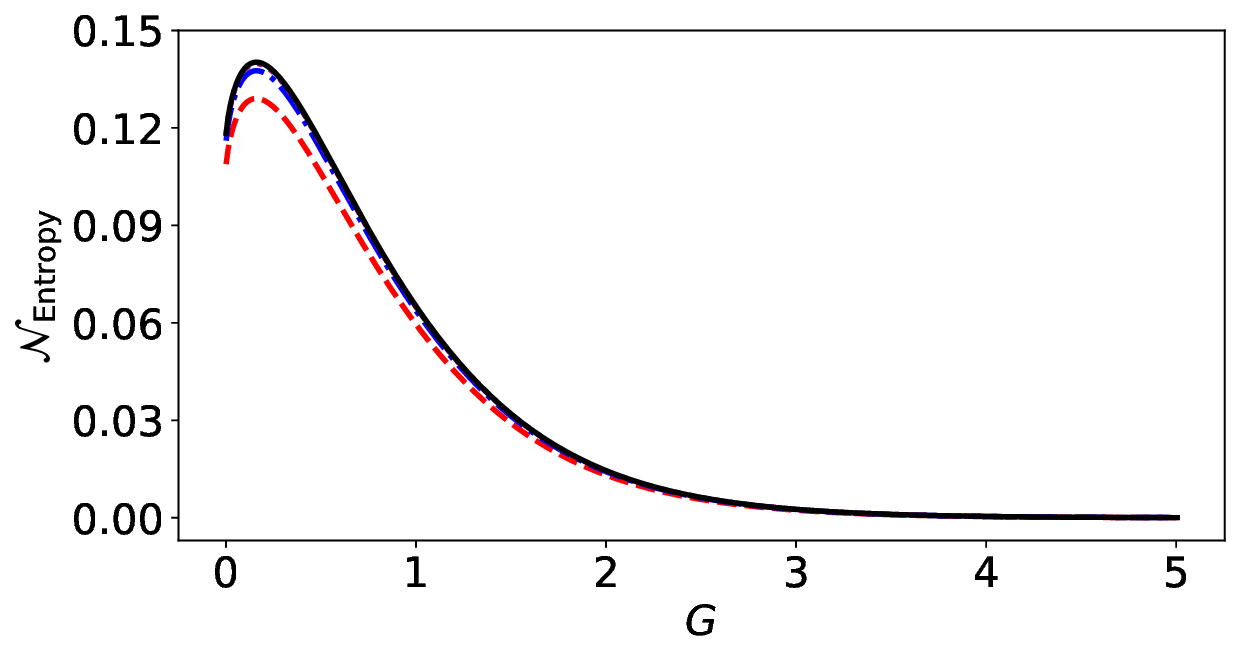}}
\subfigure[]{\includegraphics[width=0.48\textwidth]{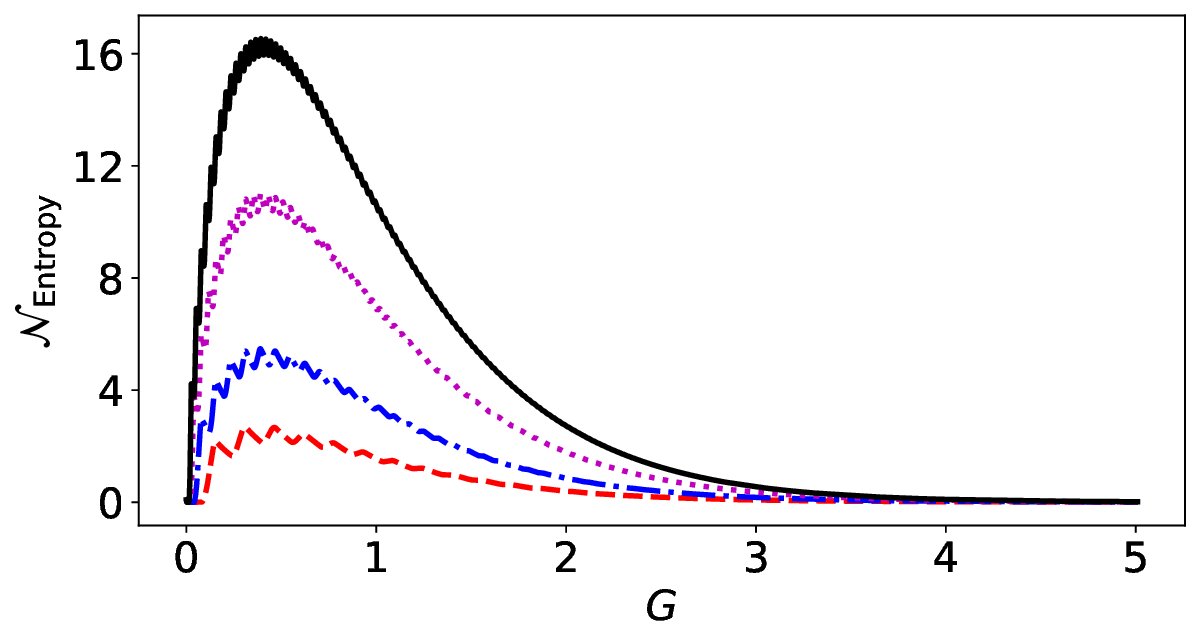}}
\caption{Same as Fig.~\ref{BLPvsG}, except that now we are looking at the non-Markovianity measure $\mathcal{N}_{\text{Entropy}}$.}
\label{entropy vs G}
\end{figure}

\begin{figure}[h!]
\centering
\includegraphics[width=0.48\textwidth]{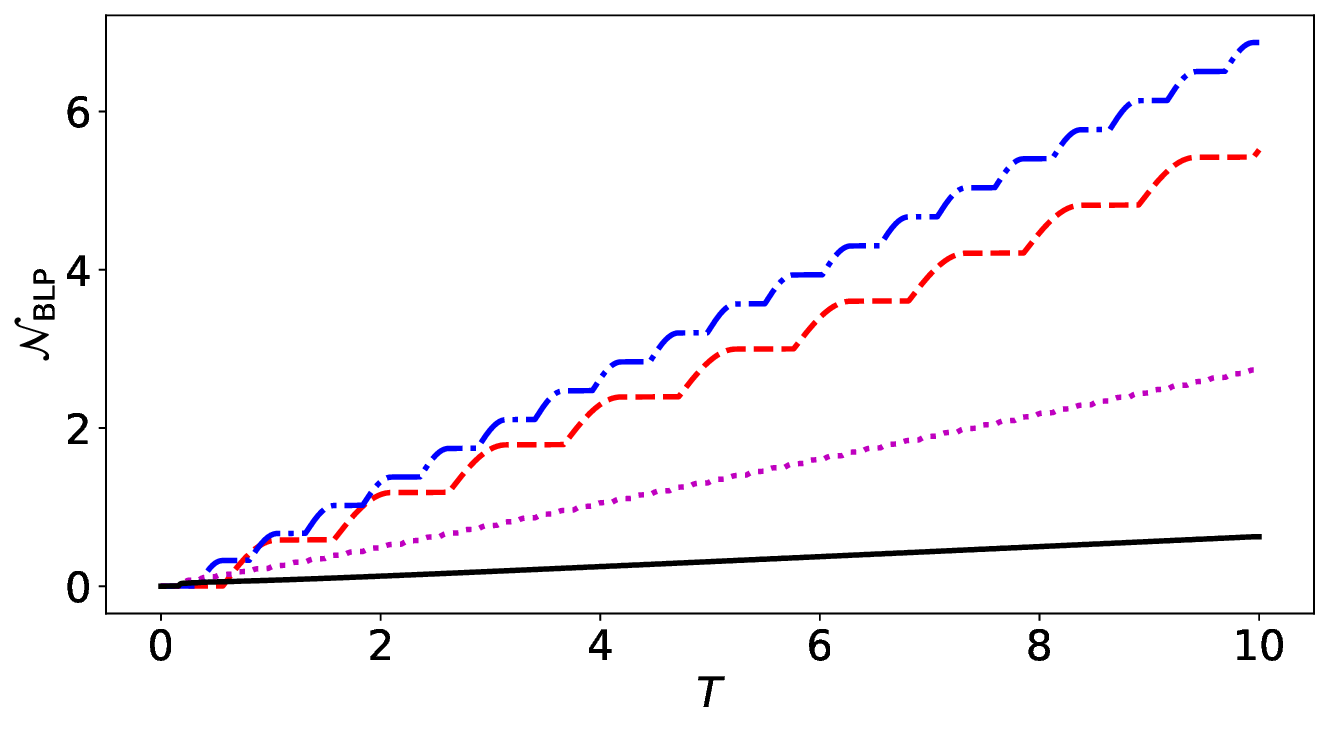}
\caption{Plot of $\mathcal{N}_{\text{BLP}}$ versus the the final time $T$ for different values of the coupling strength $G$. The red-dashed, blue dashed-dotted, magenta-dotted, and solid black curves correspond to $G = 0.5$, $1$, $3$, and $5$, respectively. Here $s=3$ and $\omega_c = 3$.}
\label{figBLPvsTdiffG}
\end{figure}

From Figs.~\ref{BLP vs s} and \ref{entropy vs s}, we also see that for large values of $s$, the non-Markovianity is negligible. This is simply because decoherence becomes too dominant as we keep on increasing $s$. Whatever state we start off from quickly evolves to the completely mixed state and there is effectively no further backflow of information, at least as quantified by $\mathcal{N}_{\text{BLP}}$ and $\mathcal{N}_{\text{Entropy}}$ (as discussed in Appendix \ref{RHPappendix}, the RHP measure of non-Markovianity is quantitatively and qualitatively different). To be precise, while it is true that $\Gamma(t)$ approaches a constant value as time increases, this constant value becomes larger and larger as we increase $s$. Consequently, we are effectively in the regime of complete decoherence. To sum up, as we increase the environment's influence, one might expect greater opportunity for information to flow back from the environment to the system, especially since the effect of the indirect interactions is also expected to increase. In fact, we have checked that the non-Markovianity condition is repeatedly satisfied for large values of $s$ too. However, one also needs to remember that decoherence also increases; consequently, the contribution to the non-Markovianity measures $\mathcal{N}_{\text{BLP}}$ and $\mathcal{N}_{\text{Entropy}}$ during each time interval when the non-Markovianity condition is satisfied becomes negligible. In other words, we can think in terms of a competition between decoherence and the indirect interaction. To further investigate this competition, we have plotted the variation of the non-Markovianity with the coupling strength (see Figs.~\ref{BLPvsG} and \ref{entropy vs G}). From these figures, two points are clear. First, due to the effect of the indirect interaction, non-Markovianity is once again found to be greatly enhanced. We have significant non-Markovianity even for small values of $G$. Second, as the coupling strength increases, the non-Markovianity initially increases, then decreases. This is simply because, as mentioned before, for larger coupling strengths the decoherence factor becomes too large. Another way to see this is to look at how the non-Markovianity changes as the time $T$ changes (see Fig.~\ref{figBLPvsTdiffG}). It is clear from this figure that as we increase the coupling strength $G$, the evolution repeatedly becomes non-Markovian, with a significant contribution to the non-Markovianity measure during each such time interval. However, if the coupling strength $G$ becomes too large, then, just as we saw for increasing $s$, the non-Markovianity decreases due to the strong decoherence - during each time-interval where the evolution is non-Markovian, the contribution to the non-Markovian measure is very small.

\begin{figure}[t!]
\centering
\subfigure[]{\includegraphics[width=0.48\textwidth]{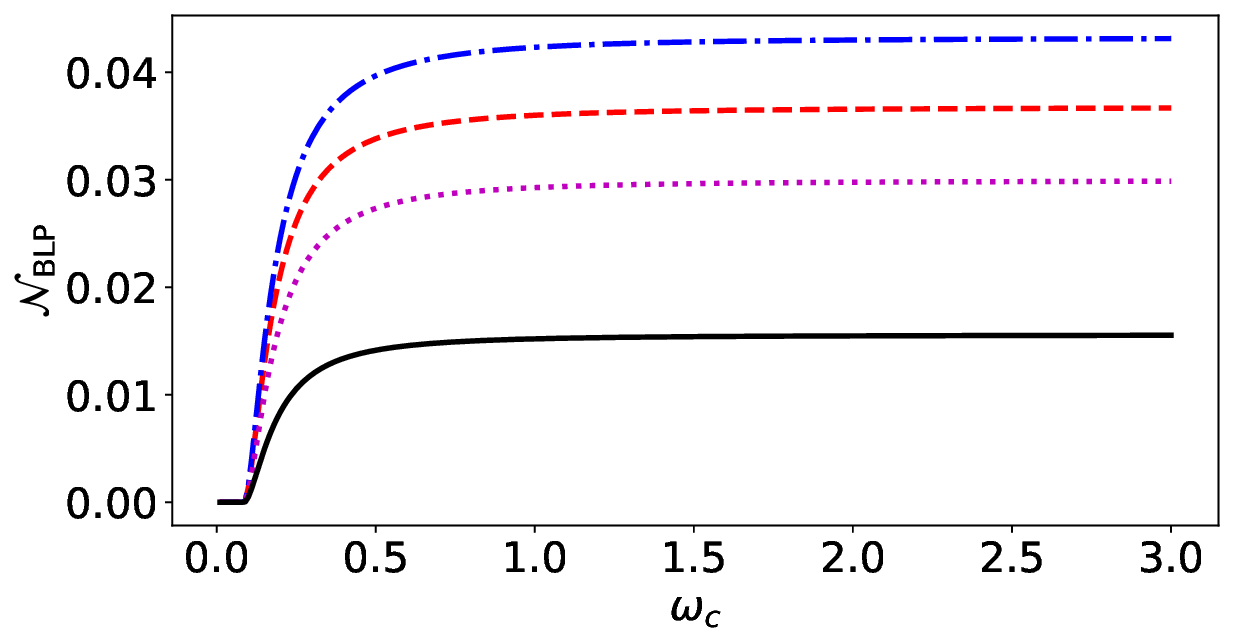}}
\subfigure[]{\includegraphics[width=0.48\textwidth]{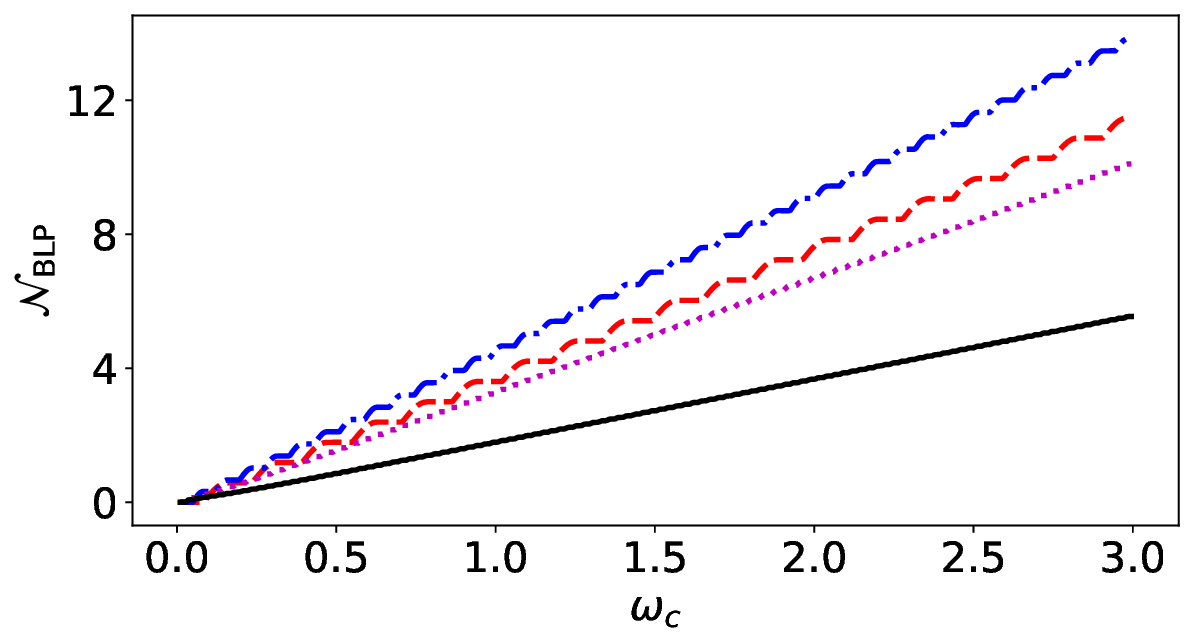}}
\caption{(a) Variation of the non-Markovianity measure $\mathcal{N}_{\text{BLP}}$ for a single TLS as a function of the cutoff frequency $\omega_c$ for four different values of the coupling strength $G$. The results correspond to $G=0.5$ (red-dashed), $G=1$ (blue dashed-dotted), $G=2$ (magenta-dotted), and $G=3$ (solid black), with $ s=3$ for all. (b) Same as (a), but with $N = 2$. The indirect interaction now comes into play.}
\label{BLP vs omega}
\end{figure}

\begin{figure}[t!]
\centering
\subfigure[]{\includegraphics[width=0.48\textwidth]{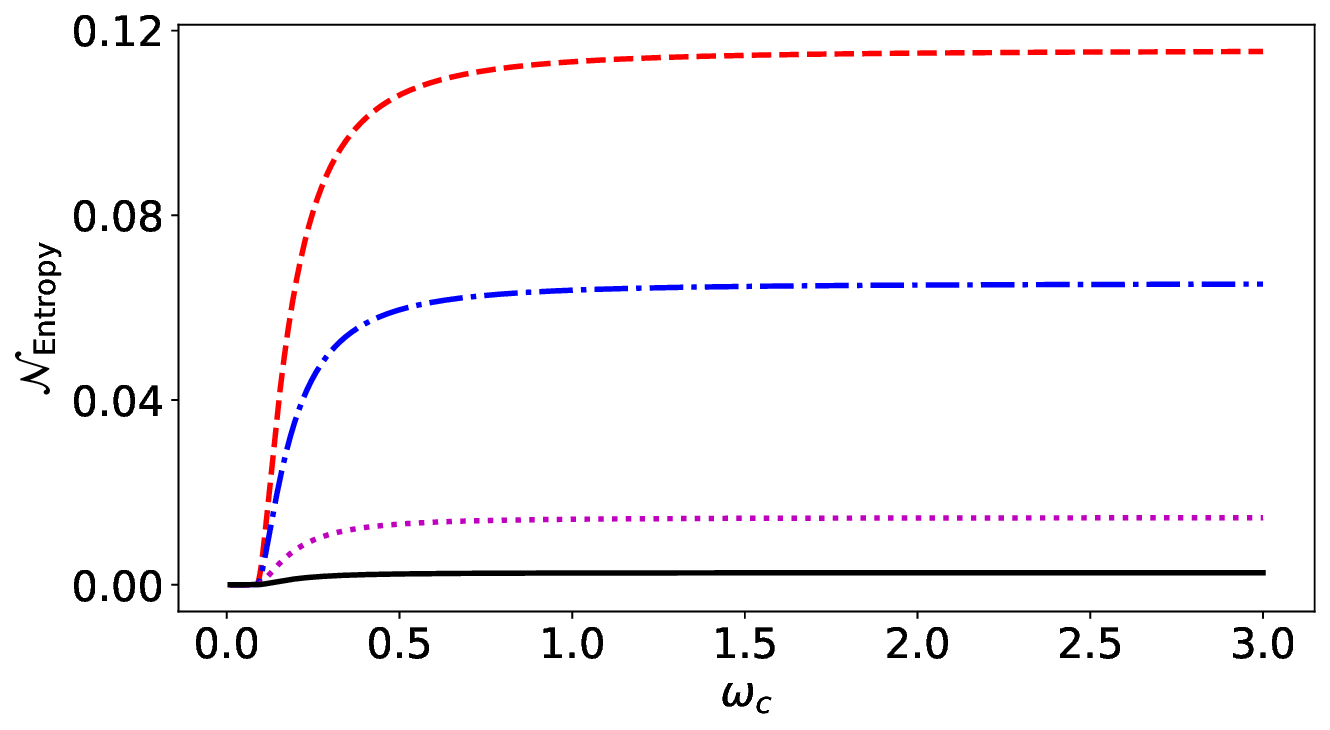}}
\subfigure[]{\includegraphics[width=0.48\textwidth]{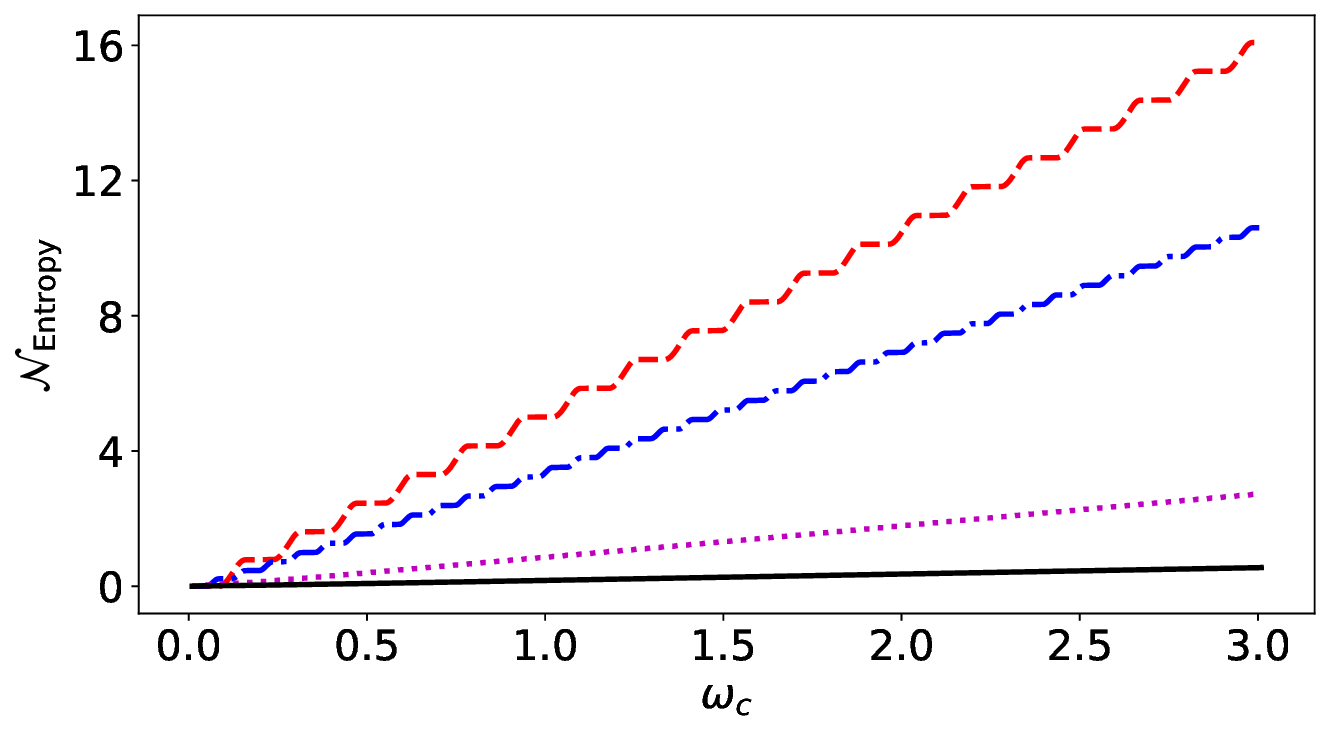}}
\caption{Same as Fig.~\ref{BLP vs omega}, except that now we are looking at the non-Markovianity measure $\mathcal{N}_{\text{Entropy}}$.}
\label{entropy vs omega}
\end{figure}

We can also see the competition between decoherence and the indirect interaction by looking at the effect of changing the cutoff frequency $\omega_c$ on the non-Markovianity. This is illustrated in Figs.~\ref{BLP vs omega} and \ref{entropy vs omega} for super-Ohmic environments. First, note that with a single two-level system and a super-Ohmic environment, the non-Markovianity saturates as the cutoff frequency increases. The behavior is completely different in the presence of the indirect interaction. As we increase the cutoff frequency, the decoherence factor increases, but then so does the indirect interaction. Since we are in the regime of incomplete decoherence here (we are using $s = 3$ in Figs.~\ref{BLP vs omega} and \ref{entropy vs omega}), the increasing indirect interaction wins - we have more intervals of time where the condition given in Eq.~\eqref{nonmarkovianitycondition} is satisfied. It is also interesting to note that the non-Markovianity is greater for $G = 1$ than $G = 0.5$ if we use $\mathcal{N}_{\text{BLP}}$ (compare the red and blue curves in Fig.~\ref{BLP vs omega}), while the opposite is true if we use $\mathcal{N}_{\text{Entropy}}$. This is consistent with the results in Figs.~\ref{BLPvsG} and \ref{entropy vs G}; the maximum value of $\mathcal{N}_{\text{Entropy}}$ is obtained at relatively smaller values of $G$. This emphasizes that while $\mathcal{N}_{\text{BLP}}$ and $\mathcal{N}_{\text{Entropy}}$ exhibit largely similar behavior, they are not identical.

\begin{figure}[t!]
\centering
\subfigure[]{\includegraphics[width=0.48\textwidth]{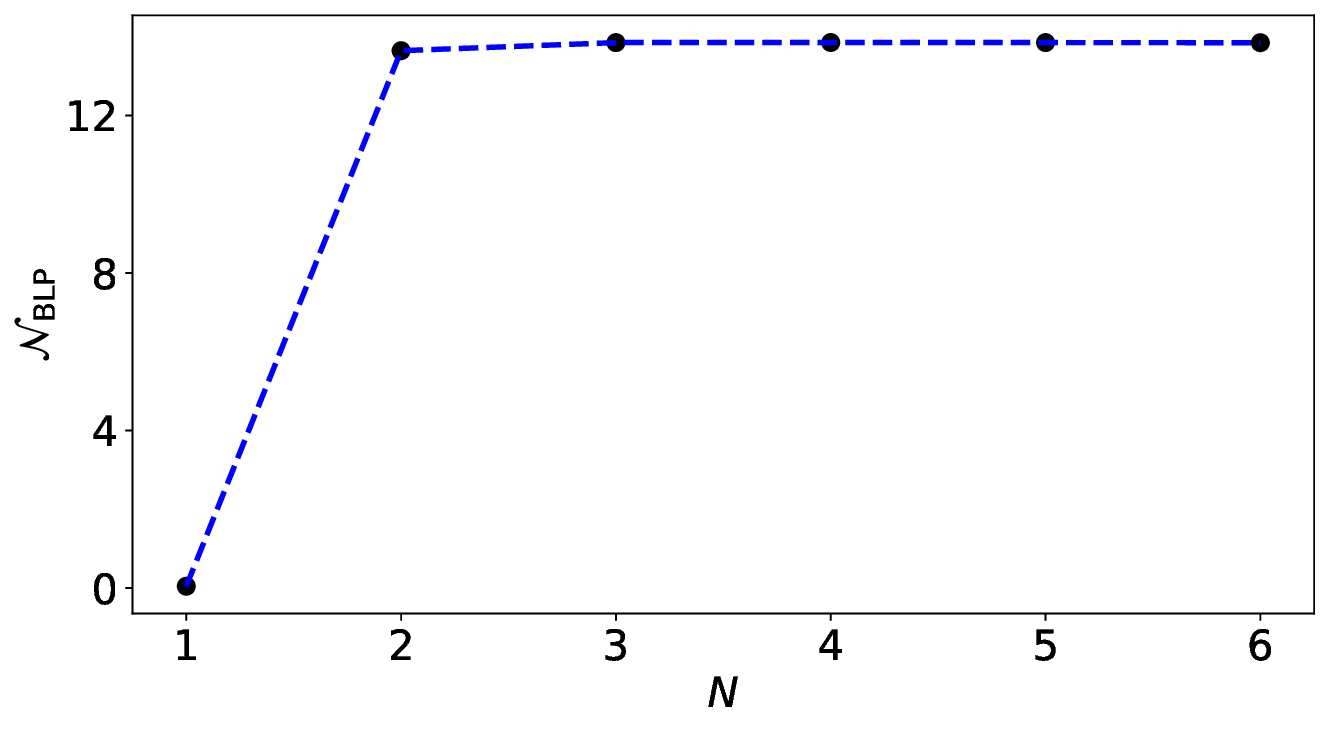}}
\subfigure[]{\includegraphics[width=0.48\textwidth]{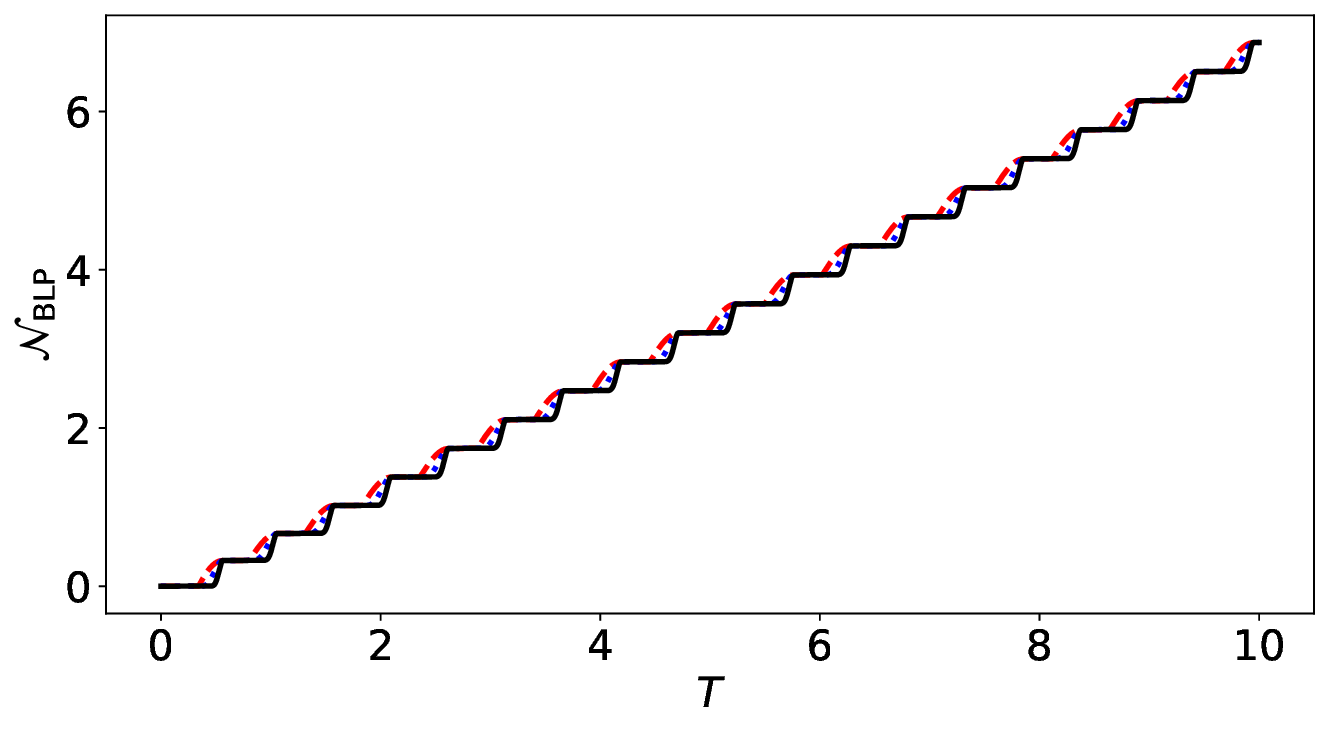}}
\caption {(a) Variation of non-Markovianity calculated via the BLP measure as the number of TLSs $N$ is varied. Here we are using $\omega_c=3$, $s=3$, $G=1$ and $T = 20$. Qualitatively similar results are obtained for $\mathcal{N}_{\text{Entropy}}$. (b) Variation of $\mathcal{N}_{\text{BLP}}$ versus the time $T$ for $N = 2$ (red-dashed), $N = 5$ (blue-dotted), and $N = 20$ (solid-black). We have again used $\omega_c=3$, $s=3$, and $G=1$. At the `treads' of the `staircase', the curves overlap.}
\label{NM vs n}
\end{figure}

It is worth pointing that when including the effect of the indirect interaction, we have considered $N = 2$. That is, we consider two TLSs interacting with a common environment, and then we trace out one TLS. Let us now examine what happens for arbitrary $N$. Results are shown in Fig.~\ref{NM vs n}(a). As expected, we see a big jump when we go from $N = 1$ to $N = 2$. For a larger value of $N$, we find that the BLP and relative entropy non-Markovianity measures, $\mathcal{N}_{\text{BLP}}$ and $\mathcal{N}_{\text{Entropy}}$, are largely independent of $N$. This can be seen by noting that we look at changes in $f(t)g(t)$ (with $f(t) = e^{-\Gamma(t)}$ and $g(t) = |(\cos[\Delta(t)])^{N - 1}|$) over the time-intervals where we have non-Markovianity. Regardless of the value of $N$, $g(t)$ cannot exceed one. To further examine this, we plot the non-Markovianity as a function of the time $T$ for different values of $N$ [see Fig.~\ref{NM vs n}(b)]. From this graph, we see that both the time-intervals over which we do have non-Markovianity and the change in the non-Markovianity over each interval is only very weakly dependent on $N$ for $N \geq 2$. We should also remark that some works consider scaling the coupling strength between the TLSs and the harmonic oscillator environment with the number of TLSs - see, for instance, Refs.~\cite{SilbeyJCP1991,EmaryPRE2003}. If we do consider the coupling strength to be $G/N$, then $\mathcal{N}_{\text{BLP}}$ obviously will in general decrease with increasing $N$ since the coupling strength significantly affects $\mathcal{N}_{\text{BLP}}$. In this work, similar to Ref.~\cite{VorrathPRL2005}, we are not scaling the coupling strength. It is also worth noting that we have shown the behavior of the non-Markovianty measure $\mathcal{N}_{\text{BLP}}$ in Fig.~\ref{NM vs n}; similar behavior is seem for $\mathcal{N}_{\text{Entropy}}$. On the other hand, the RHP measure of non-Markovianity, $\mathcal{N}_{\text{RHP}}$ shows very different behavior. We find that $\mathcal{N}_{\text{RHP}}$ increases with increasing $N$ (see Appendix \ref{RHPappendix}).

\begin{figure}[t!]
    \centering
    \subfigure[]
    {\includegraphics[width=0.48\textwidth]{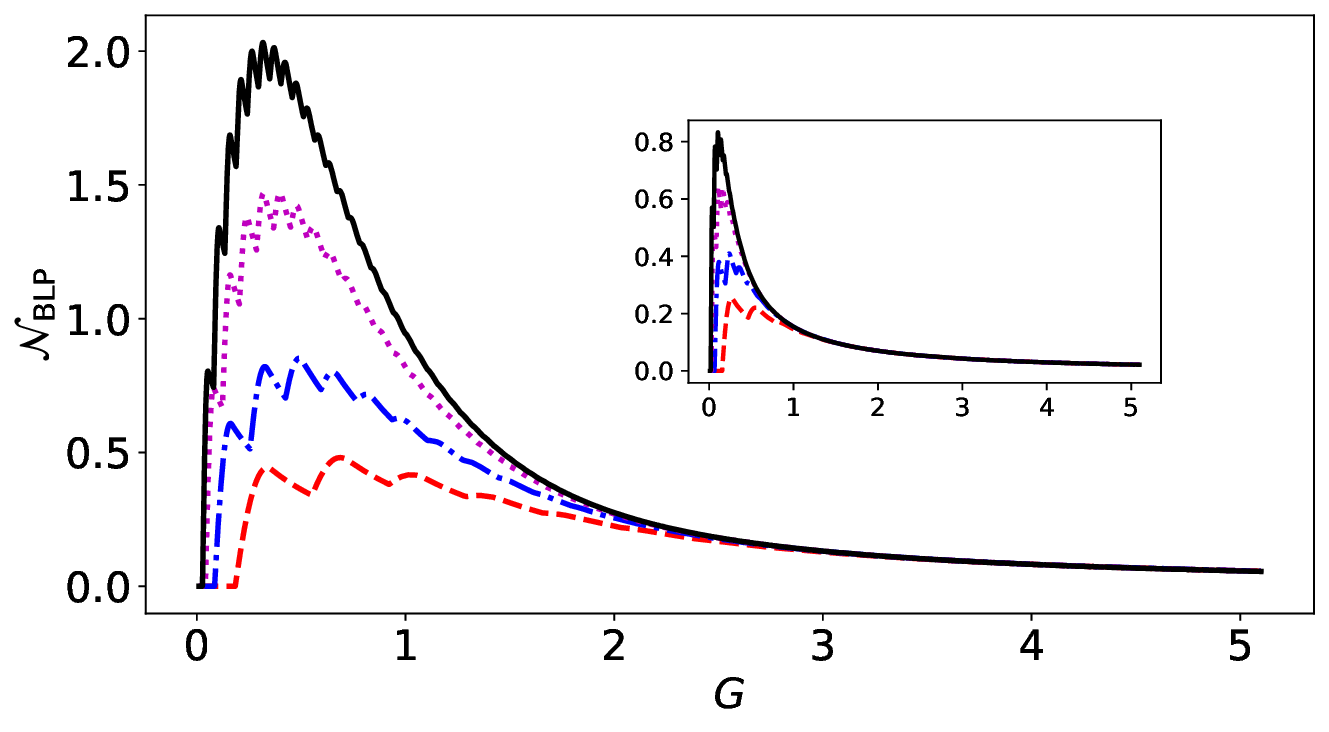}}
    \subfigure[]
    {\includegraphics[width=0.48\textwidth]{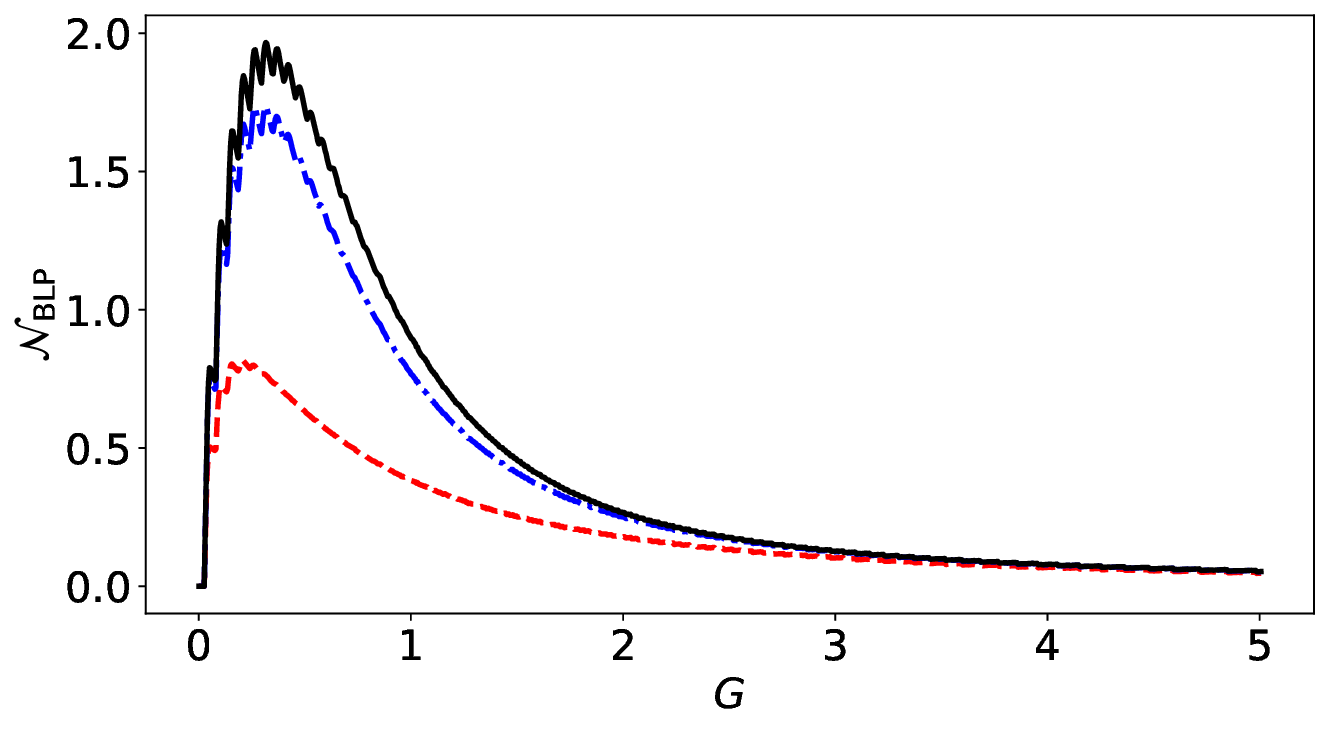}}
    \caption{(a)Variation of the non-Markovianity measure $\mathcal{N}_{\text{BLP}}$ as a function of the coupling strength $G$ for four different values of the cutoff frequency $\omega_c$. The results correspond to $\omega_c=0.5$ (red-dashed), $\omega_c=1$ (blue dashed-dotted), $\omega_c=2$ (magenta dotted), and $\omega_c=3$ (solid black). The main figure corresponds to the Ohmic case ($s=1$), while the inset considers a sub-Ohmic environment with  $s=0.5$. Here, $N = 2$ and $T = 20$. (b) Variation of $\mathcal{N}_{\text{BLP}}$ as a function of the coupling strength $G$ for $\beta = 1$ (red-dashed), $\beta = 20$ (blue dot-dashed), and $\beta = 50$ (solid black). Here we have used $\omega_c = 3$. The rest of the parameters are the same as (a). If the energy-level spacing is in the GHz regime, then $\beta = 10$ corresponds to temperature in the $\text{mK}$ regime; this is the operating temperature for superconducting qubits \cite{NoriRepProg2011}.}
    \label{ohmicsubohmicBLPchangingG}
\end{figure}

\begin{figure}[t!]
    \centering
    \subfigure[]
    {\includegraphics[width=1\linewidth]{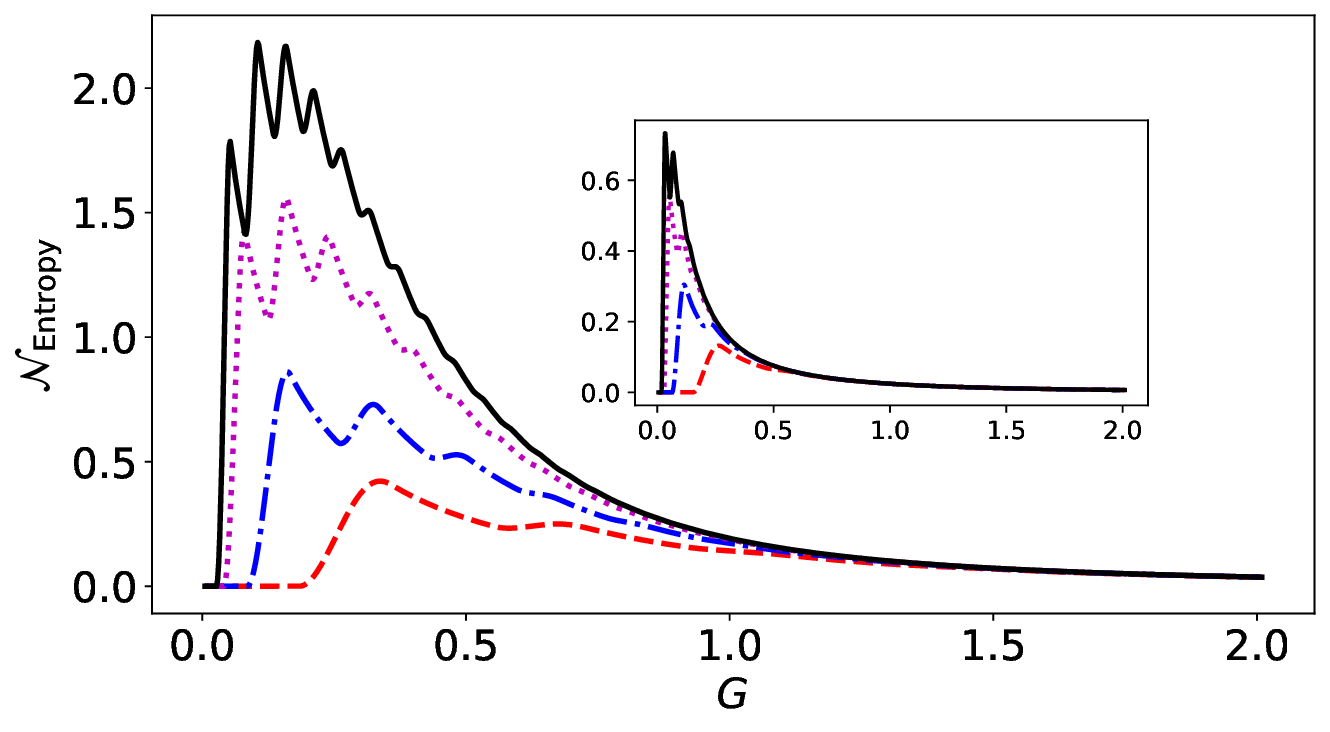}}
    \subfigure[]
    {\includegraphics[width=1\linewidth]{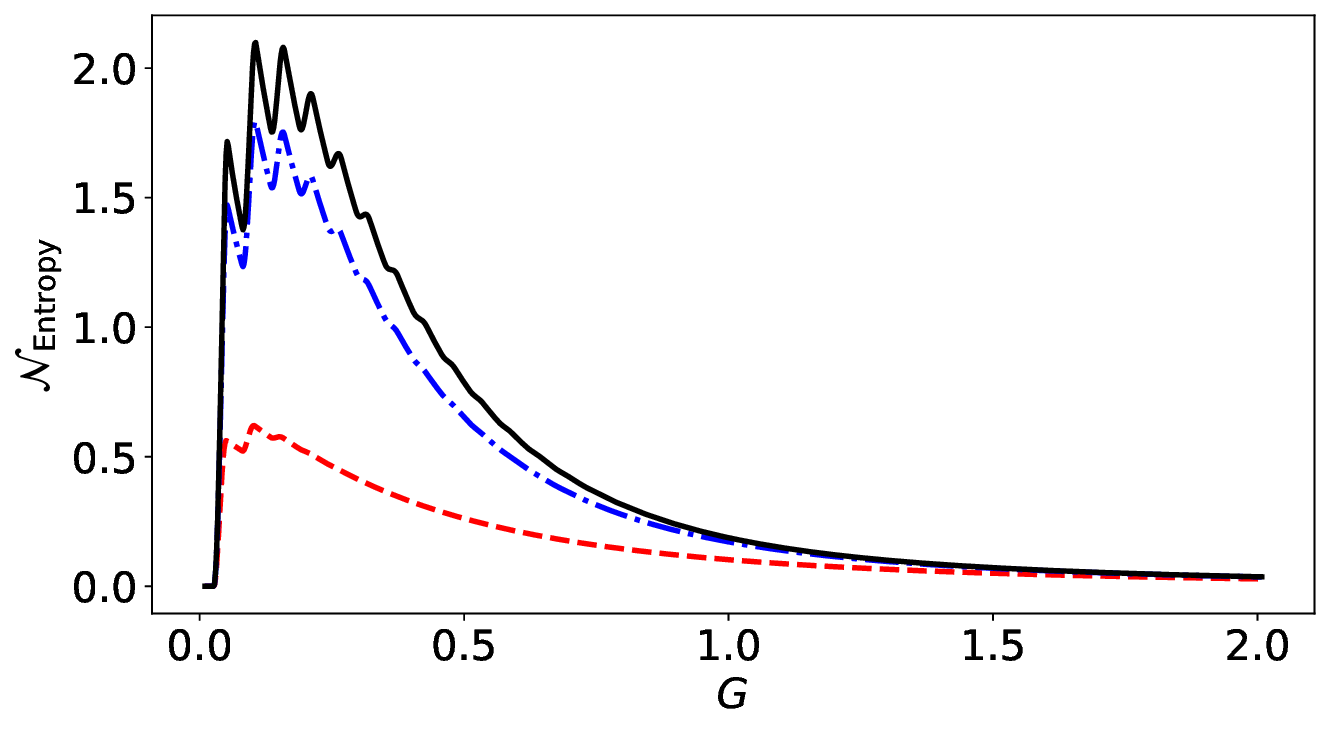}}
    \caption{Same as Fig.~\ref{ohmicsubohmicBLPchangingG}, except that we are now looking at $\mathcal{N}_{\text{Entropy}}$.}
    \label{ohmicsubohmicentropychangingG}
\end{figure}

Until now, we have primarily concentrated on a super-Ohmic environment (in particular, $s = 3$). As we have seen, for a super-Ohmic environment with $s > 1$ and zero temperature, we have incomplete decoherence in the presence of a non-vanishing indirect interaction. Consequently, the non-Markovianity measures continue to increase due to the indirect interaction. For Ohmic and sub-Ohmic environments, however, the decoherence factor $\Gamma(t)$ keeps on increasing as time increases. In fact, as we noted before, for Ohmic and sub-Ohmic environments, with a single TLS, the non-Markovianity is zero (see Figs.~\ref{BLP vs s} and \ref{entropy vs s}). This is not the case when indirect interaction is present. It is then clear that while decoherence will eventually cause effectively complete decoherence for Ohmic and sub-Ohmic environments, the indirect interaction can and does lead to some non-Markovianity before it can do so. However, the non-Markovianity no longer keeps on increasing as we keep on increasing the time $T$. Results for the non-Markovianity as a function of the coupling strength $G$ are illustrated in Figs.~\ref{ohmicsubohmicBLPchangingG}(a) and \ref{ohmicsubohmicentropychangingG}(a). The most important point from these graphs is that we clearly see significant non-Markovianity, even for small values of $G$, due to the indirect interaction in the Ohmic and sub-Ohmic regimes (in contrast, there is no non-Markovianity whatsoever when we have a single TLS and thus no indirect interaction). Also, once again, as we increase the coupling strength, the non-Markovianity increases initially, but then it goes down as the decoherence factor becomes too large. The fact that the non-Markovianity in general does not smoothly increase and then smoothly decrease as $G$ increases is again a manifestation of the competition between the decoherence and the indirect interaction. If $G$ changes by a small amount, it may be the case that now decoherence becomes more dominant; if there is a further small change in $G$, the indirect interaction becomes more dominant, and so on.

In Figs.~\ref{ohmicsubohmicBLPchangingG}(b) and \ref{ohmicsubohmicentropychangingG}(b), we have also shown what happens at finite temperatures. The effect of finite temperature shows up in the decoherence factor $\Gamma(t)$, which must now be written as (see Appendix \ref{derivation})
\[
\Gamma(t) = \int_{0}^{\infty} 
\frac{J(\omega)}{\omega^2} \left[ 1 - \cos(\omega t) \right]
\coth\!\left(\frac{\beta \omega}{2}\right) d\omega,
\]
where, as before, we choose the spectral density as \( J(\omega) = G \omega^s \omega_c^{1-s} e^{-\omega / \omega_c} \), 
and \( \beta\) is the inverse temperature. To go further, we note that we expand the hyperbolic cotangent function in a Taylor series as
\[
\coth\!\left(\frac{\beta \omega}{2}\right) 
= 1 + 2 \sum_{n=1}^{\infty} e^{-n \beta \omega}.
\]
Substituting this series into the integral allows term-wise integration. In particular, for an Ohmic environment, we get 
\[
\Gamma(t) = 
\frac{G}{2}\ln(1 + \omega_c^2 t^2) 
+ G \sum_{n=1}^{\infty} \ln\!\left[1 + \frac{t^2}{(1/\omega_c + n\beta)^2}\right].
\]
The derivative \( \dot{\Gamma}(t) \) can be computed similarly. These results allows us to evaluate the non-Markovianity measures as shown in Figs.~\ref{ohmicsubohmicBLPchangingG}(b) and \ref{ohmicsubohmicentropychangingG}(b). There are two points to note from these figures. First, as the temperature decreases, we approach our zero temperature results - compare the black curves in Figs.~\ref{ohmicsubohmicBLPchangingG}(b) and \ref{ohmicsubohmicentropychangingG}(b) with the black curves in Figs.~~\ref{ohmicsubohmicBLPchangingG}(a) and \ref{ohmicsubohmicentropychangingG}(a). Second, even for $\beta = 1$, we do observe significant non-Markovianity. This should be contrasted with the case when we do not have indirect interaction, where there is no non-Markovianity even at zero temperature. It is also worth noting what happens to $\Gamma(t)$ at long times. At finite temperatures, we find that 
\begin{equation*}
\lim_{t \to \infty} \Gamma(t) = \int_0^\infty d\omega \frac{J(\omega)}{\omega^2} \coth\left(\frac{\beta \omega}{2}\right). \\
\end{equation*}
By looking at the low-frequency behavior, it is clear that $\Gamma(t)$ now diverges unless $s > 2$.

\section{Conclusions}\label{conclu}
In conclusion, we have investigated the effect of indirect interactions induced by a common environment on non-Markovianity. We derived a general formalism to calculate the non-Markovianity of a TLS undergoing pure dephasing, accounting for the effects of indirect interactions. To apply this formalism, we considered a model of $N$ two-level systems interacting with a common harmonic oscillator environment. We conclusively showed that indirect interactions qualitatively and quantitatively affect non-Markovianity to a great extent. Moreover, even in the weak-coupling regime, indirect interactions lead to significant non-Markovianity. Because dephasing typically dominates relaxation - particularly at short times - and harmonic-oscillator environments are so prevalent, we expect our results to have broad applicability. Our findings provide deeper insights into how non-Markovian effects can arise in quantum systems, which may have implications for quantum information processing and decoherence control strategies. 

\section*{Acknowledgements}

Support from the LUMS FIF grant FIF-0952 is acknowledged.

\section*{Data Availability statement}

The data that support the findings of this article are openly available at \href{https://doi.org/10.5281/zenodo.18768495}{https://doi.org/10.5281/zenodo.18768495}.

\appendix

\section{Derivation of the reduced density matrix}
\label{derivation}

As described in the main text, the Hamiltonian of the pure dephasing model of $N$ two-level systems that we are considering is given by $H = H_{S}+H_{B}+H_{SB}$ with
$$
\begin{aligned}
H_{S} &= \frac{\omega_0}{2} \sum_{i=1}^{N} \sigma_z^{(i)}, \\
H_{B} &=\sum_{k} \omega_{k} b_{k}^{\dagger} b_{k}, \\
H_{SB} &= \sum_{i=1}^{N} \sigma_z^{(i)} \sum_{k} \left( g_{k}^* b_{k} + g_{k}b_{k}^{\dagger} \right).
\end{aligned}
$$
Given the initial state of the system plus the environment, we need to find the state of the $N$ TLSs $\rho_S(t)$ at a later time $t$. For this task, we start from 
\begin{align}
\rho_{S}(t)= \text{Tr}_{B}\left[U(t) \rho(0) U^{\dagger}(t)\right],
\end{align}
where $\rho(0)$ is the total initial state. To find the unitary time-evolution operator, we switch to the interaction picture. In other words, 
\begin{align}
U(t)= U_0(t) U_{I}(t),   
\end{align}
where $U_0(t)=e^{-i\left(H_{S}+H_{B}\right) t}$ is the free unitary time-evolution operator and $U_{I}(t)$ is the unitary operator due to system-environment interaction. In this interaction picture, the Hamiltonian becomes
\begin{align*}
H_{SB}(t) &=e^{i\left(H_{S}+H_{B}\right) t} H_{SB} e^{-i\left(H_{S}+H_{B}\right) t}.
\end{align*}\\
The Baker-Campbell-Hausdorff (BCH) identity \cite{puri2001mathematical} allows us to simplify this to 
\[
H_{SB}(t) = \sum_{i=1}^{N} \sigma_z^{(i)} \sum_{k} \left( g_{k}^{*} b_{k} e^{-i \omega_k t} + g_{k} b_{k}^{\dagger} e^{i \omega_k t} \right).
\]
To find \(U_I(t)\) corresponding to this interaction picture Hamiltonian, we use the Magnus expansion \cite{Blanes_2010}. Namely, 
\[
U_I(t) = \exp\left[\sum_{i=1}^{\infty} A_i(t)\right],
\]
where
\[
A_1(t) = -i \int_{0}^{t} dt_1 \, H_{SB}(t_1),
\]
\[
A_2(t) = -\frac{1}{2} \int_{0}^{t} dt_1 \int_{0}^{t_1} dt_2 \, \left[ H_{SB}(t_1), H_{SB}(t_2) \right],
\]
and so on. Using these expressions for $A_1$ and $A_2$, we find that  

\[
A_1(t) = \frac{1}{2}\sum_{j=1}^{N} \sigma_z^{(i)} \sum_{k} \left[ \alpha_{k}(t) b_{k}^{\dagger} - \alpha_{k}^*(t) b_{k} \right],
\]
with
\[
\alpha_{k}(t) = \frac{2 g_{k}}{\omega_{k}} \left( 1 - e^{i \omega_{k} t} \right),
\]
and 
\[
A_2(t) =-\frac{i}{4}\Delta(t)\bigg(N + 2 \sum_{i < j}^{N} \sigma_z^{(i)} \sigma_z^{(j)} \bigg),
\]
with 
\[
\Delta(t) = \sum_{k} \frac{4|g_{k}|^2}{\omega_k^2} \left[ \sin(\omega_k t) - \omega_k t \right].
\]
Now, all the higher-order terms in the Magnus expansion are zero. Therefore, putting these results together, and dropping the irrelevant term proportional to identity in $A_2$, we finally have that 
\begin{align*}
U(t) = e^{-i(H_S + H_B)t}
\exp&\Bigg[
    \frac{1}{2}\sum_{i=1}^{N}\sigma_z^{(i)}
       \sum_k (\alpha_k b_k^\dagger - \alpha_k^* b_k)
\\ 
 &- \frac{i}{2}\!
       \left(\sum_{\substack{i,j=1\\ i<j}}^{N}
       \sigma_z^{(i)} \sigma_z^{(j)}\right) \,\Delta(t)
\Bigg].
\end{align*}
It is also useful to note that $\sum_{i<j} \sigma_z^{(i)}\sigma_z^{(j)}\ket{n} = \sum_{i<j} n_i n_j \ket{n}$. 

We now go back to finding the state of the $N$ TLSs via 
\[
\rho_S(t) = \text{Tr}_B \left[ U(t) \rho(0) U^{\dagger}(t) \right].
\]
To make further progress, we use the joint eigenbasis of the $\sigma_z^{(i)}$ operators $\ket{n}$ [see Eq.~\eqref{basisstates}], where $n$ is a string of the form $n_1 n_2 \hdots n_N$ with entries $+1$ or $-1$. In other words, $n_i = \pm 1$. For convenience, define the operator $P_{nn'} = \ket{n}\bra{n'}$. Then,
\[
[\rho_S(t)]_{n'n} = \text{Tr} \left[  U(t) \rho(0) U^{\dagger}(t) \ket{n} \bra{n'} \right],
\]
with $\text{Tr}$ being the total trace. Using the cyclic invariance of the trace, 
\[
[\rho_S(t)]_{n'n} = \text{Tr} \left[\rho(0) U^{\dagger}(t) \ket{n} \bra{n'} U(t) \right].
\]
Using the time-evolution operator we have found, 
\begin{align}
\left[\rho_S(t)\right]_{n'n} &= e^{-i\omega_0  (s_{n'} - s_n) t/2} e^{-\frac{i}{4} \Delta(t) (s_{n'}^2 - s_n^2)} \notag \\ &\text{Tr}\left(\rho(0) e^{-R_{n'n}} P_{nn'} \right).
\end{align}
Here $s_n = \sum_i n_i$, and we have used the identity $\sum_{i<j} n_i n_j = \frac{1}{2}(s_n^2 - N)$. Also,
\[
R_{n'n} = \frac{1}{2} (s_n - s_{n'}) \sum_{k} \left(\alpha_{k}(t)  b_k^\dagger - \alpha_{k}^*(t) b_k\right).
\]
We now consider the total state to be a simple product state of the form \(\rho_S(0)\) and the environment state, with the environment assumed to be in thermal equilibrium. In other words,
\[
    \rho(0) = \rho_S(0) \otimes \rho_B,
\]
with 
\[
    \rho_B = \frac{e^{-\beta H_B}}{Z_B},
\]
and
\[
    Z_B = \text{Tr}_B \left[ e^{- \beta H_B} \right].
\]
Here $\beta$ is the inverse temperature. With this total state, we get 
\begin{align*}
\left[\rho_S(t)\right]_{n'n} &= \left[\rho_S(0)\right]_{n'n} e^{i\omega_0 (s_{n'} - s_n) t} e^{-\frac{i}{4} \Delta(t) (s_{n'}^2 - s_n^2)} \times \\
&\text{Tr}_B \left[ \rho_B e^{-R_{n'n} (t)} \right]. 
\end{align*}
The trace over the environment is found to be 
$$ \text{Tr}_B\left[ \rho_B e^{-R_{n'n} (t)} \right] = e^{-\frac{1}{4} (s_{n'} - s_n)^2 \Gamma(t)}, $$
where
\[
\Gamma(t) = \sum_{k} \frac{4 | g_k^2 |}{\omega_{k}^2} [1 - \cos (\omega_{k} t)] \coth \left( \frac{\beta \omega_{k}}{2} \right).
\]
We therefore finally have the density matrix for the $N$ two-level systems:
\begin{align*}
\left[\rho_S(t)\right]_{n'n} = &\left[\rho_S(0)\right]_{n'n} e^{-i\omega_0 (s_{n'} - s_n) t/2} e^{-\frac{i}{4} \Delta(t) (s_{n'}^2 - s_n^2)} \times \\
&e^{-\frac{1}{4}(s_{n'} - s_n)^2 \Gamma(t)}.  
\end{align*}
We now consider all the two-level systems to be initially in the state $\ket{+} = \frac{1}{\sqrt{2}}\left(\ket{0} + \ket{1}\right)$. By taking a partial trace over all two-level systems apart from one, we obtain the single two-level system state as   
\[
\rho_1(t) = \frac{1}{2}
\begin{pmatrix}
1 &  \eta \\
 \eta^* & 1
\end{pmatrix},
\]
where $\eta = e^{-\Gamma(t)} e^{-i\omega_0 t} \left(\cos\left[\Delta(t)\right]\right)^{N - 1}$. This partial trace is straightforward to calculate if we note that, for this partial trace, $s_{n'} - s_n = 2$. Also, $s_{n'}^2 - s_n^2 = 4\sum_{i = 2}^N n_i$, with each of these $n_i = \pm 1$ (both with probability $1/2$). If we assume that every two-level system is initially in the state $\ket{-} = \frac{1}{\sqrt{2}}(\ket{0} - \ket{1})$, we obtain instead
\[
\rho_2(t) = \frac{1}{2}
\begin{pmatrix}
1 & -\eta \\
-\eta^* & 1
\end{pmatrix}.
\]
In fact, the same state is obtained if the initial state of the two-level systems is $\ket{-++\cdots+}$, that is, the initial state of the single TLS is $\ket{-}$ while the other TLSs start from the state $\ket{+}$. This choice of initial conditions ensures that the environment seen by the single TLS remains the same, and therefore the dynamical map remains the same. The fact that the dynamics are the same follows due to the fact that the total Hamiltonian commutes with the $\sigma_z$ matrix for any spin. In particular, the off-diagonal matrix element of the single-qubit density matrix at time $t$ is given by 
$$ [\rho_1(t)]_{+1,-1} = \text{Tr} [(\ket{-1}\bra{+1} \otimes \mathds{1}_{N - 1} \otimes \mathds{1}_B) U(t) \rho(0) U^\dagger(t)].$$
Here the trace is the total trace, $\mathds{1}_{N - 1}$ is the identity operator for the auxiliary $N - 1$ TLSs, and $\mathds{1}_B$ is the identity operator in the Hilbert space of the collection of harmonic oscillators. We now define the unitary (and Hermitian) operator $Z_p = \prod_{i = 2}^N \sigma_z^{(i)}$. This is the product of the Pauli $Z$ matrices for all TLSs except the first. It then follows that 
\begin{align*}
[\rho_1(t)]_{+1,-1} = &\text{Tr} [Z_pZ_p(\ket{-1}\bra{+1} \otimes \mathds{1}_{N - 1} \otimes \mathds{1}_B) \, \times \\
&Z_pZ_p U(t)Z_pZ_p \rho(0)Z_pZ_p U^\dagger(t)Z_pZ_p].
\end{align*}
It is clear that $Z_p(\ket{-1}\bra{+1} \otimes \mathds{1}_{N - 1} \otimes \mathds{1}_B)Z_p = (\ket{-1}\bra{+1} \otimes \mathds{1}_{N - 1} \otimes \mathds{1}_B)$. Since $Z_p$ commutes with the total Hamiltonian, $Z_p U(t)Z_p = U(t)$ and $Z_p U^\dagger(t) Z_p = U^\dagger(t)$. Also, $Z_p (\ket{-+ \cdots +}\bra{-+ \cdots +} \otimes \rho_B) Z_p = \ket{-- \cdots -}\bra{--\cdots -} \otimes \rho_B$. The fact that our claim is true should now be self-evident.

\section{RHP Measure}
\label{RHPappendix}

In this appendix, we use the Rivas-Huelga-Plenio (RHP) \cite{Rivas_2014} measure of non-Markovianity to show that it too displays characteristics somewhat similar to those of the other two measures discussed in the main text. As usual, we consider a single two-level system undergoing pure dephasing. Following the notation in the main text, we know that the diagonal elements of the reduced density matrix in the $\sigma_z$ eigenbasis do not change. The off-diagonal elements can be found via 
\begin{equation*}
\rho_{+1,-1}(t) = \alpha(t)\,\rho_{+1,-1}(0),
\end{equation*}
where, as before, $\alpha(t) = f(t)g(t)e^{-i[\omega_0 t + \chi(t)]}$. Now, to compute the RHP measure, it is useful to compute the derivative. This is given by 
\begin{equation*}
\dot{\rho}_{+1,-1}(t)
= \dot{\alpha}(t)\,\rho_{+1,-1}(0)
= \frac{\dot{\alpha}(t)}{\alpha(t)}\,\rho_{+1,-1}(t).
\end{equation*}
This motivates us to define the `instantaneous complex rate'
\begin{equation*}
\lambda(t) \equiv \frac{\dot{\alpha}(t)}{\alpha(t)}
= \frac{d}{dt}\ln \alpha(t).
\end{equation*}
Given the form of $\alpha(t)$, it is easy to see that 
\begin{equation*}
\lambda(t)
= \frac{\dot{f}(t)}{f(t)}
+ \frac{\dot{g}(t)}{g(t)}
- i\,\dot{\chi}(t) - i\omega_0.
\end{equation*}
It is useful to note that 
\begin{equation*}
\Re\lambda(t)=\frac{\dot{f}(t)}{f(t)}+\frac{\dot{g}(t)}{g(t)},
\qquad
\Im\lambda(t)=-\dot{\chi}(t) - \omega_0.
\end{equation*}

\begin{figure}[t!]
\centering
\subfigure[]
{\includegraphics[width=0.48\textwidth]{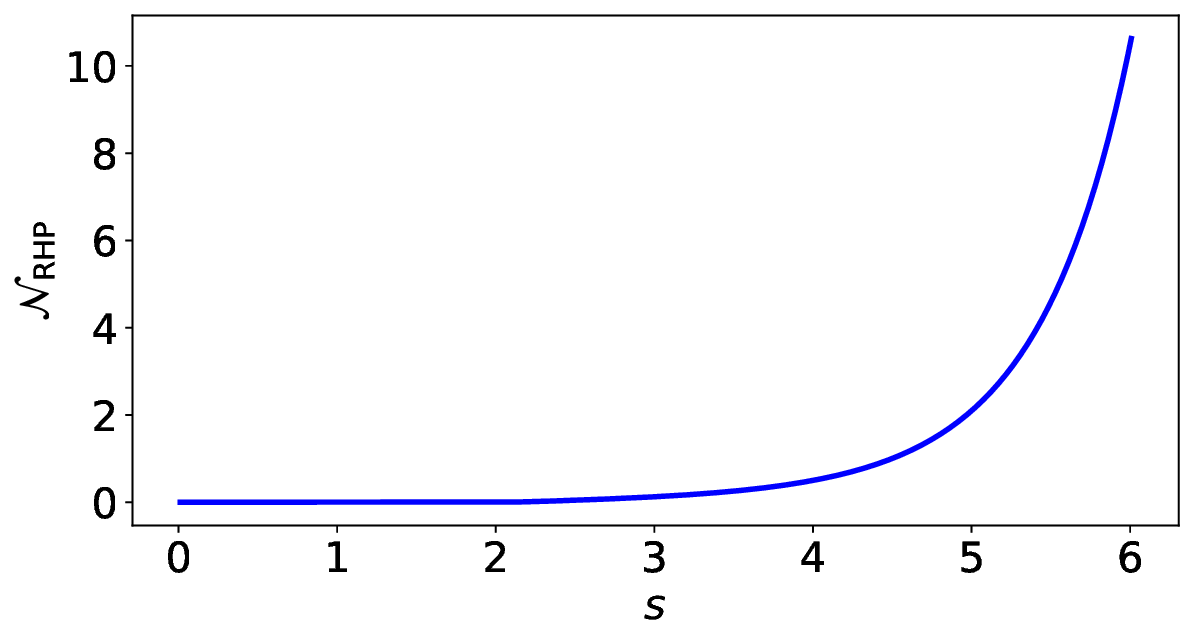}}
\subfigure[]{\includegraphics[width=0.48\textwidth]{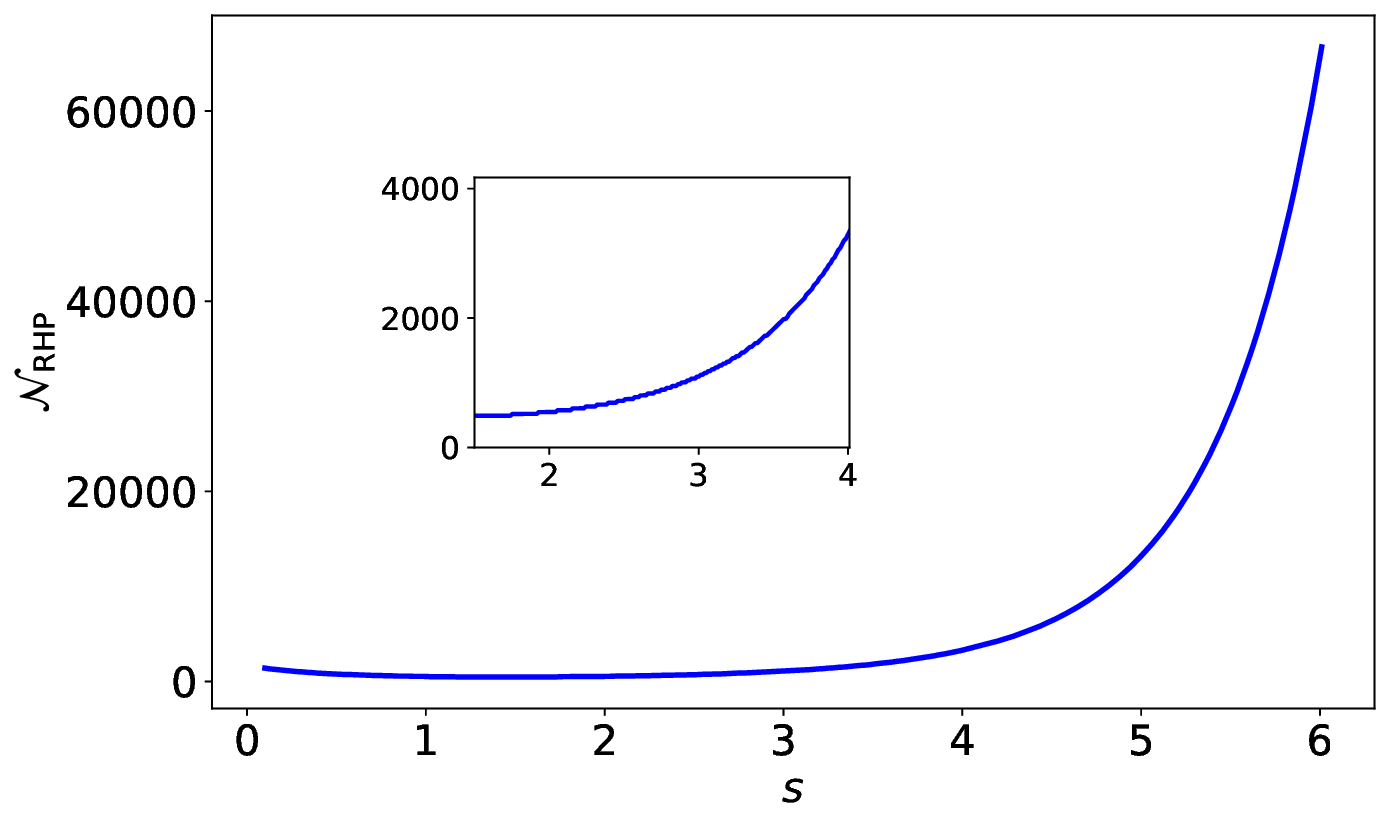}}
\caption{(a) Plot of the RHP measure of non-Markovianity, $\mathcal{N}_{\text{RHP}}$, as a function of the Ohmicity parameter $s$ for a single TLS (that is, $N = 1$). As usual, we work in dimensionless units with $\hbar = 1$, and here we set $G = 1$, $\omega_c = 3$, and $T = 20$. We are working in the zero-temperature regime. (b) Same as (a), but using two TLSs and then tracing out one of these. The dynamics are now influenced by an indirect interaction. The inset illustrates that the non-Markovianity is non-zero throughout.}
\label{RHP vs s}
\end{figure}

\begin{figure}[t!]
\centering
\subfigure[]{\includegraphics[width=0.48\textwidth]{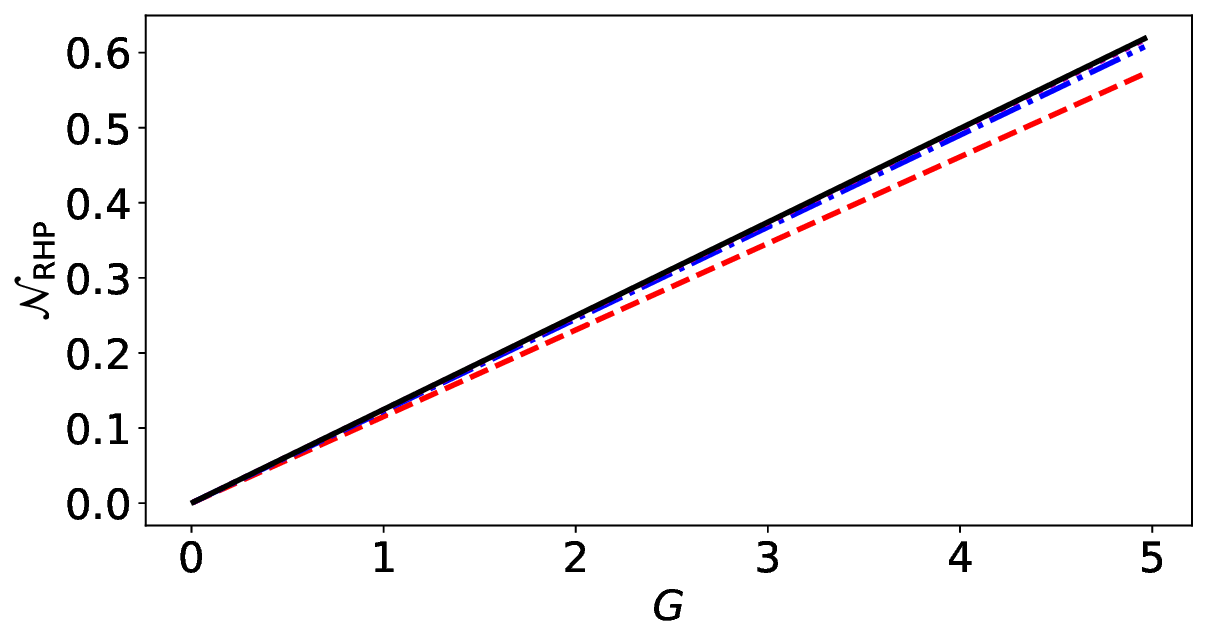}}
\subfigure[]{\includegraphics[width=0.48\textwidth]{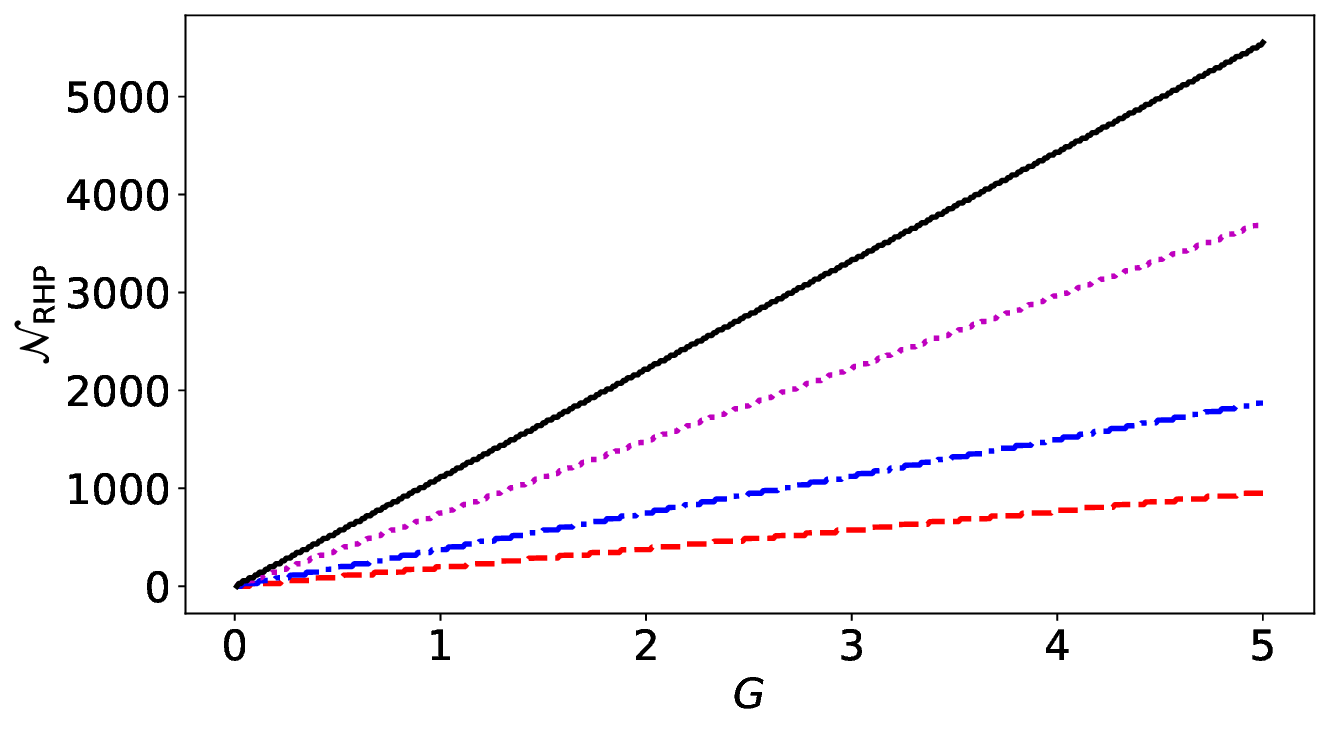}}
\caption{(a) RHP measure of non-Markovianity $\mathcal{N}_{\text{RHP}}$ for a single two-level system plotted as a function of the coupling strength $G$ for four different values of the cutoff frequency $\omega_c$. The red-dashed, blue dashed-dotted, magenta-dotted, and black-solid curves correspond to $\omega_c = 0.5$, $1$, $2$, and $3$, respectively. Here, we have used $s = 3$ and $T = 20$. (b) Same as (a), but with the indirect interaction included.}
\label{RHPvsG}
\end{figure}

We now use these results to find the RHP measure. For pure dephasing, the reduced dynamics admits a time-local master equation of the form \cite{breuer2002theory}
\begin{equation*}
\dot{\rho}(t)
= -i[H_{\mathrm{eff}}(t),\rho(t)]
+ \gamma(t)\big(\sigma_z\rho(t)\sigma_z-\rho(t)\big),
\end{equation*}
where we can write $H_{\mathrm{eff}}(t)=h(t)\sigma_z$ for some $h(t)$. From this master equation, it can be shown that 
\begin{equation*}
\dot{\rho}_{+1,-1}(t)
= -2\gamma(t)\rho_{+1,-1}(t)-2i\,h(t)\rho_{+1,-1}(t).
\end{equation*}
The instantaneous complex rate from here is thus
\begin{equation*}
\lambda(t) = -2\gamma(t)-2i\,h(t).
\end{equation*}
By equating this to the $\lambda(t)$ previously found, we find that 
\begin{equation*}
\gamma(t) = -\frac{1}{2}\left(\frac{\dot{f}(t)}{f(t)} + \frac{\dot{g}(t)}{g(t)}\right)
          = -\frac{1}{2}\frac{d}{dt}\ln\!\big[f(t)g(t)\big].
\end{equation*}
Now, non-Markovianity according to the RHP conditions occurs whenever $\gamma(t)<0$, otherwise the dynamical map is CP-divisible and we then have Markovianity according to RHP. This condition can be written equivalently as
$\frac{d}{dt}\ln\!\big[f(t)g(t)\big] > 0.$
As in the main text, we set $f(t)=e^{-\Gamma(t)}$. Then, $\gamma(t) = \dot{\Gamma}(t) - \frac{\dot{g}(t)}{g(t)}$.
The condition for non-Markovianity, according to the RHP measure, then reduces to 
\begin{equation*}
\dot{g}(t) - \dot{\Gamma}(t)\,g(t) > 0.
\end{equation*}
This is exactly the same condition obtained for the BLP measure as well as the measure based on relative entropy [see Eq.~\eqref{nonmarkovianitycondition}].

\begin{figure}[b!]
\centering
\subfigure[]{\includegraphics[width=0.48\textwidth]{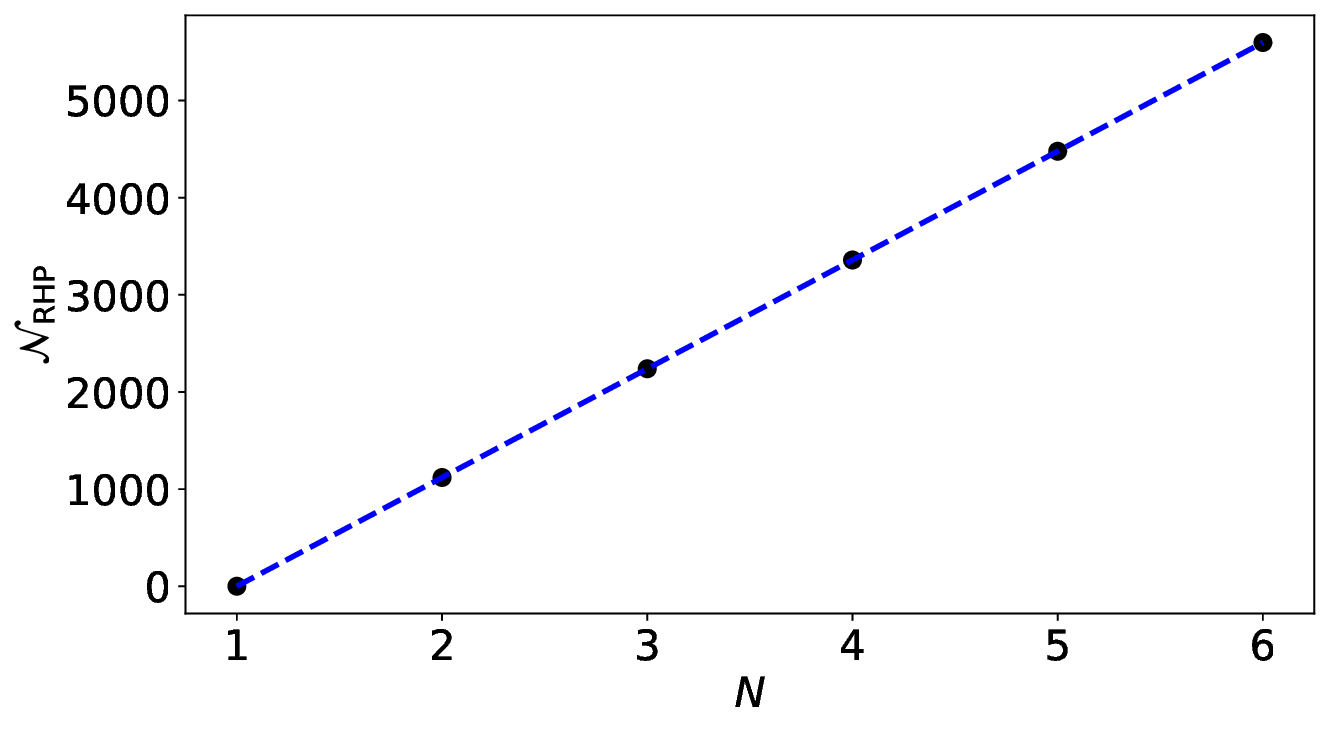}}
\caption {Variation of non-Markovianity calculated via the RHP measure as the number of TLSs $N$ is varied. Here we are using $\omega_c=3$, $s=3$, $G=1$, and $T = 20$.}
\label{NM-RHP vs n}
\end{figure}

The non-Markovianity is quantified for our pure dephasing case by using \cite{PhysRevA.90.052103} 
$$\mathcal{N}_{\text{RHP}} = -2\int_{\gamma(t) < 0} \gamma(t) \, dt. $$
We now present the numerical evaluation of the RHP non-Markovianity measure. This can be calculated since for our pure dephasing model, we can find $\gamma(t)$. We first plot the non-Markovianity measure $\mathcal{N}_{\text{RHP}}$ as a function of the Ohmicity parameter $s$; results are shown in Fig.~\ref{RHP vs s}. Once again, we see that the non-Markovianity measure can be much greater due to the indirect interactions in multiple TLSs, as compared to the single TLS case. For a single TLS, the non-Markovianity starts to increase sharply around $s = 4$ and continues to increase as $s$ increases. This is consistent with the results reported in Ref.~\cite{PhysRevA.90.052103}. Moreover, this contrasts sharply with the trends observed using $\mathcal{N}_{\text{BLP}}$ and $\mathcal{N}_{\text{Entropy}}$. This drastic difference comes about because the RHP measure effectively looks at changes in $\ln[f(t)g(t)]$ over the time intervals when we have non-Markovian evolution, while the BLP measure, for instance, looks at changes in $f(t)g(t)$ over the same time intervals. 

We next look at how $\mathcal{N}_{\text{RHP}}$ changes as the coupling strength $G$ changes. Results are shown in Fig.~\ref{RHPvsG}. Our main point remains true: the presence of the indirect interaction greatly enhances non-Markovianity. However, once again, the behavior of $\mathcal{N}_{\text{RHP}}$ is different compared to the two non-Markovianity measures discussed in the main text. Simply put, the reason for this difference is that as $G$ increases, $\Gamma(t)$ increases, while $e^{-\Gamma(t)}$ rapidly decreases. $\mathcal{N}_{\text{RHP}}$ looks at changes in the decay rate, rather than its exponential, so it keeps on increasing as $G$ increases.  

Finally, let us also note that, in contrast with $\mathcal{N}_{\text{BLP}}$ and $\mathcal{N}_{\text{Entropy}}$, $\mathcal{N}_{\text{RHP}}$ increases with $N$. This is illustrated in Fig.~\ref{NM-RHP vs n}. This increase with $N$ is expected for $\mathcal{N}_{\text{RHP}}$ since $\mathcal{N}_{\text{RHP}}$ looks at the difference in the logarithm of $f(t)g(t)$ over the time intervals when we have non-Markovian evolution. Since $g(t) = |(\cos[\Delta(t)])^{N-1}|$, it is obvious that $\mathcal{N}_{\text{RHP}}$ will increase with $N$. Interestingly, it has been shown that $\mathcal{N}_{\text{RHP}}$ could increase with an increasing number of depolarizing channels \cite{MukherjeePRA2024}; here, we have shown that $\mathcal{N}_{\text{RHP}}$ can increase via the indirect interactions from an increasing number of TLSs.


\begin{thebibliography}{56}%
\makeatletter
\providecommand \@ifxundefined [1]{%
 \@ifx{#1\undefined}
}%
\providecommand \@ifnum [1]{%
 \ifnum #1\expandafter \@firstoftwo
 \else \expandafter \@secondoftwo
 \fi
}%
\providecommand \@ifx [1]{%
 \ifx #1\expandafter \@firstoftwo
 \else \expandafter \@secondoftwo
 \fi
}%
\providecommand \natexlab [1]{#1}%
\providecommand \enquote  [1]{``#1''}%
\providecommand \bibnamefont  [1]{#1}%
\providecommand \bibfnamefont [1]{#1}%
\providecommand \citenamefont [1]{#1}%
\providecommand \href@noop [0]{\@secondoftwo}%
\providecommand \href [0]{\begingroup \@sanitize@url \@href}%
\providecommand \@href[1]{\@@startlink{#1}\@@href}%
\providecommand \@@href[1]{\endgroup#1\@@endlink}%
\providecommand \@sanitize@url [0]{\catcode `\\12\catcode `\$12\catcode
  `\&12\catcode `\#12\catcode `\^12\catcode `\_12\catcode `\%12\relax}%
\providecommand \@@startlink[1]{}%
\providecommand \@@endlink[0]{}%
\providecommand \url  [0]{\begingroup\@sanitize@url \@url }%
\providecommand \@url [1]{\endgroup\@href {#1}{\urlprefix }}%
\providecommand \urlprefix  [0]{URL }%
\providecommand \Eprint [0]{\href }%
\providecommand \doibase [0]{https://doi.org/}%
\providecommand \selectlanguage [0]{\@gobble}%
\providecommand \bibinfo  [0]{\@secondoftwo}%
\providecommand \bibfield  [0]{\@secondoftwo}%
\providecommand \translation [1]{[#1]}%
\providecommand \BibitemOpen [0]{}%
\providecommand \bibitemStop [0]{}%
\providecommand \bibitemNoStop [0]{.\EOS\space}%
\providecommand \EOS [0]{\spacefactor3000\relax}%
\providecommand \BibitemShut  [1]{\csname bibitem#1\endcsname}%
\let\auto@bib@innerbib\@empty
\bibitem [{\citenamefont {Nielsen}\ and\ \citenamefont
  {Chuang}(2010)}]{nielsen2010quantum}%
  \BibitemOpen
  \bibfield  {author} {\bibinfo {author} {\bibfnamefont {M.~A.}\ \bibnamefont
  {Nielsen}}\ and\ \bibinfo {author} {\bibfnamefont {I.~L.}\ \bibnamefont
  {Chuang}},\ }\href@noop {} {\emph {\bibinfo {title} {Quantum computation and
  quantum information}}}\ (\bibinfo  {publisher} {Cambridge University Press},\
  \bibinfo {year} {2010})\BibitemShut {NoStop}%
\bibitem [{\citenamefont {Haroche}\ and\ \citenamefont
  {Raimond}(2006)}]{haroche2014exploring}%
  \BibitemOpen
  \bibfield  {author} {\bibinfo {author} {\bibfnamefont {S.}~\bibnamefont
  {Haroche}}\ and\ \bibinfo {author} {\bibfnamefont {J.-M.}\ \bibnamefont
  {Raimond}},\ }\href@noop {} {\emph {\bibinfo {title} {Exploring the quantum:
  atoms, cavities, and photons}}}\ (\bibinfo  {publisher} {Oxford University
  Press},\ \bibinfo {address} {Oxford},\ \bibinfo {year} {2006})\BibitemShut
  {NoStop}%
\bibitem [{\citenamefont {Schlosshauer}(2007)}]{schlosshauer2007decoherence}%
  \BibitemOpen
  \bibfield  {author} {\bibinfo {author} {\bibfnamefont {M.}~\bibnamefont
  {Schlosshauer}},\ }\href@noop {} {\emph {\bibinfo {title} {Decoherence and
  the quantum-to-classical transition}}}\ (\bibinfo  {publisher} {Springer},\
  \bibinfo {address} {Berlin},\ \bibinfo {year} {2007})\BibitemShut {NoStop}%
\bibitem [{\citenamefont {Breuer}\ and\ \citenamefont
  {Petruccione}(2007)}]{breuer2002theory}%
  \BibitemOpen
  \bibfield  {author} {\bibinfo {author} {\bibfnamefont {H.-P.}\ \bibnamefont
  {Breuer}}\ and\ \bibinfo {author} {\bibfnamefont {F.}~\bibnamefont
  {Petruccione}},\ }\href@noop {} {\emph {\bibinfo {title} {The Theory of Open
  Quantum Systems}}}\ (\bibinfo  {publisher} {Oxford University Press},\
  \bibinfo {address} {Oxford},\ \bibinfo {year} {2007})\BibitemShut {NoStop}%
\bibitem [{\citenamefont {de~Vega}\ and\ \citenamefont
  {Alonso}(2017)}]{RevModPhys.89.015001}%
  \BibitemOpen
  \bibfield  {author} {\bibinfo {author} {\bibfnamefont {I.}~\bibnamefont
  {de~Vega}}\ and\ \bibinfo {author} {\bibfnamefont {D.}~\bibnamefont
  {Alonso}},\ }\bibfield  {title} {\bibinfo {title} {Dynamics of non-markovian
  open quantum systems},\ }\href {https://doi.org/10.1103/RevModPhys.89.015001}
  {\bibfield  {journal} {\bibinfo  {journal} {Rev. Mod. Phys.}\ }\textbf
  {\bibinfo {volume} {89}},\ \bibinfo {pages} {015001} (\bibinfo {year}
  {2017})}\BibitemShut {NoStop}%
\bibitem [{\citenamefont {Caruso}\ \emph
  {et~al.}(2014{\natexlab{a}})\citenamefont {Caruso}, \citenamefont
  {Giovannetti}, \citenamefont {Lupo},\ and\ \citenamefont
  {Mancini}}]{RevModPhys.86.1203}%
  \BibitemOpen
  \bibfield  {author} {\bibinfo {author} {\bibfnamefont {F.}~\bibnamefont
  {Caruso}}, \bibinfo {author} {\bibfnamefont {V.}~\bibnamefont {Giovannetti}},
  \bibinfo {author} {\bibfnamefont {C.}~\bibnamefont {Lupo}},\ and\ \bibinfo
  {author} {\bibfnamefont {S.}~\bibnamefont {Mancini}},\ }\bibfield  {title}
  {\bibinfo {title} {Quantum channels and memory effects},\ }\href
  {https://doi.org/10.1103/RevModPhys.86.1203} {\bibfield  {journal} {\bibinfo
  {journal} {Rev. Mod. Phys.}\ }\textbf {\bibinfo {volume} {86}},\ \bibinfo
  {pages} {1203} (\bibinfo {year} {2014}{\natexlab{a}})}\BibitemShut {NoStop}%
\bibitem [{\citenamefont {Bylicka}\ \emph {et~al.}(2014)\citenamefont
  {Bylicka}, \citenamefont {Chru{\'s}ci{\'n}ski},\ and\ \citenamefont
  {Maniscalco}}]{bylicka2014non}%
  \BibitemOpen
  \bibfield  {author} {\bibinfo {author} {\bibfnamefont {B.}~\bibnamefont
  {Bylicka}}, \bibinfo {author} {\bibfnamefont {D.}~\bibnamefont
  {Chru{\'s}ci{\'n}ski}},\ and\ \bibinfo {author} {\bibfnamefont
  {S.}~\bibnamefont {Maniscalco}},\ }\bibfield  {title} {\bibinfo {title}
  {Non-markovianity and reservoir memory of quantum channels: a quantum
  information theory perspective},\ }\href {https://doi.org/10.1038/srep05720}
  {\bibfield  {journal} {\bibinfo  {journal} {Sci. Rep.}\ }\textbf {\bibinfo
  {volume} {4}},\ \bibinfo {pages} {5720} (\bibinfo {year} {2014})}\BibitemShut
  {NoStop}%
\bibitem [{\citenamefont {Zhang}\ \emph {et~al.}(2012)\citenamefont {Zhang},
  \citenamefont {Lo}, \citenamefont {Xiong}, \citenamefont {Tu},\ and\
  \citenamefont {Nori}}]{PhysRevLett.109.170402}%
  \BibitemOpen
  \bibfield  {author} {\bibinfo {author} {\bibfnamefont {W.-M.}\ \bibnamefont
  {Zhang}}, \bibinfo {author} {\bibfnamefont {P.-Y.}\ \bibnamefont {Lo}},
  \bibinfo {author} {\bibfnamefont {H.-N.}\ \bibnamefont {Xiong}}, \bibinfo
  {author} {\bibfnamefont {M.~W.-Y.}\ \bibnamefont {Tu}},\ and\ \bibinfo
  {author} {\bibfnamefont {F.}~\bibnamefont {Nori}},\ }\bibfield  {title}
  {\bibinfo {title} {General non-markovian dynamics of open quantum systems},\
  }\href {https://doi.org/10.1103/PhysRevLett.109.170402} {\bibfield  {journal}
  {\bibinfo  {journal} {Phys. Rev. Lett.}\ }\textbf {\bibinfo {volume} {109}},\
  \bibinfo {pages} {170402} (\bibinfo {year} {2012})}\BibitemShut {NoStop}%
\bibitem [{\citenamefont {Wenderoth}\ \emph {et~al.}(2021)\citenamefont
  {Wenderoth}, \citenamefont {Breuer},\ and\ \citenamefont
  {Thoss}}]{PhysRevA.104.012213}%
  \BibitemOpen
  \bibfield  {author} {\bibinfo {author} {\bibfnamefont {S.}~\bibnamefont
  {Wenderoth}}, \bibinfo {author} {\bibfnamefont {H.-P.}\ \bibnamefont
  {Breuer}},\ and\ \bibinfo {author} {\bibfnamefont {M.}~\bibnamefont
  {Thoss}},\ }\bibfield  {title} {\bibinfo {title} {Non-markovian effects in
  the spin-boson model at zero temperature},\ }\href
  {https://doi.org/10.1103/PhysRevA.104.012213} {\bibfield  {journal} {\bibinfo
   {journal} {Phys. Rev. A}\ }\textbf {\bibinfo {volume} {104}},\ \bibinfo
  {pages} {012213} (\bibinfo {year} {2021})}\BibitemShut {NoStop}%
\bibitem [{\citenamefont {Shrikant}\ and\ \citenamefont
  {Mandayam}(2023)}]{10.3389/frqst.2023.1134583}%
  \BibitemOpen
  \bibfield  {author} {\bibinfo {author} {\bibfnamefont {U.}~\bibnamefont
  {Shrikant}}\ and\ \bibinfo {author} {\bibfnamefont {P.}~\bibnamefont
  {Mandayam}},\ }\bibfield  {title} {\bibinfo {title} {Quantum
  non-markovianity: Overview and recent developments},\ }\href
  {https://doi.org/10.3389/frqst.2023.1134583} {\bibfield  {journal} {\bibinfo
  {journal} {Front. Quantum Sci. Technol.}\ }\textbf {\bibinfo {volume} {2}},\
  \bibinfo {pages} {1134583} (\bibinfo {year} {2023})}\BibitemShut {NoStop}%
\bibitem [{\citenamefont {Tiwari}\ and\ \citenamefont
  {Banerjee}(2023)}]{10.3389/frqst.2023.1207552}%
  \BibitemOpen
  \bibfield  {author} {\bibinfo {author} {\bibfnamefont {D.}~\bibnamefont
  {Tiwari}}\ and\ \bibinfo {author} {\bibfnamefont {S.}~\bibnamefont
  {Banerjee}},\ }\bibfield  {title} {\bibinfo {title} {Impact of non-markovian
  evolution on characterizations of quantum thermodynamics},\ }\bibfield
  {journal} {\bibinfo  {journal} {Front. Quantum Sci. Technol.}\ }\textbf
  {\bibinfo {volume} {2}},\ \href {https://doi.org/10.3389/frqst.2023.1207552}
  {10.3389/frqst.2023.1207552} (\bibinfo {year} {2023})\BibitemShut {NoStop}%
\bibitem [{\citenamefont {Mouloudakis}\ \emph {et~al.}(2023)\citenamefont
  {Mouloudakis}, \citenamefont {Stergou},\ and\ \citenamefont
  {Lambropoulos}}]{Mouloudakis_2023}%
  \BibitemOpen
  \bibfield  {author} {\bibinfo {author} {\bibfnamefont {G.}~\bibnamefont
  {Mouloudakis}}, \bibinfo {author} {\bibfnamefont {I.}~\bibnamefont
  {Stergou}},\ and\ \bibinfo {author} {\bibfnamefont {P.}~\bibnamefont
  {Lambropoulos}},\ }\bibfield  {title} {\bibinfo {title} {Non-markovianity in
  the time evolution of open quantum systems assessed by means of quantum state
  distance},\ }\href {https://doi.org/10.1088/1402-4896/ace0de} {\bibfield
  {journal} {\bibinfo  {journal} {Phys. Scr.}\ }\textbf {\bibinfo {volume}
  {98}},\ \bibinfo {pages} {085111} (\bibinfo {year} {2023})}\BibitemShut
  {NoStop}%
\bibitem [{\citenamefont {Seneviratne}\ \emph {et~al.}(2024)\citenamefont
  {Seneviratne}, \citenamefont {Walters},\ and\ \citenamefont
  {Wang}}]{doi:10.1021/acsomega.3c09720}%
  \BibitemOpen
  \bibfield  {author} {\bibinfo {author} {\bibfnamefont {A.}~\bibnamefont
  {Seneviratne}}, \bibinfo {author} {\bibfnamefont {P.~L.}\ \bibnamefont
  {Walters}},\ and\ \bibinfo {author} {\bibfnamefont {F.}~\bibnamefont
  {Wang}},\ }\bibfield  {title} {\bibinfo {title} {Exact non-markovian quantum
  dynamics on the nisq device using kraus operators},\ }\href
  {https://doi.org/10.1021/acsomega.3c09720} {\bibfield  {journal} {\bibinfo
  {journal} {ACS Omega}\ }\textbf {\bibinfo {volume} {9}},\ \bibinfo {pages}
  {9666} (\bibinfo {year} {2024})}\BibitemShut {NoStop}%
\bibitem [{\citenamefont {Cangemi}\ \emph {et~al.}(2024)\citenamefont
  {Cangemi}, \citenamefont {Bhadra},\ and\ \citenamefont
  {Levy}}]{CANGEMI20241}%
  \BibitemOpen
  \bibfield  {author} {\bibinfo {author} {\bibfnamefont {L.~M.}\ \bibnamefont
  {Cangemi}}, \bibinfo {author} {\bibfnamefont {C.}~\bibnamefont {Bhadra}},\
  and\ \bibinfo {author} {\bibfnamefont {A.}~\bibnamefont {Levy}},\ }\bibfield
  {title} {\bibinfo {title} {Quantum engines and refrigerators},\ }\href
  {https://doi.org/https://doi.org/10.1016/j.physrep.2024.07.001} {\bibfield
  {journal} {\bibinfo  {journal} {Phys. Rep.}\ }\textbf {\bibinfo {volume}
  {1087}},\ \bibinfo {pages} {1} (\bibinfo {year} {2024})}\BibitemShut
  {NoStop}%
\bibitem [{\citenamefont {Link}\ \emph {et~al.}(2023)\citenamefont {Link},
  \citenamefont {Luoma},\ and\ \citenamefont {Strunz}}]{Link_2023}%
  \BibitemOpen
  \bibfield  {author} {\bibinfo {author} {\bibfnamefont {V.}~\bibnamefont
  {Link}}, \bibinfo {author} {\bibfnamefont {K.}~\bibnamefont {Luoma}},\ and\
  \bibinfo {author} {\bibfnamefont {W.~T.}\ \bibnamefont {Strunz}},\ }\bibfield
   {title} {\bibinfo {title} {Non-markovian quantum state diffusion for spin
  environments},\ }\href {https://doi.org/10.1088/1367-2630/aceff3} {\bibfield
  {journal} {\bibinfo  {journal} {New J. Phys.}\ }\textbf {\bibinfo {volume}
  {25}},\ \bibinfo {pages} {093006} (\bibinfo {year} {2023})}\BibitemShut
  {NoStop}%
\bibitem [{\citenamefont {Buscemi}\ \emph {et~al.}(2025)\citenamefont
  {Buscemi}, \citenamefont {Gangwar}, \citenamefont {Goswami}, \citenamefont
  {Badhani}, \citenamefont {Pandit}, \citenamefont {Mohan}, \citenamefont
  {Das},\ and\ \citenamefont {Bera}}]{PRXQuantum.6.020316}%
  \BibitemOpen
  \bibfield  {author} {\bibinfo {author} {\bibfnamefont {F.}~\bibnamefont
  {Buscemi}}, \bibinfo {author} {\bibfnamefont {R.}~\bibnamefont {Gangwar}},
  \bibinfo {author} {\bibfnamefont {K.}~\bibnamefont {Goswami}}, \bibinfo
  {author} {\bibfnamefont {H.}~\bibnamefont {Badhani}}, \bibinfo {author}
  {\bibfnamefont {T.}~\bibnamefont {Pandit}}, \bibinfo {author} {\bibfnamefont
  {B.}~\bibnamefont {Mohan}}, \bibinfo {author} {\bibfnamefont
  {S.}~\bibnamefont {Das}},\ and\ \bibinfo {author} {\bibfnamefont {M.~N.}\
  \bibnamefont {Bera}},\ }\bibfield  {title} {\bibinfo {title} {Causal and
  noncausal revivals of information: A new regime of non-markovianity in
  quantum stochastic processes},\ }\href
  {https://doi.org/10.1103/PRXQuantum.6.020316} {\bibfield  {journal} {\bibinfo
   {journal} {PRX Quantum}\ }\textbf {\bibinfo {volume} {6}},\ \bibinfo {pages}
  {020316} (\bibinfo {year} {2025})}\BibitemShut {NoStop}%
\bibitem [{\citenamefont {Cao}\ \emph {et~al.}(2020)\citenamefont {Cao} \emph
  {et~al.}}]{Cao:2020pup}%
  \BibitemOpen
  \bibfield  {author} {\bibinfo {author} {\bibfnamefont {J.}~\bibnamefont
  {Cao}} \emph {et~al.},\ }\bibfield  {title} {\bibinfo {title} {{Quantum
  biology revisited}},\ }\href {https://doi.org/10.1126/sciadv.aaz4888}
  {\bibfield  {journal} {\bibinfo  {journal} {Sci. Adv.}\ }\textbf {\bibinfo
  {volume} {6}},\ \bibinfo {pages} {eaaz4888} (\bibinfo {year}
  {2020})}\BibitemShut {NoStop}%
\bibitem [{\citenamefont {Altherr}\ and\ \citenamefont
  {Yang}(2021)}]{PhysRevLett.127.060501}%
  \BibitemOpen
  \bibfield  {author} {\bibinfo {author} {\bibfnamefont {A.}~\bibnamefont
  {Altherr}}\ and\ \bibinfo {author} {\bibfnamefont {Y.}~\bibnamefont {Yang}},\
  }\bibfield  {title} {\bibinfo {title} {Quantum metrology for non-markovian
  processes},\ }\href {https://doi.org/10.1103/PhysRevLett.127.060501}
  {\bibfield  {journal} {\bibinfo  {journal} {Phys. Rev. Lett.}\ }\textbf
  {\bibinfo {volume} {127}},\ \bibinfo {pages} {060501} (\bibinfo {year}
  {2021})}\BibitemShut {NoStop}%
\bibitem [{\citenamefont {Nakamura}\ and\ \citenamefont
  {Ankerhold}(2024{\natexlab{a}})}]{PhysRevResearch.6.033215}%
  \BibitemOpen
  \bibfield  {author} {\bibinfo {author} {\bibfnamefont {K.}~\bibnamefont
  {Nakamura}}\ and\ \bibinfo {author} {\bibfnamefont {J.}~\bibnamefont
  {Ankerhold}},\ }\bibfield  {title} {\bibinfo {title} {Gate operations for
  superconducting qubits and non-markovianity},\ }\href
  {https://doi.org/10.1103/PhysRevResearch.6.033215} {\bibfield  {journal}
  {\bibinfo  {journal} {Phys. Rev. Res.}\ }\textbf {\bibinfo {volume} {6}},\
  \bibinfo {pages} {033215} (\bibinfo {year} {2024}{\natexlab{a}})}\BibitemShut
  {NoStop}%
\bibitem [{\citenamefont {White}\ \emph {et~al.}(2020)\citenamefont {White},
  \citenamefont {Hill}, \citenamefont {Pollock}, \citenamefont {Hollenberg},\
  and\ \citenamefont {Modi}}]{white2020demonstration}%
  \BibitemOpen
  \bibfield  {author} {\bibinfo {author} {\bibfnamefont {G.~A.}\ \bibnamefont
  {White}}, \bibinfo {author} {\bibfnamefont {C.~D.}\ \bibnamefont {Hill}},
  \bibinfo {author} {\bibfnamefont {F.~A.}\ \bibnamefont {Pollock}}, \bibinfo
  {author} {\bibfnamefont {L.~C.}\ \bibnamefont {Hollenberg}},\ and\ \bibinfo
  {author} {\bibfnamefont {K.}~\bibnamefont {Modi}},\ }\bibfield  {title}
  {\bibinfo {title} {Demonstration of non-markovian process characterisation
  and control on a quantum processor},\ }\href
  {https://doi.org/10.1038/s41467-020-20113-3} {\bibfield  {journal} {\bibinfo
  {journal} {Nat. Commun.}\ }\textbf {\bibinfo {volume} {11}},\ \bibinfo
  {pages} {6301} (\bibinfo {year} {2020})}\BibitemShut {NoStop}%
\bibitem [{\citenamefont {Caruso}\ \emph
  {et~al.}(2014{\natexlab{b}})\citenamefont {Caruso}, \citenamefont
  {Giovannetti}, \citenamefont {Lupo},\ and\ \citenamefont
  {Mancini}}]{caruso2014quantum}%
  \BibitemOpen
  \bibfield  {author} {\bibinfo {author} {\bibfnamefont {F.}~\bibnamefont
  {Caruso}}, \bibinfo {author} {\bibfnamefont {V.}~\bibnamefont {Giovannetti}},
  \bibinfo {author} {\bibfnamefont {C.}~\bibnamefont {Lupo}},\ and\ \bibinfo
  {author} {\bibfnamefont {S.}~\bibnamefont {Mancini}},\ }\bibfield  {title}
  {\bibinfo {title} {Quantum channels and memory effects},\ }\href@noop {}
  {\bibfield  {journal} {\bibinfo  {journal} {Reviews of Modern Physics}\
  }\textbf {\bibinfo {volume} {86}},\ \bibinfo {pages} {1203} (\bibinfo {year}
  {2014}{\natexlab{b}})}\BibitemShut {NoStop}%
\bibitem [{\citenamefont {Prakash}\ and\ \citenamefont
  {Hebbe~Madhusudhana}(2024)}]{PhysRevResearch.6.043127}%
  \BibitemOpen
  \bibfield  {author} {\bibinfo {author} {\bibfnamefont {A.}~\bibnamefont
  {Prakash}}\ and\ \bibinfo {author} {\bibfnamefont {B.}~\bibnamefont
  {Hebbe~Madhusudhana}},\ }\bibfield  {title} {\bibinfo {title} {Characterizing
  non-markovian and coherent errors in quantum simulation},\ }\href
  {https://doi.org/10.1103/PhysRevResearch.6.043127} {\bibfield  {journal}
  {\bibinfo  {journal} {Phys. Rev. Res.}\ }\textbf {\bibinfo {volume} {6}},\
  \bibinfo {pages} {043127} (\bibinfo {year} {2024})}\BibitemShut {NoStop}%
\bibitem [{\citenamefont {Kamin}\ \emph {et~al.}(2020)\citenamefont {Kamin},
  \citenamefont {Tabesh}, \citenamefont {Salimi}, \citenamefont {Kheirandish},\
  and\ \citenamefont {Santos}}]{Kamin_2020}%
  \BibitemOpen
  \bibfield  {author} {\bibinfo {author} {\bibfnamefont {F.~H.}\ \bibnamefont
  {Kamin}}, \bibinfo {author} {\bibfnamefont {F.~T.}\ \bibnamefont {Tabesh}},
  \bibinfo {author} {\bibfnamefont {S.}~\bibnamefont {Salimi}}, \bibinfo
  {author} {\bibfnamefont {F.}~\bibnamefont {Kheirandish}},\ and\ \bibinfo
  {author} {\bibfnamefont {A.~C.}\ \bibnamefont {Santos}},\ }\bibfield  {title}
  {\bibinfo {title} {Non-markovian effects on charging and self-discharging
  process of quantum batteries},\ }\href
  {https://doi.org/10.1088/1367-2630/ab9ee2} {\bibfield  {journal} {\bibinfo
  {journal} {New J. Phys.}\ }\textbf {\bibinfo {volume} {22}},\ \bibinfo
  {pages} {083007} (\bibinfo {year} {2020})}\BibitemShut {NoStop}%
\bibitem [{\citenamefont {Chen}\ \emph {et~al.}(2024)\citenamefont {Chen},
  \citenamefont {Yang}, \citenamefont {Couvertier}, \citenamefont {Ding},
  \citenamefont {Chatterjee},\ and\ \citenamefont {Yu}}]{e26090742}%
  \BibitemOpen
  \bibfield  {author} {\bibinfo {author} {\bibfnamefont {P.}~\bibnamefont
  {Chen}}, \bibinfo {author} {\bibfnamefont {N.}~\bibnamefont {Yang}}, \bibinfo
  {author} {\bibfnamefont {A.}~\bibnamefont {Couvertier}}, \bibinfo {author}
  {\bibfnamefont {Q.}~\bibnamefont {Ding}}, \bibinfo {author} {\bibfnamefont
  {R.}~\bibnamefont {Chatterjee}},\ and\ \bibinfo {author} {\bibfnamefont
  {T.}~\bibnamefont {Yu}},\ }\bibfield  {title} {\bibinfo {title} {Chaos in
  optomechanical systems coupled to a non-markovian environment},\ }\href
  {https://www.mdpi.com/1099-4300/26/9/742} {\bibfield  {journal} {\bibinfo
  {journal} {Entropy}\ }\textbf {\bibinfo {volume} {26}} (\bibinfo {year}
  {2024})}\BibitemShut {NoStop}%
\bibitem [{\citenamefont {Roy}\ \emph {et~al.}(2024)\citenamefont {Roy},
  \citenamefont {Bera}, \citenamefont {Gupta},\ and\ \citenamefont
  {Majumdar}}]{Roy_2024}%
  \BibitemOpen
  \bibfield  {author} {\bibinfo {author} {\bibfnamefont {P.}~\bibnamefont
  {Roy}}, \bibinfo {author} {\bibfnamefont {S.}~\bibnamefont {Bera}}, \bibinfo
  {author} {\bibfnamefont {S.}~\bibnamefont {Gupta}},\ and\ \bibinfo {author}
  {\bibfnamefont {A.~S.}\ \bibnamefont {Majumdar}},\ }\bibfield  {title}
  {\bibinfo {title} {Device-independent quantum secure direct communication
  under non-markovian quantum channels},\ }\bibfield  {journal} {\bibinfo
  {journal} {Quantum Inf. Process.}\ }\textbf {\bibinfo {volume} {23}},\ \href
  {https://doi.org/10.1007/s11128-024-04397-8} {10.1007/s11128-024-04397-8}
  (\bibinfo {year} {2024})\BibitemShut {NoStop}%
\bibitem [{\citenamefont {Nakamura}\ and\ \citenamefont
  {Ankerhold}(2024{\natexlab{b}})}]{PhysRevB.109.014315}%
  \BibitemOpen
  \bibfield  {author} {\bibinfo {author} {\bibfnamefont {K.}~\bibnamefont
  {Nakamura}}\ and\ \bibinfo {author} {\bibfnamefont {J.}~\bibnamefont
  {Ankerhold}},\ }\bibfield  {title} {\bibinfo {title} {Qubit dynamics beyond
  lindblad: Non-markovianity versus rotating wave approximation},\ }\href
  {https://doi.org/10.1103/PhysRevB.109.014315} {\bibfield  {journal} {\bibinfo
   {journal} {Phys. Rev. B}\ }\textbf {\bibinfo {volume} {109}},\ \bibinfo
  {pages} {014315} (\bibinfo {year} {2024}{\natexlab{b}})}\BibitemShut
  {NoStop}%
\bibitem [{\citenamefont {Figueroa-Romero}\ \emph {et~al.}(2021)\citenamefont
  {Figueroa-Romero}, \citenamefont {Modi}, \citenamefont {Harris},
  \citenamefont {Stace},\ and\ \citenamefont {Hsieh}}]{PRXQuantum.2.040351}%
  \BibitemOpen
  \bibfield  {author} {\bibinfo {author} {\bibfnamefont {P.}~\bibnamefont
  {Figueroa-Romero}}, \bibinfo {author} {\bibfnamefont {K.}~\bibnamefont
  {Modi}}, \bibinfo {author} {\bibfnamefont {R.~J.}\ \bibnamefont {Harris}},
  \bibinfo {author} {\bibfnamefont {T.~M.}\ \bibnamefont {Stace}},\ and\
  \bibinfo {author} {\bibfnamefont {M.-H.}\ \bibnamefont {Hsieh}},\ }\bibfield
  {title} {\bibinfo {title} {Randomized benchmarking for non-markovian noise},\
  }\href {https://doi.org/10.1103/PRXQuantum.2.040351} {\bibfield  {journal}
  {\bibinfo  {journal} {PRX Quantum}\ }\textbf {\bibinfo {volume} {2}},\
  \bibinfo {pages} {040351} (\bibinfo {year} {2021})}\BibitemShut {NoStop}%
\bibitem [{\citenamefont {Gaikwad}\ \emph {et~al.}(2024)\citenamefont
  {Gaikwad}, \citenamefont {Kowsari}, \citenamefont {Brame}, \citenamefont
  {Song}, \citenamefont {Zhang}, \citenamefont {Esposito}, \citenamefont
  {Ranadive}, \citenamefont {Cappelli}, \citenamefont {Roch}, \citenamefont
  {Levenson-Falk},\ and\ \citenamefont {Murch}}]{PhysRevLett.132.200401}%
  \BibitemOpen
  \bibfield  {author} {\bibinfo {author} {\bibfnamefont {C.}~\bibnamefont
  {Gaikwad}}, \bibinfo {author} {\bibfnamefont {D.}~\bibnamefont {Kowsari}},
  \bibinfo {author} {\bibfnamefont {C.}~\bibnamefont {Brame}}, \bibinfo
  {author} {\bibfnamefont {X.}~\bibnamefont {Song}}, \bibinfo {author}
  {\bibfnamefont {H.}~\bibnamefont {Zhang}}, \bibinfo {author} {\bibfnamefont
  {M.}~\bibnamefont {Esposito}}, \bibinfo {author} {\bibfnamefont
  {A.}~\bibnamefont {Ranadive}}, \bibinfo {author} {\bibfnamefont
  {G.}~\bibnamefont {Cappelli}}, \bibinfo {author} {\bibfnamefont
  {N.}~\bibnamefont {Roch}}, \bibinfo {author} {\bibfnamefont {E.~M.}\
  \bibnamefont {Levenson-Falk}},\ and\ \bibinfo {author} {\bibfnamefont
  {K.~W.}\ \bibnamefont {Murch}},\ }\bibfield  {title} {\bibinfo {title}
  {Entanglement assisted probe of the non-markovian to markovian transition in
  open quantum system dynamics},\ }\href
  {https://doi.org/10.1103/PhysRevLett.132.200401} {\bibfield  {journal}
  {\bibinfo  {journal} {Phys. Rev. Lett.}\ }\textbf {\bibinfo {volume} {132}},\
  \bibinfo {pages} {200401} (\bibinfo {year} {2024})}\BibitemShut {NoStop}%
\bibitem [{\citenamefont {Breuer}\ \emph {et~al.}(2009)\citenamefont {Breuer},
  \citenamefont {Laine},\ and\ \citenamefont {Piilo}}]{PhysRevLett.103.210401}%
  \BibitemOpen
  \bibfield  {author} {\bibinfo {author} {\bibfnamefont {H.-P.}\ \bibnamefont
  {Breuer}}, \bibinfo {author} {\bibfnamefont {E.-M.}\ \bibnamefont {Laine}},\
  and\ \bibinfo {author} {\bibfnamefont {J.}~\bibnamefont {Piilo}},\ }\bibfield
   {title} {\bibinfo {title} {Measure for the degree of non-markovian behavior
  of quantum processes in open systems},\ }\href
  {https://doi.org/10.1103/PhysRevLett.103.210401} {\bibfield  {journal}
  {\bibinfo  {journal} {Phys. Rev. Lett.}\ }\textbf {\bibinfo {volume} {103}},\
  \bibinfo {pages} {210401} (\bibinfo {year} {2009})}\BibitemShut {NoStop}%
\bibitem [{\citenamefont {Settimo}\ \emph {et~al.}(2022)\citenamefont
  {Settimo}, \citenamefont {Breuer},\ and\ \citenamefont
  {Vacchini}}]{PhysRevA.106.042212}%
  \BibitemOpen
  \bibfield  {author} {\bibinfo {author} {\bibfnamefont {F.}~\bibnamefont
  {Settimo}}, \bibinfo {author} {\bibfnamefont {H.-P.}\ \bibnamefont
  {Breuer}},\ and\ \bibinfo {author} {\bibfnamefont {B.}~\bibnamefont
  {Vacchini}},\ }\bibfield  {title} {\bibinfo {title} {Entropic and
  trace-distance-based measures of non-markovianity},\ }\href
  {https://doi.org/10.1103/PhysRevA.106.042212} {\bibfield  {journal} {\bibinfo
   {journal} {Phys. Rev. A}\ }\textbf {\bibinfo {volume} {106}},\ \bibinfo
  {pages} {042212} (\bibinfo {year} {2022})}\BibitemShut {NoStop}%
\bibitem [{\citenamefont {Amato}\ \emph {et~al.}(2018)\citenamefont {Amato},
  \citenamefont {Breuer},\ and\ \citenamefont {Vacchini}}]{PhysRevA.98.012120}%
  \BibitemOpen
  \bibfield  {author} {\bibinfo {author} {\bibfnamefont {G.}~\bibnamefont
  {Amato}}, \bibinfo {author} {\bibfnamefont {H.-P.}\ \bibnamefont {Breuer}},\
  and\ \bibinfo {author} {\bibfnamefont {B.}~\bibnamefont {Vacchini}},\
  }\bibfield  {title} {\bibinfo {title} {Generalized trace distance approach to
  quantum non-markovianity and detection of initial correlations},\ }\href
  {https://doi.org/10.1103/PhysRevA.98.012120} {\bibfield  {journal} {\bibinfo
  {journal} {Phys. Rev. A}\ }\textbf {\bibinfo {volume} {98}},\ \bibinfo
  {pages} {012120} (\bibinfo {year} {2018})}\BibitemShut {NoStop}%
\bibitem [{\citenamefont {Laine}\ \emph {et~al.}(2010)\citenamefont {Laine},
  \citenamefont {Piilo},\ and\ \citenamefont {Breuer}}]{PhysRevA.81.062115}%
  \BibitemOpen
  \bibfield  {author} {\bibinfo {author} {\bibfnamefont {E.-M.}\ \bibnamefont
  {Laine}}, \bibinfo {author} {\bibfnamefont {J.}~\bibnamefont {Piilo}},\ and\
  \bibinfo {author} {\bibfnamefont {H.-P.}\ \bibnamefont {Breuer}},\ }\bibfield
   {title} {\bibinfo {title} {Measure for the non-markovianity of quantum
  processes},\ }\href {https://doi.org/10.1103/PhysRevA.81.062115} {\bibfield
  {journal} {\bibinfo  {journal} {Phys. Rev. A}\ }\textbf {\bibinfo {volume}
  {81}},\ \bibinfo {pages} {062115} (\bibinfo {year} {2010})}\BibitemShut
  {NoStop}%
\bibitem [{\citenamefont {Ángel Rivas}\ \emph {et~al.}(2014)\citenamefont
  {Ángel Rivas}, \citenamefont {Huelga},\ and\ \citenamefont
  {Plenio}}]{Rivas_2014}%
  \BibitemOpen
  \bibfield  {author} {\bibinfo {author} {\bibnamefont {Ángel Rivas}},
  \bibinfo {author} {\bibfnamefont {S.~F.}\ \bibnamefont {Huelga}},\ and\
  \bibinfo {author} {\bibfnamefont {M.~B.}\ \bibnamefont {Plenio}},\ }\bibfield
   {title} {\bibinfo {title} {Quantum non-markovianity: characterization,
  quantification and detection},\ }\href
  {https://doi.org/10.1088/0034-4885/77/9/094001} {\bibfield  {journal}
  {\bibinfo  {journal} {Rep. Prog. Phys.}\ }\textbf {\bibinfo {volume} {77}},\
  \bibinfo {pages} {094001} (\bibinfo {year} {2014})}\BibitemShut {NoStop}%
\bibitem [{\citenamefont {Megier}\ \emph {et~al.}(2021)\citenamefont {Megier},
  \citenamefont {Smirne},\ and\ \citenamefont
  {Vacchini}}]{PhysRevLett.127.030401}%
  \BibitemOpen
  \bibfield  {author} {\bibinfo {author} {\bibfnamefont {N.}~\bibnamefont
  {Megier}}, \bibinfo {author} {\bibfnamefont {A.}~\bibnamefont {Smirne}},\
  and\ \bibinfo {author} {\bibfnamefont {B.}~\bibnamefont {Vacchini}},\
  }\bibfield  {title} {\bibinfo {title} {Entropic bounds on information
  backflow},\ }\href {https://doi.org/10.1103/PhysRevLett.127.030401}
  {\bibfield  {journal} {\bibinfo  {journal} {Phys. Rev. Lett.}\ }\textbf
  {\bibinfo {volume} {127}},\ \bibinfo {pages} {030401} (\bibinfo {year}
  {2021})}\BibitemShut {NoStop}%
\bibitem [{\citenamefont {Mallick}\ \emph {et~al.}(2024)\citenamefont
  {Mallick}, \citenamefont {Mukherjee}, \citenamefont {Maity},\ and\
  \citenamefont {Majumdar}}]{PhysRevA.109.022247}%
  \BibitemOpen
  \bibfield  {author} {\bibinfo {author} {\bibfnamefont {B.}~\bibnamefont
  {Mallick}}, \bibinfo {author} {\bibfnamefont {S.}~\bibnamefont {Mukherjee}},
  \bibinfo {author} {\bibfnamefont {A.~G.}\ \bibnamefont {Maity}},\ and\
  \bibinfo {author} {\bibfnamefont {A.~S.}\ \bibnamefont {Majumdar}},\
  }\bibfield  {title} {\bibinfo {title} {Assessing non-markovian dynamics
  through moments of the choi state},\ }\href
  {https://doi.org/10.1103/PhysRevA.109.022247} {\bibfield  {journal} {\bibinfo
   {journal} {Phys. Rev. A}\ }\textbf {\bibinfo {volume} {109}},\ \bibinfo
  {pages} {022247} (\bibinfo {year} {2024})}\BibitemShut {NoStop}%
\bibitem [{\citenamefont {Maity}\ \emph {et~al.}(2020)\citenamefont {Maity},
  \citenamefont {Bhattacharya},\ and\ \citenamefont {Majumdar}}]{Maity_2020}%
  \BibitemOpen
  \bibfield  {author} {\bibinfo {author} {\bibfnamefont {A.~G.}\ \bibnamefont
  {Maity}}, \bibinfo {author} {\bibfnamefont {S.}~\bibnamefont
  {Bhattacharya}},\ and\ \bibinfo {author} {\bibfnamefont {A.~S.}\ \bibnamefont
  {Majumdar}},\ }\bibfield  {title} {\bibinfo {title} {Detecting
  non-markovianity via uncertainty relations},\ }\href
  {https://doi.org/10.1088/1751-8121/ab7135} {\bibfield  {journal} {\bibinfo
  {journal} {J. Phys. A: Math. Theor.}\ }\textbf {\bibinfo {volume} {53}},\
  \bibinfo {pages} {175301} (\bibinfo {year} {2020})}\BibitemShut {NoStop}%
\bibitem [{\citenamefont {Lu}\ \emph {et~al.}(2010)\citenamefont {Lu},
  \citenamefont {Wang},\ and\ \citenamefont {Sun}}]{PhysRevA.82.042103}%
  \BibitemOpen
  \bibfield  {author} {\bibinfo {author} {\bibfnamefont {X.-M.}\ \bibnamefont
  {Lu}}, \bibinfo {author} {\bibfnamefont {X.}~\bibnamefont {Wang}},\ and\
  \bibinfo {author} {\bibfnamefont {C.~P.}\ \bibnamefont {Sun}},\ }\bibfield
  {title} {\bibinfo {title} {Quantum fisher information flow and non-markovian
  processes of open systems},\ }\href
  {https://doi.org/10.1103/PhysRevA.82.042103} {\bibfield  {journal} {\bibinfo
  {journal} {Phys. Rev. A}\ }\textbf {\bibinfo {volume} {82}},\ \bibinfo
  {pages} {042103} (\bibinfo {year} {2010})}\BibitemShut {NoStop}%
\bibitem [{\citenamefont {Rashid}\ \emph {et~al.}(2024)\citenamefont {Rashid},
  \citenamefont {Lone},\ and\ \citenamefont {Ganai}}]{Rashid_2024}%
  \BibitemOpen
  \bibfield  {author} {\bibinfo {author} {\bibfnamefont {M.}~\bibnamefont
  {Rashid}}, \bibinfo {author} {\bibfnamefont {M.~Q.}\ \bibnamefont {Lone}},\
  and\ \bibinfo {author} {\bibfnamefont {P.~A.}\ \bibnamefont {Ganai}},\
  }\bibfield  {title} {\bibinfo {title} {Quantum non-markovianity of a qubit in
  presence of state dependent bath},\ }\href
  {https://doi.org/10.1088/1402-4896/ad31ee} {\bibfield  {journal} {\bibinfo
  {journal} {Phys. Scr.}\ }\textbf {\bibinfo {volume} {99}},\ \bibinfo {pages}
  {045117} (\bibinfo {year} {2024})}\BibitemShut {NoStop}%
\bibitem [{\citenamefont {Abiuso}\ \emph {et~al.}(2023)\citenamefont {Abiuso},
  \citenamefont {Scandi}, \citenamefont {De~Santis},\ and\ \citenamefont
  {Surace}}]{abiuso2023characterizing}%
  \BibitemOpen
  \bibfield  {author} {\bibinfo {author} {\bibfnamefont {P.}~\bibnamefont
  {Abiuso}}, \bibinfo {author} {\bibfnamefont {M.}~\bibnamefont {Scandi}},
  \bibinfo {author} {\bibfnamefont {D.}~\bibnamefont {De~Santis}},\ and\
  \bibinfo {author} {\bibfnamefont {J.}~\bibnamefont {Surace}},\ }\bibfield
  {title} {\bibinfo {title} {Characterizing (non-) markovianity through fisher
  information},\ }\href {https://doi.org/10.21468/SciPostPhys.15.1.014}
  {\bibfield  {journal} {\bibinfo  {journal} {SciPost Phys.}\ }\textbf
  {\bibinfo {volume} {15}},\ \bibinfo {pages} {014} (\bibinfo {year}
  {2023})}\BibitemShut {NoStop}%
\bibitem [{\citenamefont {El~Anouz}\ \emph {et~al.}(2020)\citenamefont
  {El~Anouz}, \citenamefont {El~Allati},\ and\ \citenamefont
  {Metwally}}]{el2020different}%
  \BibitemOpen
  \bibfield  {author} {\bibinfo {author} {\bibfnamefont {K.}~\bibnamefont
  {El~Anouz}}, \bibinfo {author} {\bibfnamefont {A.}~\bibnamefont
  {El~Allati}},\ and\ \bibinfo {author} {\bibfnamefont {N.}~\bibnamefont
  {Metwally}},\ }\bibfield  {title} {\bibinfo {title} {Different indicators for
  markovian and non-markovian dynamics},\ }\href
  {https://doi.org/10.1016/j.physleta.2019.126122} {\bibfield  {journal}
  {\bibinfo  {journal} {Physics Letters A}\ }\textbf {\bibinfo {volume}
  {384}},\ \bibinfo {pages} {126122} (\bibinfo {year} {2020})}\BibitemShut
  {NoStop}%
\bibitem [{\citenamefont {Addis}\ \emph {et~al.}(2014)\citenamefont {Addis},
  \citenamefont {Bylicka}, \citenamefont {Chru\ifmmode \acute{s}\else
  \'{s}\fi{}ci\ifmmode~\acute{n}\else \'{n}\fi{}ski},\ and\ \citenamefont
  {Maniscalco}}]{PhysRevA.90.052103}%
  \BibitemOpen
  \bibfield  {author} {\bibinfo {author} {\bibfnamefont {C.}~\bibnamefont
  {Addis}}, \bibinfo {author} {\bibfnamefont {B.}~\bibnamefont {Bylicka}},
  \bibinfo {author} {\bibfnamefont {D.}~\bibnamefont {Chru\ifmmode
  \acute{s}\else \'{s}\fi{}ci\ifmmode~\acute{n}\else \'{n}\fi{}ski}},\ and\
  \bibinfo {author} {\bibfnamefont {S.}~\bibnamefont {Maniscalco}},\ }\bibfield
   {title} {\bibinfo {title} {Comparative study of non-markovianity measures in
  exactly solvable one- and two-qubit models},\ }\href
  {https://doi.org/10.1103/PhysRevA.90.052103} {\bibfield  {journal} {\bibinfo
  {journal} {Phys. Rev. A}\ }\textbf {\bibinfo {volume} {90}},\ \bibinfo
  {pages} {052103} (\bibinfo {year} {2014})}\BibitemShut {NoStop}%
\bibitem [{\citenamefont {Addis}\ \emph {et~al.}(2013)\citenamefont {Addis},
  \citenamefont {Haikka}, \citenamefont {McEndoo}, \citenamefont
  {Macchiavello},\ and\ \citenamefont {Maniscalco}}]{addis2013two}%
  \BibitemOpen
  \bibfield  {author} {\bibinfo {author} {\bibfnamefont {C.}~\bibnamefont
  {Addis}}, \bibinfo {author} {\bibfnamefont {P.}~\bibnamefont {Haikka}},
  \bibinfo {author} {\bibfnamefont {S.}~\bibnamefont {McEndoo}}, \bibinfo
  {author} {\bibfnamefont {C.}~\bibnamefont {Macchiavello}},\ and\ \bibinfo
  {author} {\bibfnamefont {S.}~\bibnamefont {Maniscalco}},\ }\bibfield  {title}
  {\bibinfo {title} {Two-qubit non-markovianity induced by a common
  environment},\ }\href {https://doi.org/10.1103/PhysRevA.87.052109} {\bibfield
   {journal} {\bibinfo  {journal} {Phys. Rev. A}\ }\textbf {\bibinfo {volume}
  {87}},\ \bibinfo {pages} {052109} (\bibinfo {year} {2013})}\BibitemShut
  {NoStop}%
\bibitem [{\citenamefont {Mirza}\ and\ \citenamefont
  {Chaudhry}(2024)}]{mirza2024improving}%
  \BibitemOpen
  \bibfield  {author} {\bibinfo {author} {\bibfnamefont {A.~R.}\ \bibnamefont
  {Mirza}}\ and\ \bibinfo {author} {\bibfnamefont {A.~Z.}\ \bibnamefont
  {Chaudhry}},\ }\bibfield  {title} {\bibinfo {title} {Improving the estimation
  of environment parameters via a two-qubit scheme},\ }\href
  {https://doi.org/10.1038/s41598-024-57150-7} {\bibfield  {journal} {\bibinfo
  {journal} {Sci. Rep.}\ }\textbf {\bibinfo {volume} {14}},\ \bibinfo {pages}
  {6803} (\bibinfo {year} {2024})}\BibitemShut {NoStop}%
\bibitem [{\citenamefont {Camati}\ \emph {et~al.}(2020)\citenamefont {Camati},
  \citenamefont {Santos},\ and\ \citenamefont {Serra}}]{SerraPRA2020}%
  \BibitemOpen
  \bibfield  {author} {\bibinfo {author} {\bibfnamefont {P.~A.}\ \bibnamefont
  {Camati}}, \bibinfo {author} {\bibfnamefont {J.~F.~G.}\ \bibnamefont
  {Santos}},\ and\ \bibinfo {author} {\bibfnamefont {R.~M.}\ \bibnamefont
  {Serra}},\ }\bibfield  {title} {\bibinfo {title} {Employing non-markovian
  effects to improve the performance of a quantum otto refrigerator},\ }\href
  {https://doi.org/10.1103/PhysRevA.102.012217} {\bibfield  {journal} {\bibinfo
   {journal} {Phys. Rev. A}\ }\textbf {\bibinfo {volume} {102}},\ \bibinfo
  {pages} {012217} (\bibinfo {year} {2020})}\BibitemShut {NoStop}%
\bibitem [{\citenamefont {Aiache}\ \emph {et~al.}(2024)\citenamefont {Aiache},
  \citenamefont {Seida}, \citenamefont {El~Anouz},\ and\ \citenamefont
  {El~Allati}}]{AllatiPRE2024}%
  \BibitemOpen
  \bibfield  {author} {\bibinfo {author} {\bibfnamefont {Y.}~\bibnamefont
  {Aiache}}, \bibinfo {author} {\bibfnamefont {C.}~\bibnamefont {Seida}},
  \bibinfo {author} {\bibfnamefont {K.}~\bibnamefont {El~Anouz}},\ and\
  \bibinfo {author} {\bibfnamefont {A.}~\bibnamefont {El~Allati}},\ }\bibfield
  {title} {\bibinfo {title} {Non-markovian enhancement of nonequilibrium
  quantum thermometry},\ }\href {https://doi.org/10.1103/PhysRevE.110.024132}
  {\bibfield  {journal} {\bibinfo  {journal} {Phys. Rev. E}\ }\textbf {\bibinfo
  {volume} {110}},\ \bibinfo {pages} {024132} (\bibinfo {year}
  {2024})}\BibitemShut {NoStop}%
\bibitem [{\citenamefont {Aiache}\ \emph {et~al.}(2025)\citenamefont {Aiache},
  \citenamefont {Ullah}, \citenamefont {M\"ustecapl\ifmmode \imath \else \i
  \fi{}o\ifmmode~\breve{g}\else \u{g}\fi{}lu},\ and\ \citenamefont
  {Allati}}]{AllatiPRA2025}%
  \BibitemOpen
  \bibfield  {author} {\bibinfo {author} {\bibfnamefont {Y.}~\bibnamefont
  {Aiache}}, \bibinfo {author} {\bibfnamefont {A.}~\bibnamefont {Ullah}},
  \bibinfo {author} {\bibfnamefont {O.~E.}\ \bibnamefont {M\"ustecapl\ifmmode
  \imath \else \i \fi{}o\ifmmode~\breve{g}\else \u{g}\fi{}lu}},\ and\ \bibinfo
  {author} {\bibfnamefont {A.~E.}\ \bibnamefont {Allati}},\ }\bibfield  {title}
  {\bibinfo {title} {Quantum metrology of a structured reservoir},\ }\href
  {https://doi.org/10.1103/ypp6-5bt9} {\bibfield  {journal} {\bibinfo
  {journal} {Phys. Rev. A}\ }\textbf {\bibinfo {volume} {111}},\ \bibinfo
  {pages} {062619} (\bibinfo {year} {2025})}\BibitemShut {NoStop}%
\bibitem [{\citenamefont {Kurt}(2023)}]{KurtEntropy2023}%
  \BibitemOpen
  \bibfield  {author} {\bibinfo {author} {\bibfnamefont {A.}~\bibnamefont
  {Kurt}},\ }\bibfield  {title} {\bibinfo {title} {Interplay between
  non-markovianity of noise and dynamics in quantum systems},\ }\href
  {https://doi.org/10.3390/e25030501} {\bibfield  {journal} {\bibinfo
  {journal} {Entropy}\ }\textbf {\bibinfo {volume} {25}},\ \bibinfo {pages}
  {501} (\bibinfo {year} {2023})}\BibitemShut {NoStop}%
\bibitem [{\citenamefont {Rivas}\ \emph {et~al.}(2010)\citenamefont {Rivas},
  \citenamefont {Huelga},\ and\ \citenamefont {Plenio}}]{RivasPRL2010}%
  \BibitemOpen
  \bibfield  {author} {\bibinfo {author} {\bibfnamefont {A.}~\bibnamefont
  {Rivas}}, \bibinfo {author} {\bibfnamefont {S.~F.}\ \bibnamefont {Huelga}},\
  and\ \bibinfo {author} {\bibfnamefont {M.~B.}\ \bibnamefont {Plenio}},\
  }\bibfield  {title} {\bibinfo {title} {Entanglement and non-markovianity of
  quantum evolutions},\ }\href {https://doi.org/10.1103/PhysRevLett.105.050403}
  {\bibfield  {journal} {\bibinfo  {journal} {Phys. Rev. Lett.}\ }\textbf
  {\bibinfo {volume} {105}},\ \bibinfo {pages} {050403} (\bibinfo {year}
  {2010})}\BibitemShut {NoStop}%
\bibitem [{\citenamefont {Chaudhry}\ and\ \citenamefont
  {Gong}(2013)}]{ChaudhryPRA2013}%
  \BibitemOpen
  \bibfield  {author} {\bibinfo {author} {\bibfnamefont {A.~Z.}\ \bibnamefont
  {Chaudhry}}\ and\ \bibinfo {author} {\bibfnamefont {J.}~\bibnamefont
  {Gong}},\ }\bibfield  {title} {\bibinfo {title} {Amplification and
  suppression of system-bath-correlation effects in an open many-body system},\
  }\href {https://doi.org/10.1103/PhysRevA.87.012129} {\bibfield  {journal}
  {\bibinfo  {journal} {Phys. Rev. A}\ }\textbf {\bibinfo {volume} {87}},\
  \bibinfo {pages} {012129} (\bibinfo {year} {2013})}\BibitemShut {NoStop}%
\bibitem [{\citenamefont {Suárez}\ and\ \citenamefont
  {Silbey}(1991)}]{SilbeyJCP1991}%
  \BibitemOpen
  \bibfield  {author} {\bibinfo {author} {\bibfnamefont {A.}~\bibnamefont
  {Suárez}}\ and\ \bibinfo {author} {\bibfnamefont {R.}~\bibnamefont
  {Silbey}},\ }\bibfield  {title} {\bibinfo {title} {Properties of a
  macroscopic system as a thermal bath},\ }\href
  {https://doi.org/10.1063/1.461190} {\bibfield  {journal} {\bibinfo  {journal}
  {J. Chem. Phys.}\ }\textbf {\bibinfo {volume} {95}},\ \bibinfo {pages} {9115}
  (\bibinfo {year} {1991})}\BibitemShut {NoStop}%
\bibitem [{\citenamefont {Emary}\ and\ \citenamefont
  {Brandes}(2003)}]{EmaryPRE2003}%
  \BibitemOpen
  \bibfield  {author} {\bibinfo {author} {\bibfnamefont {C.}~\bibnamefont
  {Emary}}\ and\ \bibinfo {author} {\bibfnamefont {T.}~\bibnamefont
  {Brandes}},\ }\bibfield  {title} {\bibinfo {title} {Chaos and the quantum
  phase transition in the dicke model},\ }\href
  {https://doi.org/10.1103/PhysRevE.67.066203} {\bibfield  {journal} {\bibinfo
  {journal} {Phys. Rev. E}\ }\textbf {\bibinfo {volume} {67}},\ \bibinfo
  {pages} {066203} (\bibinfo {year} {2003})}\BibitemShut {NoStop}%
\bibitem [{\citenamefont {Vorrath}\ and\ \citenamefont
  {Brandes}(2005)}]{VorrathPRL2005}%
  \BibitemOpen
  \bibfield  {author} {\bibinfo {author} {\bibfnamefont {T.}~\bibnamefont
  {Vorrath}}\ and\ \bibinfo {author} {\bibfnamefont {T.}~\bibnamefont
  {Brandes}},\ }\bibfield  {title} {\bibinfo {title} {Dynamics of a large spin
  with strong dissipation},\ }\href
  {https://doi.org/10.1103/PhysRevLett.95.070402} {\bibfield  {journal}
  {\bibinfo  {journal} {Phys. Rev. Lett.}\ }\textbf {\bibinfo {volume} {95}},\
  \bibinfo {pages} {070402} (\bibinfo {year} {2005})}\BibitemShut {NoStop}%
\bibitem [{\citenamefont {Buluta}\ \emph {et~al.}(2011)\citenamefont {Buluta},
  \citenamefont {Ashhab},\ and\ \citenamefont {Nori}}]{NoriRepProg2011}%
  \BibitemOpen
  \bibfield  {author} {\bibinfo {author} {\bibfnamefont {I.}~\bibnamefont
  {Buluta}}, \bibinfo {author} {\bibfnamefont {S.}~\bibnamefont {Ashhab}},\
  and\ \bibinfo {author} {\bibfnamefont {F.}~\bibnamefont {Nori}},\ }\bibfield
  {title} {\bibinfo {title} {Natural and artificial atoms for quantum
  computation},\ }\href {https://doi.org/10.1088/0034-4885/74/10/104401}
  {\bibfield  {journal} {\bibinfo  {journal} {Rep. Prog. Phys.}\ }\textbf
  {\bibinfo {volume} {74}},\ \bibinfo {pages} {104401} (\bibinfo {year}
  {2011})}\BibitemShut {NoStop}%
\bibitem [{\citenamefont {Puri}(2001)}]{puri2001mathematical}%
  \BibitemOpen
  \bibfield  {author} {\bibinfo {author} {\bibfnamefont {R.~R.}\ \bibnamefont
  {Puri}},\ }\href@noop {} {\emph {\bibinfo {title} {Mathematical methods of
  quantum optics}}}\ (\bibinfo  {publisher} {Springer},\ \bibinfo {address}
  {Berlin},\ \bibinfo {year} {2001})\BibitemShut {NoStop}%
\bibitem [{\citenamefont {Blanes}\ \emph {et~al.}(2010)\citenamefont {Blanes},
  \citenamefont {Casas}, \citenamefont {Oteo},\ and\ \citenamefont
  {Ros}}]{Blanes_2010}%
  \BibitemOpen
  \bibfield  {author} {\bibinfo {author} {\bibfnamefont {S.}~\bibnamefont
  {Blanes}}, \bibinfo {author} {\bibfnamefont {F.}~\bibnamefont {Casas}},
  \bibinfo {author} {\bibfnamefont {J.~A.}\ \bibnamefont {Oteo}},\ and\
  \bibinfo {author} {\bibfnamefont {J.}~\bibnamefont {Ros}},\ }\bibfield
  {title} {\bibinfo {title} {A pedagogical approach to the magnus expansion},\
  }\href {https://doi.org/10.1088/0143-0807/31/4/020} {\bibfield  {journal}
  {\bibinfo  {journal} {Eur. J. Phys.}\ }\textbf {\bibinfo {volume} {31}},\
  \bibinfo {pages} {907} (\bibinfo {year} {2010})}\BibitemShut {NoStop}%
\bibitem [{\citenamefont {Mukherjee}\ \emph {et~al.}(2024)\citenamefont
  {Mukherjee}, \citenamefont {Mallick}, \citenamefont {Yanamandra},
  \citenamefont {Bhattacharya},\ and\ \citenamefont
  {Maity}}]{MukherjeePRA2024}%
  \BibitemOpen
  \bibfield  {author} {\bibinfo {author} {\bibfnamefont {S.}~\bibnamefont
  {Mukherjee}}, \bibinfo {author} {\bibfnamefont {B.}~\bibnamefont {Mallick}},
  \bibinfo {author} {\bibfnamefont {S.}~\bibnamefont {Yanamandra}}, \bibinfo
  {author} {\bibfnamefont {S.}~\bibnamefont {Bhattacharya}},\ and\ \bibinfo
  {author} {\bibfnamefont {A.~G.}\ \bibnamefont {Maity}},\ }\bibfield  {title}
  {\bibinfo {title} {Interplay between the hilbert-space dimension of a control
  system and the memory induced by a quantum switch},\ }\href
  {https://doi.org/10.1103/PhysRevA.110.042624} {\bibfield  {journal} {\bibinfo
   {journal} {Phys. Rev. A}\ }\textbf {\bibinfo {volume} {110}},\ \bibinfo
  {pages} {042624} (\bibinfo {year} {2024})}\BibitemShut {NoStop}%
\end{thebibliography}
\end{document}